\documentclass[thmsa,letterpaper]{article}
\usepackage{sw20lart}

\input{tcilatex}
\begin{document}

\author{Sawa Manoff\thanks{%
Permanent address: Bulgarian Academy of Sciences, Institute for Nuclear
Research and Nuclear Energy, Department of Theoretical\ Physics, Blvd.
Tzarigradsko Chaussee 72, 1784 Sofia, Bulgaria. E-mail address:
smanov@inrne.acad.bg} \\
Bogoliubov Laboratory of Theoretical Physics\\
Joint Institute for Nuclear Research\\
141980 Dubna, Russia}
\title{Spaces with contravariant and covariant affine connections and metrics}
\date{\_ . \_}
\maketitle

\begin{abstract}
\textit{The theory of spaces with different (not only by sign) contravariant
and covariant affine connections and metrics [}$(\overline{L}_n,g)$\textit{%
-spaces] is worked out within the framework of the tensor analysis over
differentiable manifolds and in a volume necessary for the further
considerations of the kinematics of vector fields and the Lagrangian theory
of tensor fields over }$(\overline{L}_n,g)$\textit{-spaces. The possibility
of introducing different (not only by sign) affine connections for
contravariant and covariant tensor fields over differentiable manifolds with
finite dimensions is discussed. The action of the deviation operator, having
an important role for deviation equations in gravitational physics, is
considered for the case of contravariant and covariant vector fields over
differentiable manifolds with different affine connections (called }$%
\overline{L}_n$\textit{-spaces). A deviation identity for contravariant
vector fields is obtained. The notions covariant, contravariant, covariant
projective and contravariant projective metric are introduced in (}$%
\overline{L}_n,g$\textit{)-spaces. The action of the covariant and the Lie
differential operator on the different type of metrics is found. The notions
of symmetric covariant and contravariant (Riemannian) connection are
determined and presented by means of the covariant and contravariant metric
and the corresponding torsion tensors. The different types of relative
tensor fields (tensor densities) as well as the invariant differential
operators acting on them are considered. The invariant volume element and
its properties under the action of different differential operators are
investigated.}
\end{abstract}

\tableofcontents

\section{Introduction}

In the present review, the differentiable manifolds with different (not only
by sign) contravariant and covariant affine connections and metrics [\textit{%
spaces with contravariant and covariant affine connections} and metrics, $(%
\overline{L}_n,g)$-spaces] are considered as models of the space-time. On
the grounds of the differential-geometric structures of the $(\overline{L}%
_n,g)$-spaces the \textit{kinematics of vector fields} and the \textit{%
dynamics of tensor fields} has been worked out as useful tools in
mathematical models for description of physical interactions and especially
the gravitational interaction in the modern gravitational physics. The
general results found for differentiable manifolds with different (not only
by sign) contravariant and covariant affine connections and metrics can be
specialized for spaces with one affine connection and a metric [the s. c. $%
(L_n,g)$-spaces] as well as for (pseudo) Riemannian spaces with or without
torsion [the s.c. $U_n$- and $V_n$-spaces]. The most results are given
either in index-free form or in a co-ordinate, or in a non-co-ordinate
basis. The main objects taken in such type of investigations can be given in
the following scheme

\begin{center}
\begin{eqnarray*}
&&\,\,\,\,\,\,\,\,\,\,\,\,\,\,\,\,\,\,\,\,\,\,\,\,\,\,{\bf Spaces\,with\,contravariant\,and\,covariant} \\
&&\ \ \ {\bf \ \,\,\,\,\,\,\,\,\,\,\,\,\,\,\,\,\,\,\,\,\,\,\,\,\,\,\,\,}\text{{\bf \thinspace \thinspace \thinspace }}{\bf affine\,connections\,and\,metrics} \\
&&\ \ \text{{\bf \thinspace \thinspace \thinspace \thinspace \thinspace
\thinspace \thinspace \thinspace \thinspace \thinspace \thinspace \thinspace
\thinspace \thinspace \thinspace \thinspace \thinspace \thinspace \thinspace
\thinspace \thinspace \thinspace \thinspace \thinspace \thinspace \thinspace
\thinspace \thinspace \thinspace \thinspace \thinspace \thinspace \thinspace
\thinspace \thinspace \thinspace \thinspace \thinspace \thinspace \thinspace
\thinspace \thinspace \thinspace \thinspace \thinspace \thinspace \thinspace
differential operators}} \\
&&\ \
\,\,\,\,\,\,\,\,\,\,\,\,\,\,\,\,\,\,\,\,\,\,\,\,\,\,\,\,\,\,\,\,\,\,\,\,\,\,\,\,\,\,\,\text{{\it covariant differential operator, }} \\
&&\ \
\,\,\,\,\,\,\,\,\,\,\,\,\,\,\,\,\,\,\,\,\,\,\,\,\,\,\,\,\,\,\,\,\,\,\,\,\,\,\,\,\,\,\,\,\,\,\,\,\,\,\,\,\,\,\,\,\text{{\it Lie differential operator,}}
\\
&&\ \
\,\,\,\,\,\,\,\,\,\,\,\,\,\,\,\,\,\,\,\,\,\,\,\,\,\,\,\,\,\,\,\,\,\,\,\,\,\,\,\,\,\,\,\,\,\,\,\,\,\,\,\,\,\,\,\,\,\,\text{{\it \ operator of curvature,}}
\\
&&\ \
\,\,\,\,\,\,\,\,\,\,\,\,\,\,\,\,\,\,\,\,\,\,\,\,\,\,\,\,\,\,\,\,\,\,\,\,\,\,\,\,\,\,\,\,\,\,\,\,\,\,\,\,\,\,\,\,\,\,\,\,\,\,\,\text{{\it deviation
operator,}} \\
&&\;\,\,\,\,\,\,\,\,\,\,\,\,\,\;\;\;\,\,\,\,\,\,\,\,\,\,\,\,\,\,\,\,\,\,\,\,\,\,\,\,\;\;\;\;\;\;\;\;\;\;\;\;\;\;\text{{\it extension operator,}} \\
&&\ \ \text{{\it \thinspace \thinspace \thinspace \thinspace \thinspace
\thinspace \thinspace \thinspace \thinspace \thinspace \thinspace \thinspace
\thinspace \thinspace \thinspace \thinspace \thinspace \thinspace \thinspace
\thinspace \thinspace \thinspace \thinspace \thinspace \thinspace \thinspace
\thinspace \thinspace \thinspace \thinspace \thinspace \thinspace \thinspace
\thinspace \thinspace \thinspace \thinspace \thinspace \thinspace \thinspace
\thinspace \thinspace \thinspace \thinspace \thinspace \thinspace \thinspace
\thinspace \thinspace \thinspace \thinspace \thinspace \thinspace \thinspace
\thinspace }{\bf affine connections, metrics,}} \\
&&\text{{\bf special tensor fields, tensor densities, invariant volume
element}}
\end{eqnarray*}
%

$\mid $

\begin{eqnarray*}
&&\
\,\,\,\,\,\,\,\,\,\,\,\,\,\,\,\,\,\,\,\,\,\,\,\,\,\,\,\,\,\,\,\,\,\,\,\,\,\,\,\,\,\,\,\,\,\,{\bf \,Kinematic\,characteristics\ } \\
&&\
\,\,\,\,\,\,\,\,\,\,\,\,\,\,\,\,\,\,\,\,\,\,\,\,\,\,\,\,\,\,\,\,\,\,\,\,\,\,\,{\bf of\,contravariant\,vector\,fields} \\
&&\ \,\,\,\,\,\,\,\,\,\,\,\,\,\,{\bf \,}\text{{\bf relative velocity}{\it \
(shear, rotation and expansion velocities),}} \\
&&\ \text{{\bf relative acceleration}{\it \ (shear, rotation and expansion
accelerations),}} \\
&&\ \,\,\text{{\it \thinspace \thinspace \thinspace \thinspace \thinspace
\thinspace \thinspace \thinspace \thinspace \thinspace \thinspace \thinspace
\thinspace \thinspace \thinspace \thinspace \thinspace \thinspace \thinspace
\thinspace \thinspace \thinspace \thinspace \thinspace \thinspace \thinspace
\thinspace \thinspace \thinspace \thinspace \thinspace \thinspace \thinspace
\thinspace \thinspace \thinspace \thinspace \thinspace \thinspace \thinspace
\thinspace \thinspace \thinspace \thinspace \thinspace \thinspace \thinspace
\thinspace \thinspace \thinspace \thinspace \thinspace \thinspace \thinspace
\thinspace \thinspace \thinspace \thinspace \thinspace \thinspace \thinspace
\thinspace }{\bf \thinspace deviation equations,}} \\
&&\,\,\,\,\,\,\,\,\,\,\,\,\,\,\,\,\,\,\,\,\,\,\,\,\,\,\,\,\,\,\,\,\,\,\,\,\,\,\,\,\,\text{{\bf geodesic and auto-paralel equations}{\it , }} \\
&&\
\,\,\,\,\,\,\,\,\,\,\,\,\,\,\,\,\,\,\,\,\,\,\,\,\,\,\,\,\,\,\,\,\,\,\,\,\,\,\,\,\,\,\,\,\,\,\,\,\,\,\,\,\,\,\,\,\,\text{{\bf Fermi-Walker transports}{\it ,}} \\
&&\
\,\,\,\,\,\,\,\,\,\,\,\,\,\,\,\,\,\,\,\,\,\,\,\,\,\,\,\,\,\,\,\,\,\,\,\,\,\,\,\,\,\,\,\,\,\,\,\,\,\,\,\,\,\,\,\,\,\,\,\,\,\,\,\,\,\text{{\bf conformal
transports }}
\end{eqnarray*}
%

$\mid $

\begin{eqnarray*}
&&{\bf Lagrangian\,theory\,of\,tensor\,fields} \\
&&\,\,\,\,\,\,\,\,\,\,\,\,\,\,\,\,\,\,\,\,\,\,\,\,\,\,\,\,\,\,\,\text{{\bf \thinspace Lagrangian density}{\it ,}} \\
&&\ \,\,\,\,\,\,\,\,\,\,\,\,\,\,\,\,\,\,\,\,\,\,\,\,\,\,\,\,\,\text{{\bf variational principles}{\it ,}} \\
&&\ \,\,\,\,\,\,\,\,\,\,\,\,\,\,\,\,\,\,\text{{\bf Euler-Lagrange's equations}{\it ,}} \\
&&\ \,\,\,\,\,\,\,\,\,\,\,\,\,\,\,\,\,\,\,\,\text{{\bf energy-momentum
tensors}{\it \thinspace }}\,\,\,\,\,
\end{eqnarray*}
%
\end{center}

In this review we consider only the elements of the first part of the above
scheme related to spaces with contravariant and covariant affine connections
and metrics. The review appears as an introduction to the theory of the $(%
\overline{L}_n,g)$-spaces. It contains formulas necessary for the
development of the mechanics of tensor fields and for constructing
mathematical models of different dynamical systems described by the use of
the main objects under consideration. The general results found for
differentiable manifolds with different (not only by sign) contravariant and
covariant affine connections and metrics can be specialized for spaces with
one affine connection and a metric [the s. c. $(L_n,g)$-spaces] as well as
for (pseudo) Riemannian spaces with or without torsion [the s.c. $U_n$- and $%
V_n$-spaces]. The most results are given either in index-free form or in a
co-ordinate, or in a non-co-ordinate basis. This has been done to facilitate
the reader in choosing the right form of the results for his own further
considerations. The main conclusions are summarized in the last section.

The $(\overline{L}_n,g)$-spaces have interesting properties which could be
of use in the theoretical physics and especially in the theoretical
gravitational physics. In these type of spaces the introduction of a
contravariant non-symmetric affine connection for contravariant tensor
fields and the introduction of a symmetric (Riemannian,\ Levi-Civita
connection) for covariant tensor fields is possible. On this grounds we can
consider flat spaces [$(\overline{M}_n,g)$-spaces] with predetermined
torsion for the contravariant vector fields and with torsion-free connection
for the covariant vector fields. In analogous way these type of structures
could be induced in (pseudo) Riemannian spaces [$(\overline{V}_n,g)$-spaces].

\subsection{Space-time geometry and differential geometry}

In the last few years new attempts \cite{Hehl-4} - \cite{Hehl-5} have been
made to revive the ideas of Weyl \cite{Eddington}, \cite{Schrodinger} for
using manifolds with independent affine connection and metric (spaces with
affine connection and metric) as a model of space-time in the theory of
gravitation \cite{Hehl-5}. In such spaces the connection for co-tangent
vector fields (as dual to the tangent vector fields) differs from the
connection for the tangent vector fields only by sign. The last fact is due
to the definition of dual vector bases in dual vector spaces over points of
a manifold, which is a trivial generalization of the definition of dual
bases of algebraic dual vector spaces from the multi-linear algebra \cite
{Greub-1} - \cite{Bishop}. On the one hand, the hole modern differential
geometry is built as a rigorous logical structure having as one of its main
assumption the canonical definition for dual bases of algebraic dual vector
spaces (with equal dimensions) \cite{Choquet-Bruhat}. On the other hand, the
possibility of introducing a non-canonical definition for dual bases of
algebraic dual vector spaces (with equal dimensions) has been pointed out by
many mathematicians \cite{Kobayashi} who have not exploited this possibility
for further evolution of the differential-geometric structures and its
applications. The canonical definition of dual bases of dual spaces is so
naturally embedded in the ground of the differential geometry that no need
has occurred for changing it \cite{Matsushima} - \cite{Norden}. But the last
time evolution of the mathematical models for describing the gravitational
interaction on a classical level shows a tendency to generalizations, using
spaces with affine connection and metric, which can also be generalized
using the freedom of the differential-geometric preconditions. It has been
proved that an affine connection, which in a point or over a curve in
Riemannian spaces can vanish (a fact leading to the principle of equivalence
in ETG), can also vanish under a special choice of the basic system in a
space with affine connection and metric \cite{von der Heyde} - \cite{Hartley}%
. The last fact shows that the equivalence principle in the ETG could be
considered only as a physical interpretation of a corollary of the
mathematical apparatus used in this theory. Therefore, \textit{every
differentiable manifold with affine connection and metric can be used as a
model for space-time in which the equivalence principle holds}. But if the
manifold has two different (not only by sign) connections for tangent and
cotangent vector fields, the situation changes and is worth being
investigated.

The basic notions in the differential geometry related to the notions
considered in this review are defined for the most part in textbook and
monographs on differential geometry [see for example \cite{Yano} - \cite
{Shirokov}].

\section{Algebraic dual vector spaces. Contraction operator}

The notion of algebraic dual vector space can be introduced in a way \cite
{Efimov} in which the two vector spaces (the considered and its dual vector
space) are two independent (finite) vector spaces with equal dimensions.

Let $X$ and $X^{*}$ be two vector spaces with equal dimensions dim $X=$ dim $%
X^{*}$ $=n$. Let $S$ be an operator (mapping) such that to every pair of
elements $u\in X$ and $p\in X^{*}$ sets an element of the field $K$ ($R$ or $%
C$), i.e. 
\begin{equation}  \label{Ch 4 1.17}
S:(u,p)\rightarrow z\in K\text{ , }u\in X\text{ , }p\in X^{*}\text{ .}
\end{equation}

\begin{definition}
{\bf \ } The operator (mapping) $S$ is called {\it contraction} {\it operator }$S$ if it is a bilinear symmetric mapping, i.e. if it fulfils the
following conditions:
\end{definition}
%

(a) $S(u,p_1+p_2)=S(u,p_1)+S(u,p_2)$ , $\forall u\in X$, $\forall p_i\in
X^{*}$ , $i=1,2$ .

(b) $S(u_1+u_2,p)=S(u_1,p)+S(u_2,p)$ , $\forall u_i\in X$ , $i=1,2$ , $%
\forall p\in X^{*}$ .

(c) $S(\alpha u,p)=S(u,\alpha p)=\alpha S(u,p)$ , $\alpha \in K$ .

(d) Nondegeneracy: if $u_1,...,u_n$ are linear independent in $X$ and $%
S(u_1,p)=0$, ..., $S(u_n,p)=0$, then the $p$ is the null element in $X^{*%
\text{ }}$. In an analogous way, if $p_1,...,p_n$ are linear independent in $%
X^{*}$ and $S(u,p_1)=0$, ... , $S(u,p_n)=0$, then $u$ is the null element in 
$X$.

(e) Symmetry: $S(u,p)=S(p,u)$ , $\forall u\in X$ , $\forall p\in X^{*}$ .

Let $e_1,...,e_n$ be an arbitrary basis in $X$, and let $e^1,...,e^n$ be an
arbitrary basis in $X^{*}$. Let $u=u^ie_i\in X$ and $p=p_ke^k\in X^{*}$.
From the properties (a) - (c) it follows that 
\begin{equation}  \label{Ch 4 1.18}
S(u,p)=f^k\text{ }_i.u^i.p_k\text{ ,}
\end{equation}

\noindent where 
\begin{equation}
f^k\text{ }_i=S(e_i,e^k)=S(e^k,e_i)\in K\text{ .}  \label{Ch 4 1.19}
\end{equation}

In this way the result of the action of the contraction operator $S$ is
expressed in terms of a bilinear form. The property non-degeneracy (d) means
the non-degeneracy of the bilinear form. The result $S(u,p)$ can be defined
in different ways by giving arbitrary numbers $f^k$ $_i\in K$ for which the
condition det $(f^k$ $_i)\neq 0$ and, at the same time, the conditions (a) -
(d) are fulfilled.

\begin{definition}
{\bf \ }(Mutually) {\it algebraic dual vector spaces.} The spaces $X$ and $X^{*}$ are called (mutually) dual spaces if an contraction operator acting
on them is given and they are considered together with this operator [i.e. $(X,X^{*},S)$ with dim $X=n=$dim $X^{*}$ defines the two (mutually) dual
spaces $X$ and $X^{*}$].
\end{definition}
%

The definition for (mutually) algebraic dual spaces allows for a given
vector space $X$ an infinite number of vector spaces $X^{*}$ (in different
ways dual to $X$) to be constructed. In order to avoid this non-uniqueness
Efimov and Rosendorn \cite{Efimov} introduced the notion equivalence between
dual vector spaces [which is an additional condition to the definition of
(mutually) dual spaces].

\begin{definition}
{\bf \ }{\it Equivalent dual to }$X$ {\it vector spaces}. Let $X_1^{*}$ and $X_2^{*}$ be two $n$-dimensional vector spaces, dual to $X$. If a linear
isomorphism exists between them, such that 
\begin{equation}
S(u,p)=S(u,p^{\prime })\text{ , }\forall u\in X\text{ , }\forall p\in X_1^{*}\text{ , }p^{\prime }\in X_2^{*}\text{ ,}  \label{Ch 4 1.20}
\end{equation}
where $p^{\prime }$ is the element of $X_2^{*}$, corresponding to $p$ of $X_1^{*}$ by means of the linear isomorphism, then $X_1^{*}$ and $X_2^{*}$
are called equivalent dual to $X$ vector spaces.
\end{definition}
%

\begin{proposition}
{\bf \ }All linear (vector) spaces, dual to a given vector space $X$, are
equivalent to each other.
\end{proposition}
%

For proving this proposition, it is enough to be shown that if for $X$ and $%
X^{*}$ is given an arbitrary $S$, then for an arbitrary basis $%
e_1,...,e_n\in X$ one can find an unique dual to it basis $e^1,...,e^n$ in
the space $X^{*}$, i.e. $e^1,...,e^n\in X^{*}$ can be found in an unique way
so that $S(e_i,e^k)=f^k{}_i,$ where $f^k\,_i\in K$ are preliminary given
numbers \cite{Manoff-6}. The proof is analogous to the proof given by Efimov
and Rosendorn \cite{Efimov} for the case $S=C:C(e_k,e^i)=g_k^i$, $%
\,\,\,g_k^i=1$ for $k=i$, $g_k^i=0$ for $k\neq i$. $C(e_k,e^i)=g_k^i$ means
that the dual to $\{e_k\}$ basic vector field $e^i$ is orthogonal to all
basic vectors $e_k$ for which $k\neq i$. The contraction operator $C$ is the
corresponding to the canonical approach mapping 
\begin{equation}  \label{Ch 4 1.33}
C(u,p)=C(p,u)=p(u)=p_i.u^i\text{ .}
\end{equation}

The new definition of algebraic dual spaces is as a matter in fact
corresponding to that in the common approach. Only the dual basic vector $%
e^i $ is not orthogonal to the basic vectors $e_k:$ $S(e_k,e^i)=f^i\,_k\neq
g_k^i $. It is enough to be noticed that for an arbitrary element $p\in
X^{*} $ the corresponding linear form 
\begin{equation}  \label{Ch 4 1.31}
S(u,p)=p_i.u^{\overline{i}}=p_i.f^i\,_k.u^k=p_{\overline{i}}.u^i
\end{equation}

\noindent is given, where $p_1,...,p_n$ are the constant components of a
given vector $p\in X^{*}$. The last equality can be written also in the form 
\begin{equation}
S(u,p)=S(p,u)=p(u)=p_i.u^{\overline{i}}\text{ .}  \label{Ch 4 1.32}
\end{equation}

\begin{remark}
{\bf \ }The generalization of the notion of algebraic dual spaces for the
case of vector fields over a differentiable manifold is a trivial one. The
vector fields are considered as sections of vector bundles over a manifold.
The vector bases become dependent on the points of the manifold and the
numbers $f^i$ $_j$ are considered as functions over the manifold.
\end{remark}
\begin{remark}
{\bf \ }If the basic vectors in the tangential space $T_x(M)$ at a point $x$
of a manifold $M$ (dim $M=n$) are the co-ordinate vector fields $\partial _i$
and in the dual vector space (the co-tangent space) $T_x^{*}(M)$ the basis \{$dx^k$\} is defined as a dual to the basis \{$\partial _i$\}, where $dx^k$
are the differentials of the co-ordinates $x^k$ of the point $x$ in a given
chart, then $S(\partial _i,dx^k)=f^k\,_i$ [$f^k{}_i\in C^r(M)$]. After
multiplication of the last equality with $f_k\,^l$ and taking into account
the relation $f^k\,_i.f_k\,^l=g_i^l$ the condition follows $S(\partial
_i,f_k $ $^l.dx^k)=g_i^l$, which is equivalent to the result of the action
of the contraction operator $C$ over the vectors $\partial _i$ and $e^l$,
where $e^l=f^l{}_k.dx^k$. The new vectors $e^l$ are not in the general case
co-ordinate differentials of the co-ordinates $x^l$ at $x\in M$. They would
be differentials of new co-ordinates $x^{l^{\prime }}=x^{l^{\prime }}(x^k)$
if the relation $dx^{l^{\prime }}=A_k$ $^{l^{\prime }}.dx^k$ is connected
with the condition $e^l=dx^{l^{\prime }}$ and $x^{l^{\prime }}=\smallint
dx^{l^{\prime }}$. In analogous way, for the case, when $S(f_l$ $^i.\partial
_i,dx^k)=g_l^k$ the new vectors $e_l=f_l$ $^i.\partial _i$ in the general
case are not again co-ordinate vector fields $\partial _{i^{\prime }}$. $e_l$
would be again co-ordinate vector fields if by changing the charts (the
co-ordinates) at a point $x\in M$ the condition $f_l$ $^i=\frac{\partial x^i}{\partial x^{l^{\prime }}}$ is fulfilled.
\end{remark}
%

Thus, the definition of algebraic dual vector fields over manifolds by means
of the contraction operator $S$ as a generalization of the contraction
operator $C$ allows considerations including functions $f^i$ $_j(x^k)$
instead of the Kronecker symbol $g_j^i$.

The contraction operator $S$ can be easily generalized to a \textit{%
multilinear} contraction operator $S$.

\section{Contravariant and covariant affine connections. Covariant
differential operator}

\subsection{Affine connection. Covariant differential operator}

The notion affine connection can be defined in different ways but in all
definitions a linear mapping is given, which to a given vector of a vector
space over a point $x$ of a manifold $M$ juxtaposes a corresponding vector
from the same vector space at this point. The corresponding vector is
identified as vector of the vector space over another point of the manifold $%
M$. The way of identification is called \textit{transport} from one point to
another point of the manifold.

Vector and tensor fields over a differentiable manifold are provided with
the structure of a linear (vector) space by defining the corresponding
operations at every point of the manifold.

\begin{definition}
{\bf \ }{\it Affine connection }over a differentiable manifold $M$. Let $V(M) $ (dim $M=n$) be the set of all (smooth) vector fields over the
manifold $M$. The mapping $\nabla :V(M)\times V(M)\rightarrow V(M)$, by
means of $\nabla (u,w)\rightarrow \nabla _uw$,\thinspace \thinspace
\thinspace \thinspace \thinspace \thinspace \thinspace \thinspace \thinspace
\thinspace \thinspace \thinspace $u,w\in V(M)$, with $\nabla _u$ as a
covariant differential operator along the vector field $u$ (s. the
definition below), is called {\it affine connection} over the manifold $M$.
\end{definition}
%

\begin{definition}
A{\bf \ }{\it covariant differential operator (along the vector field }$u${\it ).} The linear differential operator (mapping) $\nabla _u$ with the
following properties
\end{definition}
%

(a) $\nabla _u(v+w)=\nabla _uv+\nabla _uw$ ,\thinspace \thinspace \thinspace
\thinspace \thinspace $u,v,w\in V(M)$,

(b) $\nabla _u(f.v)=(uf).v+f.\nabla _uv$ ,\thinspace \thinspace \thinspace
\thinspace \thinspace $f\in C^r(M)$ , $\,\,r\geq 1$ ,

(c) $\nabla _{u+v}w=\nabla _uw+\nabla _vw$ ,

(d) $\nabla _{fu}v$ $=f.\nabla _uv$ ,

(e) $\nabla _uf=uf$ , $f\in C^r(M)$ , $r\geq 1$ ,

(f) $\nabla _u(v\otimes w)=\nabla _uv\otimes w+v\otimes \nabla _uw$ (Leibniz
rule), $\otimes $ is the sign for the tensor product,

is called \textit{covariant differential operator along the vector field }$u$%
.

The result of the action of the covariant differential operator $\nabla _uv$
is often called \textit{covariant derivative of the vector field }$v$ along
the vector field $u$.

In a given chart (co-ordinate system), the determination of $\nabla
_{e_\alpha }e_\beta $ in the basis \{$e_\alpha $\} defines the components $%
\nabla $$_{\beta \gamma }^\alpha $ of the affine connection $\nabla $ 
\begin{equation}  \label{Ch 4 2.3}
\nabla _{e_\alpha }e_\beta =\nabla _{\alpha \beta }^\gamma .e_\gamma \text{
,\thinspace \thinspace \thinspace \thinspace \thinspace \thinspace
\thinspace \thinspace \thinspace \thinspace \thinspace \thinspace \thinspace
\thinspace \thinspace \thinspace }\alpha ,\beta ,\gamma =1,...,n\text{ .}
\end{equation}

$\{\nabla _{\alpha \beta }^\gamma \}$ have the transformation properties of
a linear differential geometric object \cite{Yano}, \cite{Schouten}.

\begin{definition}
{\it Space with affine connection. }Differentiable manifold $M$, provided
with affine connection $\nabla $, i.e. the pair $(M,\nabla )$, is called
space with affine connection.
\end{definition}
%

\subsection{Contravariant and covariant affine connections}

The action of the covariant differential operator on a contravariant
(tangential) co-ordinate basic vector field $\partial _i$ over $M$ along
another contravariant co-ordinate basic vector field $\partial _j$ is
determined by the affine connection $\nabla =\Gamma $ with components $%
\Gamma _{ij}^k$ in a given chart (co-ordinate system) defined through 
\begin{equation}  \label{Ch 4 2.4}
\nabla _{\partial _j}\partial _i=\Gamma _{ij}^k.\partial _k\text{ .}
\end{equation}

For a non-coordinate contravariant basis $e_\alpha \in T(M)$ , $T(M)=\cup
_{x\in M}T_x(M)$, 
\begin{equation}  \label{Ch 4 2.5}
\nabla _{e_\beta }e_\alpha =\Gamma _{\alpha \beta }^\gamma .e_\gamma \text{ .%
}
\end{equation}

\begin{definition}
{\it Contravariant affine connection}. The affine connection $\nabla =\Gamma 
$ induced by the action of the covariant differential operator on
contravariant vector fields is called contravariant affine connection.
\end{definition}
%

The action of the covariant differential operator on a covariant (dual to
contravariant) basic vector field $e^\alpha $ [$e^\alpha \in T^{*}(M)$ , $%
T^{*}(M)=\cup _{x\in M}T_x^{*}(M)$] along a contravariant basic
(non-co-ordinate) vector field $e_\beta $ is determined by the affine
connection $\nabla =P$ with components $P_{\beta \gamma }^\alpha $ defined
through 
\begin{equation}  \label{Ch 4 2.6}
\nabla _{e_\beta }e^\alpha =P_{\gamma \beta }^\alpha .e^\gamma \text{ .}
\end{equation}

For a co-ordinate covariant basis $dx^i$%
\begin{equation}  \label{Ch 4 2.7}
\nabla _{\partial _j}dx^i=P_{kj}^i.dx^k\text{ .}
\end{equation}

\begin{definition}
{\it Covariant affine connection}. The affine connection $\nabla =\,$$P$
induced by the action of the covariant differential operator on covariant
vector fields is called covariant affine connection.
\end{definition}
%

\begin{definition}
{\it Space with contravariant and covariant affine connections} ($\overline{L}_n$-space){\it .} The differentiable manifold provided with contravariant
affine connection $\Gamma $ and covariant affine connection $P$ is called
space with contravariant and covariant affine connections.
\end{definition}
%

The connection between the two connections $\Gamma $ and $P$ is based on the
connection between the two dual spaces $T(M)$ and $T^{*}(M)$, which on its
side is based on the existence of the contraction operator $S$. Usually 
\textit{commutation relations} are required between the contraction operator
and the covariant differential operator in the form 
\begin{equation}  \label{Ch 4 2.8}
S\circ \nabla _u=\nabla _u\circ S\text{ .}
\end{equation}

If the last operator equality in the form $\nabla _{\partial _k}\circ
S=S\circ \nabla _{\partial _k}$ is used for acting on the tensor product $%
dx^i\otimes \partial _j$ of two basic vector fields $dx^i\in T^{*}(M)$ and $%
\partial _j\in T(M)$, then the relation follows 
\begin{equation}  \label{Ch 4 2.12}
f^i\text{ }_{j,k}=\Gamma _{jk}^l.f^i\text{ }_l+P_{lk}^i.f^l\text{ }_j\text{
, \thinspace \thinspace \thinspace \thinspace \thinspace \thinspace
\thinspace \thinspace \thinspace \thinspace \thinspace \thinspace \thinspace
\thinspace \thinspace \thinspace \thinspace \thinspace \thinspace \thinspace 
}f^i\text{ }_{j,k}:=\partial _kf^i\text{ }_j\text{ (in a co-ordinate basis).}
\end{equation}

The last equality can be considered from two different points of view:

1. If $P_{jk}^i(x^l)$ and $\Gamma _{jk}^i(x^l)$ are given as functions of
co-ordinates in $M$, then the equality appears as a system of equations for
the unknown functions $f^i\,_j(x^l)$. The solutions of these equations
determine the action of the contraction operator $S$ on the basic vector
fields for given components of both connections. The integrability
conditions for the equations can be written in the form 
\begin{equation}  \label{Ch 4 2.13}
R^m\text{ }_{jkl}.f^i\text{ }_m+P^i\text{ }_{mkl}.f^m\text{ }_j=0\text{ ,}
\end{equation}

\noindent where $R^m$ $_{jkl}$ are the components of the contravariant
curvature tensor, constructed by means of the contravariant affine
connection $\Gamma $, and $P_{\,\,\,mkl}^i$ are the components of the
covariant curvature tensor, constructed by means of the covariant affine
connection $P$, where $[R(\partial _i,\partial _j)]dx^k=P^k$ $_{lij}.dx^l$, $%
[R(\partial _i,\partial _j)]\partial _k=R^l$ $_{kij}.\partial _l$ , $%
R(\partial _i,\partial _j)=\nabla _{\partial _i}\nabla _{\partial _j}-\nabla
_{\partial _j}\nabla _{\partial _i}$.

2. If $f^i$ $_j(x^l)$ are given as functions of the co-ordinates in $M$,
then the conditions for $f^i$ $_j$ $\det $er$\min $e the connection between
the components of the contravariant affine connection $\Gamma $ and the
components of the covariant affine connection $P$ on the grounds of the
predetermined action of the contraction operator $S$ on basic vector fields.

If $S=C$, i.e. $f^i$ $_j=g_j^i$ , then the conditions for $f^i$ $_j$ are
fulfilled for every $P=-\Gamma $, i.e. 
\begin{equation}  \label{Ch 4 2.14}
P_{jk}^i=-\Gamma _{jk}^i\text{ .}
\end{equation}

This fact can be formulated as the following proposition

\begin{proposition}
{\bf \ }$S=C$ is a sufficient condition for $P=-\Gamma $ ($P_{jk}^i=-\Gamma
_{jk}^i$).
\end{proposition}
%

\begin{corollary}
If $P\neq -\Gamma $, then $S\neq C$, i.e. if the covariant affine connection 
$P$ has to be different from the contravariant affine connection $\Gamma $
not only by sign, then the contraction operator $S$ has to be different from
the canonical contraction operator $C$ (if $S$ commutes with the covariant
differential operator).
\end{corollary}
%

The corollary allows the introduction of different (not only by sign)
contravariant and covariant connections by using contraction operator $S$,
different from the canonical contraction operator $C$.

\begin{example}
If $f^i$ $_j=e^\varphi .g_j^i$, where $\varphi \in C^r(M)$, $\varphi \neq 0$, then $P_{jk}^i=-\Gamma _{jk}^i+\varphi _{,k}.g_j^i$.
\end{example}
%

\subsection{Covariant derivatives of contravariant tensor fields}

The \textit{action} of a covariant differential operator along a
contravariant vector field $u$ is called \textit{transport along a
contravariant vector field} $u$ (or \textit{transport along} $u$).

The \textit{result of the action} of the covariant differential operator on
a tensor field is called \textit{covariant derivative} of this tensor field.

The result $\nabla _uV$ of the action of $\nabla _u$ on a contravariant
tensor field $V$ is called \textit{covariant derivative of a contravariant
tensor field} $V$ \textit{along a contravariant vector field} $u$ (or 
\textit{covariant derivative of }$V$\textit{\ along }$u$).

The action of the covariant differential operator on contravariant tensor
fields with rank $>1$ can be determined in a trivial manner on the grounds
of the Leibniz rule which the operator obeys. Then the action of the
operator $\nabla _{\partial _j}$ on a tensor basis $\partial _A=\partial
_{j_1}\otimes ...\otimes \partial _{j_l}$ can be written in the form 
\[
\begin{array}{c}
\nabla _{\partial _j}\partial _A=\nabla _{\partial _j}[\partial
_{j_1}\otimes ...\otimes \partial _{j_l}]=(\nabla _{\partial _j}\partial
_{j_1}\otimes \partial _{j_2}...\otimes \partial _{j_l})+ \\ 
+\,\,(\partial _{j_1}\otimes \nabla _{\partial _j}\partial _{j_2}\otimes
...\otimes \partial _{j_l})+...+(\partial _{j_1}\otimes ...\otimes \nabla
_{\partial _j}\partial _{j_l})= \\ 
=\Gamma _{j_1j}^{i_1}.\partial _{i_1}\otimes ...\otimes \partial
_{j_l}+...+\Gamma _{j_lj}^{i_l}.\partial _{j_1}\otimes ...\otimes \partial
_{i_l}= \\ 
=(%
\sum_{k=1}^lg_{j_k}^i.g_m^{i_k}.g_{j_1}^{i_1}.g_{j_2}^{i_2}...g_{j_{k-1}}^{i_{k-1}}.g_{j_{k+1}}^{i_{k+1}}...g_{j_l}^{i_l}.).\Gamma _{ij}^m.(\partial _{i_1}\otimes ...\otimes \partial _{i_l})%
\text{ .}
\end{array}
\]

If we introduce the abbreviations 
\begin{equation}  \label{I.2.-19}
S_{Am}\,^{Bi}=-%
\sum_{k=1}^lg_{j_k}^i.g_m^{i_k}.g_{j_1}^{i_1}.g_{j_2}^{i_2}...g_{j_{k-1}}^{i_{k-1}}.g_{j_{k+1}}^{i_{k+1}}...g_{j_l}^{i_l}%
\text{ \thinspace \thinspace ,\thinspace }
\end{equation}
\begin{equation}  \label{I.2.-20}
\Gamma _{Aj}^B=-S_{Am}\,^{Bi}.\Gamma _{ij}^m\text{ ,\thinspace \thinspace
\thinspace \thinspace \thinspace \thinspace \thinspace \thinspace \thinspace
\thinspace \thinspace \thinspace \thinspace }A=j_1...j_l\text{ ,\thinspace
\thinspace \thinspace }B=i_1...i_l\text{ ,\thinspace }
\end{equation}

\noindent then $\nabla _{\partial _j}\partial _A$ can be written in the form 
\begin{equation}
\nabla _{\partial _j}\partial _A=\Gamma _{Aj}^B.\partial
_B=-S_{Am}\,^{Bi}.\Gamma _{ij}^m.\partial _B\text{ .}  \label{I.2.-21}
\end{equation}

The quantities $S_{Am}$ $^{Bi}$ obey the following relations

(a) $S_{Bi}$ $^{Aj}.S_{Ak}$ $^{Cl}=-g_i^l.S_{Bk}$ $^{Cj}$ , $\dim $ $M=n$, $%
l=1,...,N$ ,

(b) $S_{Bi}$ $^{Bj}=-N.n^{N-1}.g_i^j$ ,

(c) $S_{Bi}$ $^{Ai}=-N.g_B^A$ ,

\noindent where 
\begin{equation}
g_B^A=g_{i_1}^{j_1}...g_{i_{m-1}}^{j_{m-1}}.g_{i_m}^{j_m}.g_{i_{m+1}}^{j_{m+1}}...g_{i_l}^{j_l}%
\text{ }  \label{Ch 4 2.22}
\end{equation}

\noindent is defined as \textit{multi-Kronecker symbol} of rank $l$%
\begin{equation}
\begin{array}{c}
g_B^A=1\text{ }i_k=j_k\text{ (for all }k\text{ simultaneously)} \\ 
=0\text{ }i_k\neq j_k\text{ , }k=1,.................,l\text{ .}
\end{array}
\label{Ch 4 2.23}
\end{equation}

The covariant derivative along a contravariant vector field $u$ of a
contravariant tensor field $V=V^A.\partial _A$ can be written in a
co-ordinate basis as 
\begin{equation}  \label{I.2.-22}
\nabla _uV=(V^A\,_{,i}+\Gamma _{Bi}^A.V^B).u^i.\partial
_A=V^A\,_{;i}.u^i.\partial _A\text{ ,}
\end{equation}

\noindent where 
\begin{equation}
V^A\,_{;i}=V^A\,_{,i}+\Gamma _{Bi}^A.V^B  \label{I.2.-23}
\end{equation}

\noindent is called first covariant derivative of the components $V^A$ of
the contravariant tensor field $V$ along a contravariant co-ordinate basic
vector field $\partial _i$%
\begin{equation}
\nabla _{\partial _i}V=V^A\,_{;i}.\partial _A\text{ .}  \label{I.2.-24}
\end{equation}

In an analogous way we find for the second covariant derivative $\nabla _\xi
\nabla _uV$%
\[
\nabla _\xi \nabla _uV=(V^A\,_{;j;i}.u^j+V^A\,_{;j}.u^j\,_{;i}).\xi
^i.\partial _A=(V^A\,_{;j}.u^j)_{;i}.\xi ^i.\partial _A\text{ ,} 
\]

\noindent where 
\begin{equation}
V^A\,\,_{;j;i}=(V^A\,_{;j})_{,i}+\Gamma _{Bi}^A.V^B\,_{;j}-\Gamma
_{ji}^k.V^A\,_{;k}\text{ }  \label{I.2.-25}
\end{equation}

\noindent is the second covariant derivative of the components $V^A$ of the
contravariant vector field $V$. Here 
\begin{equation}
\nabla _\xi \nabla _uV-\nabla _u\nabla _\xi
V=[(V^A\,_{;i;j}-V^A\,_{;j;i}).u^i.\xi ^j+V^A\,_{;j}.(u^j\,_{;i}.\xi ^i-\xi
^j\,_{;i}.u^i)].\partial _A\text{ .}  \label{I.2.-26}
\end{equation}

\subsection{Covariant derivatives of covariant tensor fields}

In analogous way the covariant derivative of a covariant vector field can be
written in the form 
\begin{equation}  \label{Ch 4 2.26}
\begin{array}{c}
\nabla _up=(p_{i,j}+P_{ij}^k.p_k).u^j.dx^i=p_{i;j}.u^j.dx^i \text{ ,
\thinspace \thinspace \thinspace \thinspace \thinspace \thinspace \thinspace
\thinspace \thinspace \thinspace \thinspace }p\in T^{*}(M)\text{ ,} \\ 
\text{(in a co-ordinate basis).}
\end{array}
\end{equation}

The action of the covariant differential operator on covariant tensor fields
with rank $>1$ is generalized in a trivial manner on the grounds of the
Leibniz rule, which holds for this operator. Then the action of the operator 
$\nabla _{\partial _j}$ on the basis $dx^A=dx^{j_1}\otimes ...\otimes
dx^{j_l}$ can be written in the form 
\begin{equation}  \label{Ch 4 2.28}
\nabla _{\partial _j}dx^B=P_{Aj}^B.dx^A=-S_{Am}\text{ }^{Bi}.P_{ij}^m.dx^A%
\text{ ,}
\end{equation}

\noindent where $P_{Aj}^B=-S_{Am}$ $^{Bi}.P_{ij}^m$.

The covariant derivative of a covariant tensor field $W=W_A.dx^A=W_B.e^B$
can be written in the form 
\begin{equation}  \label{Ch 4 2.29}
\begin{array}{c}
\nabla _uW=(W_{A,j}+P_{Aj}^B.W_B).u^j.dx^A=W_{A;j}.u^j.dx^A \text{,} \\ 
\text{(in a co-ordinate basis).}
\end{array}
\end{equation}

The form of the covariant derivative of a mixed tensor field follows from
the form of the derivative of contravariant and covariant basic tensor
fields, and the Leibniz rule 
\begin{equation}  \label{Ch 4 2.31}
\begin{array}{c}
\nabla _uK=\nabla _u(K^A \text{ }_B.\partial _A\otimes dx^B)=K^A\text{ }%
_{B;j}.u^j.\partial _A\otimes dx^B= \\ 
(K^A \text{ }_{B,j}+\Gamma _{Cj}^A.K^C\text{ }_B+P_{Bj}^D.K^A\text{ }%
_D).u^j.\partial _A\otimes dx^B\text{ ,} \\ 
\text{(in a co-ordinate basis).}
\end{array}
\end{equation}

If the \textit{Kronecker tensor} is defined in the form 
\begin{equation}  \label{Ch 4 2.33}
Kr=g_j^i.\partial _i\otimes dx^j=g_\beta ^\alpha .e_\alpha \otimes e^\beta 
\text{ ,}
\end{equation}

\noindent then the components of the contravariant and the covariant affine
connection differ from each other by the components of the covariant
derivative of the Kronecker tensor, i.e. 
\begin{equation}
\Gamma _{jk}^i+P_{jk}^i=g_{j;k}^i\text{ , \thinspace \thinspace \thinspace
\thinspace \thinspace \thinspace \thinspace \thinspace \thinspace \thinspace
\thinspace \thinspace \thinspace \thinspace \thinspace \thinspace \thinspace
\thinspace \thinspace \thinspace }\Gamma _{\beta \gamma }^\alpha +P_{\beta
\gamma }^\alpha =g_{\beta /\gamma }^\alpha \text{ .}  \label{Ch 4 2.34}
\end{equation}

\begin{remark}
{\bf \ }In the special case, when $S=C$ and in the canonical approach $g_{j;k}^i=0$ ($g_{\beta /\gamma }^\alpha =0$).
\end{remark}
%

\section{Lie differential operator}

The Lie differential operator $\pounds _\xi $ along the contravariant vector
field $\xi $ appears as an other operator, which can be constructed by means
of a contravariant vector field. Its definition can be considered as a
generalization of the notion Lie derivative of tensor fields \cite
{Slebodzinski}, \cite{Yano}, \cite{Kobayashi}, \cite{Lightman}.

\begin{definition}
\textbf{\ }$\pounds _\xi $ := \textit{Lie differential operator along the
contravariant vector field\ }$\xi $ with the following properties:
\end{definition}

(a) $\pounds _\xi :V\rightarrow \overline{V}=\pounds _\xi V$ , $V,\overline{V%
}\in \otimes ^l(M)$ .

(b) $\pounds _\xi :W\rightarrow \overline{W}=\pounds _\xi W$ , $W,\overline{W%
}\in \otimes _k(M)$ .

(c) $\pounds _\xi :K\rightarrow \overline{K}=\pounds _\xi K$ ,\thinspace
\thinspace \thinspace $K$, $\,\overline{K}\in \otimes ^l$ $_k(M)$ .

(d) Linear operator with respect to tensor fields,

$\pounds _\xi (\alpha .V_1+\beta .V_2)=\alpha .\pounds _\xi V_1+\beta
.\pounds _\xi V_2$ , $\alpha ,\beta \in F(R$ or $C)$ , $V_i\in \otimes ^l(M)$
, $i=1,2$,

$\pounds _\xi (\alpha .W_1+\beta .W_2)=\alpha .\pounds _\xi W_1+\beta
.\pounds _\xi W_2$ , $W_i\in \otimes _k(M)$ , $i=1,2$,

$\pounds _\xi (\alpha .K_1+\beta .K_2)=\alpha .\pounds _\xi K_1+\beta
.\pounds _\xi K_2$ , $K_i\in \otimes ^l$ $_k(M)$ , $i=1,2$.

(e) Linear operator with respect to the contravariant field $\xi $ ,

$\pounds _{\alpha .\xi \,+\,\beta .u}=\alpha .\pounds _\xi +\beta .\pounds
_u $ , $\alpha ,\beta \in F(R$ or $C),$ $\xi ,u\in T(M)$ .

(f) Differential operator, obeying the Leibniz rule,

$\pounds _\xi (S\otimes U)=\pounds _\xi S\otimes U+S\otimes \pounds _\xi U$
, $S\in \otimes ^m$ $_q(M)$ , $U\in \otimes ^k$ $_l(M)$ .

(g) Action on function $f\in C^r(M)$ , $r\geq 1$ ,

$\pounds _\xi f=\xi f$ , $\xi \in T(M)$ .

(h) Action on contravariant vector field,

$\pounds _\xi u=[\xi ,u]$ , $\xi ,u\in T(M)$ , $[\xi ,u]=\xi \circ u-u\circ
\xi $ ,

$\pounds _\xi e_\alpha =[\xi ,e_\alpha ]=-\,(e_\alpha \xi ^\beta -\xi
^\gamma .C_{\gamma \alpha }$ $^\beta ).e_\beta $ ,

$\pounds _{e_\alpha }e_\beta =[e_\alpha ,e_\beta ]=C_{\alpha \beta }$ $%
^\gamma .e_\gamma $ , $C_{a\beta }$ $^\gamma \in C^r(M)$ ,

$\pounds _\xi \partial _i=-\xi ^j$ $_{,i}.\partial _j$ , $\pounds _{\partial
_i}\partial _j=[\partial _i,\partial _j]=0$ .

(i) \textit{Action on covariant basic vector field},

$\pounds _\xi e^\alpha =k^\alpha $ $_\beta (\xi ).e^\beta $ , $\pounds
_{e_\gamma }e^\alpha =k^\alpha $ $_{\beta \gamma }.e^\beta $ ,

$\pounds _\xi dx^i=k^i$ $_j(\xi ).dx^j$ , $\pounds _{\partial _k}dx^i=k^i$ $%
_{jk}.dx^j$ .

The action of the Lie differential operator on a covariant basic vector
field is determined by its action on a contravariant basic vector field and
the commutation relations between the Lie differential operator and the
contraction operator $S$.

\subsection{Lie derivatives of contravariant tensor fields}

The Lie differential operator $\pounds _\xi $ along a contravariant vector
field $\xi $ appears as an other operator which can be constructed by means
of a contravariant vector field. It is an operator mapping a contravariant
tensor field $V$ in a contravariant tensor field $\widetilde{V}=\pounds _\xi
V$.

The \textit{action of the Lie differential operator} along a contravariant
vector field $\xi $ is called \textit{dragging-along the contravariant
vector field} $\xi $ (or \textit{dragging-along} $\xi $).

The\textit{\ result of the action }($\pounds _\xi V$)\textit{\ of the Lie
differential operator} $\pounds _\xi $ on $V$ is called \textit{Lie
derivative of the contravariant tensor field }$V$\textit{\ along a
contravariant vector field} $\xi $ (or \textit{Lie derivative of }$V$\textit{%
\ along} $\xi $).

The \textit{commutator of two Lie differential operators} 
\begin{equation}  \label{I.3.-1}
\lbrack \pounds _\xi ,\pounds _u]=\pounds _\xi \circ \pounds _u-\pounds
_u\circ \pounds _\xi
\end{equation}

\noindent has the properties:

(a) Action on a function 
\[
\begin{array}{c}
\lbrack \pounds _\xi ,\pounds _u]f=(\pounds _\xi \circ \pounds _u-\pounds
_u\circ \pounds _\xi )f=[\xi ,u]f=(\pounds _\xi u)f=[\nabla _\xi ,\nabla
_u]f \text{ ,} \\ 
\text{ \thinspace }f\in C^r(M)\text{ ,\thinspace \thinspace \thinspace
\thinspace }r\geq 2\text{ .}
\end{array}
\]

(b) Action on a contravariant vector field 
\[
\begin{array}{c}
\lbrack \pounds _\xi ,\pounds _u]v=(\pounds _\xi \circ \pounds _u-\pounds
_u\circ \pounds _\xi )v=\pounds _\xi \pounds _uv-\pounds _u\pounds _\xi v=
\\ 
=\pounds _v\pounds _u\xi =-\pounds _{\pounds _u\xi }v=\pounds _{\pounds _\xi
u}v\text{ .}
\end{array}
\]

(c) The Jacobi identity 
\begin{equation}  \label{I.3.-2}
<[\pounds _\xi ,[\pounds _u,\pounds _v]]>\,\equiv [\pounds _\xi ,[\pounds
_u,\pounds _v]]+[\pounds _v,[\pounds _\xi ,\pounds _u]]+[\pounds _u,[\pounds
_v,\pounds _\xi ]]\equiv 0\text{ .}
\end{equation}

The Lie derivative of a contravariant vector field 
\begin{equation}  \label{I.3.-3}
\pounds _\xi u=[\xi ,u]=(\pounds _\xi u^i).\partial _i=(\xi
^k.u^i\,_{,k}-u^k.\xi ^i\,_{,k}).\partial _i\text{ ,}\,
\end{equation}

\noindent where 
\begin{equation}
\pounds _\xi u^i=\xi ^k.u^i\,_{,k}-u^k.\xi ^i\,_{,k}  \label{I.3.-4}
\end{equation}

\noindent is called \textit{Lie derivative of the components} $u^i$ of a
vector field $u$ along a contravariant vector field $\xi $ (or Lie
derivative of the components $u^i$ along $\xi $) \textit{in a co-ordinate
basis}.

In a non-co-ordinate basis the Lie derivative can be written in an analogous
way as in a co-ordinate basis 
\begin{equation}  \label{I.3.-11}
\pounds _\xi u=[\xi ,u]=(\pounds _\xi u^\alpha ).e_\alpha =(\xi ^\beta
.e_\beta u^\alpha -u^\beta .e_\beta \xi ^\alpha +C_{\beta \gamma }\,^\alpha
.\xi ^\beta .u^\gamma ).e_\alpha \text{ ,}\,
\end{equation}

\noindent where 
\begin{equation}
\pounds _\xi u^\alpha =\xi ^\beta .e_\beta u^\alpha -u^\beta .e_\beta \xi
^\alpha +C_{\beta \gamma }\,^\alpha .\xi ^\beta .u^\gamma  \label{I.3.-12}
\end{equation}

\noindent is called \textit{Lie derivative of the components }$u^\alpha $%
\textit{\ of the contravariant vector field }$u$\textit{\ along a
contravariant vector field }$\xi $\textit{\ in a non-co-ordinate basis} (or 
\textit{Lie derivative of the components }$u^\alpha $\textit{\ along} $\xi $%
). $\pounds _{e_\beta }u$ can be written in the form 
\begin{equation}
\pounds _{e_\beta }u=(e_\beta u^\alpha -C_{\gamma \beta }\,^\alpha .u^\gamma
).e_\alpha =u^\alpha \,_{//\beta }.e_\alpha =-\pounds _ue_\beta =(\pounds
_{e_\beta }u^\alpha ).e_\alpha \text{ ,}  \label{I.3.-13}
\end{equation}

\noindent where 
\begin{equation}
\pounds _{e_\beta }u^\alpha =u^\alpha \,_{//\beta }=e_\beta u^\alpha
-C_{\gamma \beta }\,^\alpha .u^\gamma \text{ .}  \label{I.3.-14}
\end{equation}

The second Lie derivative $\pounds _\xi \pounds _uv$ will have in a
non-co-ordinate basis the form 
\begin{equation}  \label{I.3.-15}
\pounds _\xi \pounds _uv=[\xi ^\beta .e_\beta (\pounds _uv^\alpha )-(\pounds
_uv^\beta ).e_\beta \xi ^\alpha -C_{\gamma \beta }\,^\alpha .(\pounds
_uv^\gamma ).\xi ^\beta ].e_\alpha =(\pounds _\xi \pounds _uv^\alpha
).e_\alpha \text{ ,}
\end{equation}

\noindent where $\pounds _\xi \pounds _uv^\alpha =\xi ^\beta .e_\beta
(\pounds _uv^\alpha )-(\pounds _uv^\beta ).e_\beta \xi ^\alpha -C_{\gamma
\beta }\,^\alpha .(\pounds _uv^\gamma ).\xi ^\beta $ is called \textit{%
second Lie derivative of the components }$v^\alpha $\textit{\ along }$u$%
\textit{\ and }$\xi $\textit{\ in a non-co-ordinate basis}.

The action of the Lie differential operator on a contravariant tensor field
with rank $k>1$ can be generalized on the basis of the validity of the
Leibniz rule under the action of this operator on the bases of the tensor
fields.

The result of the action of the operator $\pounds _\xi $ on a basis $%
\partial _A=\partial _{j_1}\otimes ...\otimes \partial _l$ can be found by
the use of the already known relation $\pounds _\xi \partial _{j_k}=-\pounds
_{\partial _{j_k}}\xi =-\xi ^m\,\,_{,j_k}.\partial _m$. Then $\pounds _\xi
\partial _A=S_{Am}\,^{Bn}.\xi ^m\,_{,n}.\partial _B$, and 
\begin{equation}  \label{I.3.-19}
\pounds _\xi V=\pounds _\xi (V^A.\partial _A)=(\pounds _\xi V^A).\partial
_A=(\xi ^k.V^A\,_{,k}+S_{Bk}\,^{Al}.V^B.\xi ^k\,_{,l}).\partial _A\text{ ,}
\end{equation}

where $\pounds _\xi V^A=\xi ^k.V^A\,_{,k}+S_{Bk}\,^{Al}.V^B.\xi ^k\,_{,l}$
is the \textit{Lie derivative of the components} $V^A$ \textit{of a
contravariant tensor field }$V$\textit{\ along a contravariant vector field }%
$\xi $\textit{\ in a co-ordinate basis} (or \textit{Lie derivative of the
components }$V^A$\textit{\ along }$\xi $\textit{\ in a co-ordinate basis}).

For $\pounds _\xi \pounds _uV$ we obtain 
\begin{equation}  \label{I.3.-21}
\pounds _\xi \pounds _uV=(\pounds _\xi \pounds _uV^A).\partial _A=[\xi
^k(\pounds _uV^A)_{,k}+S_{Bk}\,^{Al}.(\pounds _uV^B).\xi ^k\,_{,l}].\partial
_A\text{ ,}
\end{equation}

\noindent where $\pounds _\xi \pounds _uV^A=\xi ^k(\pounds
_uV^A)_{,k}+S_{Bk}\,^{Al}.(\pounds _uV^B).\xi ^k\,_{,l}$ is called second
Lie derivative of the components $V^A$ along $u$ and $\xi $ in a co-ordinate
basis.

The result of the action of the Lie differential operator $\pounds _\xi $ on
a non-co-ordinate basis $e_A$ can be found in an analogous way as that for a
co-ordinate basis. Since 
\begin{equation}  \label{I.3.-23}
\pounds _\xi e_\beta =-\xi ^\alpha \,_{//\beta }.e_\alpha \text{ ,\thinspace
\thinspace \thinspace \thinspace \thinspace }\xi ^\alpha \,_{//\beta
}=e_\beta \xi ^\alpha -C_{\gamma \beta }\,^\alpha .\xi ^\gamma \text{ ,}
\end{equation}
\begin{equation}  \label{I.3.-24}
\begin{array}{c}
\pounds _\xi e_A=\pounds _\xi [e_{\alpha _1}\otimes ...\otimes e_{\alpha
_l}]=(\pounds _\xi e_{\alpha _1}\otimes e_{\alpha _2}...\otimes e_{\alpha
_l})+ \\ 
+\,\,(e_{\alpha _1}\otimes \pounds _\xi e_{\alpha _2}\otimes ...\otimes
e_{\alpha _l})+...+(e_{\alpha _1}\otimes ...\otimes \pounds _\xi e_{\alpha
_l})= \\ 
=S_{A\alpha }\,^{B\beta }.\xi ^\alpha \,_{//\beta }.e_B \text{ ,\thinspace
\thinspace \thinspace \thinspace \thinspace } \\ 
A=\alpha _1...\alpha _l\text{ ,\thinspace \thinspace \thinspace \thinspace }%
B=\beta _1...\beta _l\text{ ,\thinspace \thinspace \thinspace }e_B=e_{\beta
_1}\otimes ...\otimes e_{\beta _l}\text{ ,}
\end{array}
\end{equation}

\noindent then 
\begin{equation}
\begin{array}{c}
\pounds _\xi e_A=S_{A\alpha }\,^{B\beta }.\xi ^\alpha \,_{//\beta }.e_B\text{
and }\pounds _\xi V=\pounds _\xi (V^A.e_A)=(\pounds _\xi V^A).e_A= \\ 
=(\xi ^\alpha .e_\alpha V^A+S_{B\alpha }\,^{A\beta }.V^B.\xi ^\alpha
\,_{//\beta }).e_A\text{ .}
\end{array}
\label{I.3.-25}
\end{equation}

The explicit form of the expression $S_{B\alpha }\,^{A\beta }.\xi ^\alpha
\,_{//\beta }$ can be given as 
\begin{equation}  \label{I.3.-26}
S_{B\alpha }\,^{A\beta }.\xi ^\alpha \,_{//\beta }=S_{B\alpha }\,^{A\beta
}.e_\beta \xi ^\alpha -S_{B\alpha }\,^{A\beta }.C_{\gamma \beta }\,^\alpha
.\xi ^\gamma \text{ ,}
\end{equation}

\noindent and if we introduce the abbreviations 
\begin{equation}
C_{B\gamma }\,^A=S_{B\alpha }\,^{A\beta }.C_{\gamma \beta }\,^\alpha
=-S_{B\alpha }\,^{A\beta }.C_{\beta \gamma }\,^\alpha \text{ ,}
\label{I.3.-27}
\end{equation}
\begin{equation}
S_{B\alpha }\,^{A\beta }.\xi ^\alpha \,_{//\beta }=S_{B\alpha }\,^{A\beta
}.e_\beta \xi ^\alpha -C_{B\alpha }\,^A.\xi ^\alpha \text{ ,}
\label{I.3.-28}
\end{equation}

\noindent then $\pounds _\xi V^A$ can be written in the forms 
\begin{equation}
\begin{array}{c}
\pounds _\xi V^A=\xi ^\alpha .e_\alpha V^A+S_{B\alpha }\,^{A\beta }.V^B.\xi
^\alpha \,_{//\beta }= \\ 
=\xi ^\alpha .e_\alpha V^A+S_{B\alpha }\,^{A\beta }.V^B.(e_\beta \xi ^\alpha
-C_{\gamma \beta }\,^\alpha .\xi ^\gamma )= \\ 
=\xi ^\alpha .(e_\alpha V^A-S_{B\beta }\,^{A\gamma }.V^B.C_{\alpha \gamma
}\,^\beta )+S_{B\beta }\,^{A\gamma }.V^B.e_\gamma \xi ^\beta = \\ 
=\xi ^\alpha .V^A\,_{//\alpha }+S_{B\alpha }\,^{A\beta }.V^B.e_\beta \xi
^\alpha \text{ .}
\end{array}
\label{I.3.-29}
\end{equation}

$\pounds _\xi V^A$ is called \textit{Lie derivative of the components }$V^A$%
\textit{\ of a contravariant tensor field }$V$\textit{\ along }$\xi $\textit{%
\ in a non-co-ordinate basis}. Here 
\begin{equation}  \label{I.3.-30}
V^A\,_{//\alpha }=e_\alpha V^A-S_{B\beta }\,^{A\gamma }.V^B.C_{\alpha \gamma
}\,^\beta =e_\alpha V^A-C_{B\alpha }\,^A.V^B\text{ .}
\end{equation}

In a non-co-ordinate basis the relations are valid 
\begin{equation}  \label{I.3.-33}
\pounds _{e_\alpha }V=V^A\,_{//\alpha }.e_A\text{ ,\thinspace \thinspace
\thinspace \thinspace \thinspace \thinspace \thinspace }\pounds _{e_\alpha
}e_A=-\,\,\,C_{A\alpha }\,^B.e_B\text{ .}
\end{equation}

The quantity 
\begin{equation}  \label{I.3.-34}
S_{B\alpha }\,^{A\beta
}=-\sum_{k=1}^lg_{j_1}^{i_1}...g_{j_{k-1}}^{i_{k-1}}.g_\alpha
^{i_k}.g_{j_k}^\beta .g_{j_{k+1}}^{i_{k+1}}...g_{j_l}^{i_l}\text{ ,}
\end{equation}

\noindent where $l=1,...,N$, $B=j_1...j_l$, $A=i_1...i_l$, is the \textit{%
multi-contraction symbol with rank} $N$.

\subsection{Connections between the covariant and the Lie differentiations}

The action of the covariant differential operator and the action of the Lie
differential operator on functions are identified with the action of the
contravariant vector field in the construction of both operators. The
contravariant vector field acts as a differential operator on functions over
a differentiable manifold $M$%
\[
\nabla _\xi f=\xi f=\pounds _\xi f=\xi ^i.\partial _if=\xi ^\alpha .e_\alpha
f\text{ , \thinspace \thinspace \thinspace }f\in C^r(M)\text{ ,\thinspace
\thinspace \thinspace \thinspace }\xi \in T(M)\text{ .} 
\]

If we compare the Lie derivative with the covariant derivative of a
contravariant vector field in a non-co-ordinate (or co-ordinate) basis 
\begin{equation}  \label{I.4.-1}
\begin{array}{c}
\pounds _\xi u=(\pounds _\xi u^\alpha ).e_\alpha =(\xi ^\beta .e_\beta
u^\alpha -u^\beta .e_\beta \xi ^\alpha +C_{\beta \gamma }\,^\alpha .\xi
^\beta .u^\gamma ).e_\alpha \text{ ,}\, \\ 
\nabla _\xi u=(u^\alpha \,_{/\beta }.\xi ^\beta ).e_\alpha =(\xi ^\beta
.e_\beta u^\alpha +\Gamma _{\gamma \beta }^\alpha .u^\gamma .\xi ^\beta
).e_\alpha \text{ ,}
\end{array}
\end{equation}

\noindent we will see that both expressions have a common term of the type $%
\xi u^\alpha =\xi ^\beta .e_\beta u^\alpha $ allowing a relation between the
two derivatives.

After substituting $e_\beta u^\alpha $ and $e_\beta \xi ^\alpha $ from the
equalities $e_\beta u^\alpha =u^\alpha \,_{/\beta }-\Gamma _{\gamma \beta
}^\alpha .u^\gamma $ and $e_\beta \xi ^\alpha =\xi ^\alpha \,_{/\beta
}-\Gamma _{\gamma \beta }^\alpha .\xi ^\gamma $ in the expression for $%
\pounds _\xi u$ we obtain 
\begin{equation}  \label{I.4.-2}
\begin{array}{c}
\pounds _\xi u^\alpha =u^\alpha \,_{/\beta }.\xi ^\beta -\xi ^\alpha
\,_{/\beta }.u^\beta -T_{\beta \gamma }\,^\alpha .\xi ^\beta .u^\gamma = \\ 
=u^\alpha \,_{/\beta }.\xi ^\beta -(\xi ^\alpha \,_{/\beta }-T_{\beta \gamma
}\,^\alpha .\xi ^\gamma ).u^\beta \text{ ,}
\end{array}
\end{equation}

\noindent where 
\begin{equation}
T_{\beta \gamma }\,^\alpha =\Gamma _{\gamma \beta }^\alpha -\Gamma _{\beta
\gamma }^\alpha -C_{\beta \gamma }\,^\alpha =-T_{\gamma \beta }\,^\alpha 
\text{ ,}  \label{I.4.-3}
\end{equation}
\begin{equation}
\begin{array}{c}
\pounds _\xi u=(\pounds _\xi u^\alpha ).e_\alpha =(u^\alpha \,_{/\beta }.\xi
^\beta -\xi ^\alpha \,_{/\beta }.u^\beta -T_{\beta \gamma }\,^\alpha .\xi
^\beta .u^\gamma ).e_\alpha = \\ 
=\nabla _\xi u-\nabla _u\xi -T(\xi ,u)\text{ ,}
\end{array}
\label{I.4.-4}
\end{equation}

\noindent with 
\begin{equation}
T(\xi ,u)=T_{\beta \gamma }\,^\alpha .\xi ^\beta .u^\gamma .e_\alpha
=-T(u,\xi )\text{ , \ \ \ \ \ \ \ \ \ \ \ \ }T(e_\beta ,e_\gamma )=T_{\beta
\gamma }\,^\alpha .e_\alpha \text{ .}  \label{I.4.-5}
\end{equation}

The contravariant vector field $T(\xi ,u)$ is called \textit{(contravariant)
torsion vector field} (or \textit{torsion vector field}).

If we use the equality following from the expression for $\pounds _uv^\beta $%
\begin{equation}  \label{I.4.-8}
v^\beta \,_{/\gamma }.u^\gamma -u^\beta \,_{/\gamma }.v^\gamma =\pounds
_uv^\beta +T_{\alpha \gamma }\,^\beta .u^\alpha .v^\gamma \text{ }
\end{equation}

\noindent in the expression 
\[
\nabla _u\nabla _v\xi -\nabla _v\nabla _u\xi =[(\xi ^\alpha \,_{/\beta
/\gamma }-\xi ^\alpha \,_{/\gamma /\beta }).v^\beta .u^\gamma +\xi ^\alpha
\,_{/\beta }.(v^\beta \,_{/\gamma }.u^\gamma -u^\beta \,_{/\gamma }.v^\gamma
)].e_\alpha \,\text{ ,} 
\]

\noindent then 
\begin{equation}
\begin{array}{c}
\nabla _u\nabla _v\xi -\nabla _v\nabla _u\xi =[(\xi ^\alpha \,_{/\beta
/\gamma }-\xi ^\alpha \,_{/\gamma /\beta }).v^\beta .u^\gamma +\xi ^\alpha
\,_{/\beta }.(\pounds _uv^\beta +T_{\gamma \delta }\,^\beta .u^\gamma
.v^\delta )].e_\alpha = \\ 
=[(\xi ^\alpha \,_{/\beta /\gamma }-\xi ^\alpha \,_{/\gamma /\beta
}).v^\beta .u^\gamma +\xi ^\alpha \,_{/\beta }.(\pounds _uv^\beta +T^\beta
(u,v)].e_\alpha \text{ ,} \\ 
T^\beta (u,v)=T_{\delta \gamma }\,^\beta .u^\delta .v^\gamma \text{
,\thinspace \thinspace \thinspace \thinspace \thinspace \thinspace }\nabla
_{T(u,v)}\xi =\xi ^\alpha \,_{/\beta }.T_{\gamma \delta }\,^\beta .u^\gamma
.v^\delta .e_\alpha \text{ ,} \\ 
\nabla _u\nabla _v\xi -\nabla _v\nabla _u\xi -\nabla _{\pounds _uv}\xi =(\xi
^\alpha \,_{/\beta /\gamma }-\xi ^\alpha \,_{/\gamma /\beta }).v^\beta
.u^\gamma .e_\alpha +\nabla _{T(u,v)}\xi \text{ ,}
\end{array}
\label{I.4.-10}
\end{equation}

\noindent or 
\begin{equation}
\begin{array}{c}
\nabla _u\nabla _v\xi -\nabla _v\nabla _u\xi -\nabla _{\pounds _uv}\xi
-\nabla _{T(u,v)}\xi =(\xi ^\alpha \,_{/\beta /\gamma }-\xi ^\alpha
\,_{/\gamma /\beta }).v^\beta .u^\gamma .e_\alpha \text{ ,} \\ 
\nabla _{e_\gamma }\nabla _{e_\beta }\xi -\nabla _{e_\beta }\nabla
_{e_\gamma }\xi -\nabla _{\pounds _{e_\gamma }e_\beta }\xi -\nabla
_{T(e_\gamma ,e_\beta )}\xi =(\xi ^\alpha \,_{/\beta /\gamma }-\xi ^\alpha
\,_{/\gamma /\beta }).e_\alpha \text{ .}
\end{array}
\label{I.4.-12}
\end{equation}

In a co-ordinate basis the contravariant torsion vector will have the form 
\begin{equation}  \label{I.4.-13}
T(\xi ,u)=T_{kl}\,^i.\xi ^k.u^l.\partial _i=(\Gamma _{lk}^i-\Gamma
_{kl}^i).\xi ^k.u^l.\partial _i\text{ ,}
\end{equation}
\begin{equation}  \label{I.4.-14}
T_{kl}\,^i=\Gamma _{lk}^i-\Gamma _{kl}^i\text{ ,\thinspace \thinspace
\thinspace \thinspace \thinspace \thinspace \thinspace \thinspace \thinspace 
}T(\partial _k,\partial _l)=T_{kl}\,^i.\partial _i\text{ .}
\end{equation}

The Lie derivative $\pounds _\xi u$ can be now written as 
\begin{equation}  \label{I.4.-15}
\begin{array}{c}
\pounds _\xi u=(\pounds _\xi u^i).\partial _i=(u^i\,_{;k}.\xi ^k-u^k.\xi
^i\,_{;k}-T_{kl}\,^i.\xi ^k.u^l).\partial _i \text{ ,} \\ 
\pounds _\xi u^i=u^i\,_{;k}.\xi ^k-u^k.\xi ^i\,_{;k}-T_{kl}\,^i.\xi ^k.u^l%
\text{ .}
\end{array}
\end{equation}

The connection between the covariant derivative and the Lie derivative of a
contravariant tensor field can be found in an analogous way as in the case
of a contravariant vector field.

\subsection{Lie derivative of covariant basic vector fields}

\subsubsection{Lie derivative of covariant co-ordinate basic vector fields}

The commutation relations between the Lie differential operator $\pounds
_\xi $ and the contraction operator $S$ in the case of basic co-ordinate
vector fields can be written in the form 
\begin{equation}  \label{Ch 4 3.1}
\begin{array}{c}
\pounds _\xi \circ S(dx^i\otimes \partial _j)=S\circ \pounds _\xi
(dx^i\otimes \partial _j) \text{ ,} \\ 
\pounds _\xi \circ S(e^\alpha \otimes e_\beta )=S\circ \pounds _\xi
(e^\alpha \otimes e_\beta )\text{ ,}
\end{array}
\end{equation}

\noindent where 
\begin{equation}
\begin{array}{c}
\pounds _\xi \circ S(dx^i\otimes \partial _j)=\xi f^i\text{\thinspace }_j=f^i%
\text{ }_{j,k}.\xi ^k\text{ ,} \\ 
S\circ \pounds _\xi (dx^i\otimes \partial _j)=S(\pounds _\xi dx^i\otimes
\partial _j)+S(dx^i\otimes \pounds _\xi \partial _j)\text{ .}
\end{array}
\label{Ch 4 3.2}
\end{equation}

By means of the non-degenerate inverse matrix $(f^i$ $_j)^{-1}=(f_j$ $^i)$
and the connections $f^i$ $_k.f_j$ $^k=g_j^i$ , $f^k$ $_i.f_k$ $^j=g_i^j$ ,
after multiplication of the equality for $k^i\,_l(\xi )$ with $f_m$ $^j$ and
summation over $j$, the explicit form for $k^i\,_j(\xi )$ is obtained in the
form 
\begin{equation}  \label{Ch 4 3.5}
k^i\text{ }_j(\xi )=f_j\text{ }^l.\xi ^k\text{ }_{,l}.f^i\text{ }_k+f_j\text{
}^l.f^i\text{ }_{l,k}.\xi ^k\text{ .}
\end{equation}

For $\pounds _{\partial _k}dx^i=k^i$ $_j(\partial _k).dx^j=k^i$ $_{jk}.dx^j$
it follows the corresponding form 
\begin{equation}  \label{Ch 4 3.6}
\begin{array}{c}
\pounds _{\partial _k}dx^i=k^i \text{ }_{jk}.dx^j=f_j\text{ }^l.f^i\text{ }%
_{l,k}.dx^j\text{ ,} \\ 
k^i\text{ }_{jk}=f_j\text{ }^l.f^i\text{ }_{l,k}\text{ .}
\end{array}
\end{equation}

On the other hand, from the commutation relations between $S$ and the
covariant differential operator $\nabla _\xi $, the connection between the
partial derivatives of $f^i\,_j$ and the components of the contravariant and
covariant connections $\Gamma $ and $P$ follows in the form 
\begin{equation}  \label{Ch 4 3.7}
f^i\text{ }_{l,k}=P_{mk}^i.f^m\text{ }_l+\Gamma _{lk}^m.f^i\text{ }_m\text{ .%
}
\end{equation}

After substituting the last expression in the expressions for $k^i\,_j(\xi )$
and for $k^i\,_{jk}$, the corresponding quantities are obtained in the forms 
\begin{equation}  \label{Ch 4 3.8}
k^i\text{ }_j(\xi )=f_j\text{ }^l.\xi ^k\text{ }_{,l}.f^i\text{ }%
_k+(P_{jk}^i+f_j\text{ }^l.\Gamma _{lk}^m.f^i\text{ }_m).\xi ^k\text{ ,}
\end{equation}
\[
k^i\text{ }_j(\partial _k)=k^i\text{ }_{jk}=P_{jk}^i+f_j\text{ }^l.\Gamma
_{lk}^m.f^i\text{ }_m\text{ ,} 
\]
\begin{equation}  \label{Ch 4 3.9}
\pounds _\xi dx^i=[f_j\text{ }^l.\xi ^k\text{ }_{,l}.f^i\text{ }%
_k+(P_{jk}^i+f_j\text{ }^l.\Gamma _{lk}^m.f^i\text{ }_m).\xi ^k].dx^j\text{ ,%
}
\end{equation}
\begin{equation}  \label{Ch 4 3.10}
\begin{array}{c}
\pounds _{\partial _k}dx^i=k^i \text{ }_{jk}.dx^j= \\ 
=(P_{jk}^i+f_j\text{ }^l.\Gamma _{lk}^m.f^i\text{ }_m).dx^j\text{ .}
\end{array}
\end{equation}

If we introduce the abbreviations 
\begin{equation}  \label{Ch 4 3.11}
\xi _{\text{ }}^{\overline{i}}\text{ }_{,\underline{j}}=f^i\text{ }_k.\xi ^k%
\text{ }_{,l}.f_j\text{ }^l\text{ ,\thinspace \thinspace \thinspace
\thinspace \thinspace \thinspace \thinspace \thinspace \thinspace \thinspace
\thinspace \thinspace \thinspace \thinspace \thinspace \thinspace \thinspace
\thinspace \thinspace }\Gamma _{\underline{j}k}^{\overline{i}}=f_j\text{ }%
^l.\Gamma _{lk}^m.f^i\text{ }_m\text{ ,}
\end{equation}

\noindent then the Lie derivatives of covariant co-ordinate basic vector
fields $dx^i$ along the contravariant vector fields $\xi $ and $\partial _k$
can be written in the forms 
\begin{equation}
\pounds _\xi dx^i=[\xi _{\text{ }}^{\overline{i}}\text{ }_{,\underline{j}%
}+(P_{jk}^i+\Gamma _{\underline{j}k}^{\overline{i}}).\xi ^k].dx^j\text{ ,
\thinspace \thinspace \thinspace \thinspace \thinspace \thinspace \thinspace
\thinspace \thinspace \thinspace \thinspace \thinspace \thinspace \thinspace
\thinspace \thinspace \thinspace }\pounds _{\partial
_k}dx^i=(P_{jk}^i+\Gamma _{\underline{j}k}^{\overline{i}}).dx^j\text{ .}
\label{Ch 4 3.12}
\end{equation}

\subsubsection{Lie derivative of covariant non-co-ordinate basic vector
fields}

Analogous to the case of covariant co-ordinate basic vector fields the Lie
derivatives of covariant non-co-ordinate basic vector fields can be obtained
in the form 
\begin{equation}  \label{Ch 4 3.15}
\begin{array}{c}
\pounds _\xi e^\alpha =[\xi ^{\overline{\alpha }}\text{ }_{//\underline{%
\beta }}+(P_{\beta \gamma }^\alpha +\Gamma _{\underline{\beta }\gamma }^{%
\overline{\alpha }}).\xi ^\gamma ].e^\beta = \\ 
=[e_{\underline{\beta }}\xi ^{\overline{\alpha }}+(P_{\beta \gamma }^\alpha
+\Gamma _{\underline{\beta }\gamma }^{\overline{\alpha }}+C_{\underline{%
\beta }\gamma }\text{ }^{\overline{\alpha }}).\xi ^\gamma ].e^\beta \text{ ,}
\end{array}
\end{equation}
\begin{equation}  \label{Ch 4 3.16}
\pounds _{e_\gamma }e^\alpha =(P_{\beta \gamma }^\alpha +\Gamma _{\underline{%
\beta }\gamma }^{\overline{\alpha }}+C_{\underline{\beta }\gamma }\text{ }^{%
\overline{\alpha }}).e^\beta \text{ ,}
\end{equation}

\noindent where 
\begin{equation}
\begin{array}{c}
\xi ^{\overline{\alpha }}\text{ }_{//\underline{\beta }}=f^\alpha \text{ }%
_\gamma .\xi ^\gamma \text{ }_{//\delta }.f_\beta \text{ }^\delta =f^\alpha 
\text{ }_\gamma .(e_\delta \xi ^\gamma ).f_\beta \text{ }^\delta +f^\alpha 
\text{ }_\gamma .C_{\delta \sigma }\text{ }^\gamma .f_\beta \text{ }^\delta
.\xi ^\sigma = \\ 
=e_{\underline{\beta }}\xi ^{\overline{\alpha }}+C_{\underline{\beta }\sigma
}\text{ }^{\overline{\alpha }}.\xi ^\sigma \text{ ,} \\ 
e_{\underline{\beta }}\xi ^{\overline{\alpha }}=f^\alpha \text{ }_\gamma
.(e_\delta \xi ^\gamma ).f_\beta \text{ }^\delta \text{ ,\thinspace
\thinspace \thinspace \thinspace \thinspace \thinspace \thinspace \thinspace
\thinspace \thinspace \thinspace \thinspace \thinspace \thinspace \thinspace
\thinspace \thinspace \thinspace \thinspace \thinspace \thinspace \thinspace 
}C_{\underline{\beta }\sigma }\text{ }^{\overline{\alpha }}=f^\alpha \text{ }%
_\gamma .C_{\delta \sigma }\text{ }^\gamma .f_\beta \text{ }^\delta \text{ ,}
\\ 
\Gamma _{\underline{\beta }\gamma }^{\overline{\alpha }}=f_\beta \text{ }%
^\delta .\Gamma _{\delta \gamma }^\sigma .f^\alpha \text{ }_\sigma \text{ .}
\end{array}
\label{Ch 4 3.17}
\end{equation}

\subsection{Lie derivatives of covariant tensor fields}

The action of the Lie differential operator on covariant vector and tensor
fields is determined by its action on covariant basic vector fields and on
the functions over $M$.

In a co-ordinate basis the Lie derivative of a covariant vector field $p$
along a contravariant vector field $\xi $ can be written in the forms 
\begin{equation}  \label{Ch 4 3.18}
\begin{array}{c}
\pounds _\xi p=\pounds _\xi (p_i.dx^i)=(\pounds _\xi p_i).dx^i= \\ 
=[p_{i,k}.\xi ^k+p_j.\xi ^{\overline{j}}\text{ }_{,\underline{i}%
}+p_j.(P_{ik}^j+\Gamma _{\underline{i}k}^{\overline{j}}).\xi ^k].dx^i= \\ 
=[p_{i;k}.\xi ^k+\xi ^{\overline{k}}\text{ }_{;\underline{i}}.p_k+T_{k%
\underline{i}}^{\,\,\,\,\,\overline{j}}.p_j.\xi ^k].dx^i\text{ ,}
\end{array}
\end{equation}

\noindent where 
\begin{equation}
\begin{array}{c}
\xi ^{\overline{j}}\text{ }_{;\underline{i}}=f^j\text{ }_k.\xi ^k\text{ }%
_{;l}.f_i\text{ }^l\text{ ,\thinspace \thinspace \thinspace \thinspace
\thinspace \thinspace \thinspace \thinspace \thinspace \thinspace \thinspace
\thinspace \thinspace }T_{k\underline{i}}^{\,\,\,\,\,\overline{j}}=f^j\text{ 
}_l.T_{km}^{\,\,\,\,\,\,\,l}.f_i\text{ }^m\text{ ,} \\ 
T_{ki}^{\,\,\,\,\,\,\,j}=\Gamma _{ik}^j-\Gamma _{ki}^j\text{ , \thinspace
\thinspace \thinspace \thinspace \thinspace \thinspace \thinspace \thinspace
\thinspace \thinspace \thinspace \thinspace (in a co-ordinate basis).}
\end{array}
\label{Ch 4 3.19}
\end{equation}

In a non-co-ordinate basis the Lie derivative $\pounds _\xi p$ has the forms 
\begin{equation}  \label{Ch 4 3.20}
\begin{array}{c}
\pounds _\xi p=\pounds _\xi (p_\alpha .e^\alpha )=(\pounds _\xi p_\alpha
).e^\alpha = \\ 
\{(e_\gamma p_\alpha +P_{\alpha \gamma }^\beta .p_\beta ).\xi ^\gamma
+p_\beta .[e_{\underline{\alpha }}\xi ^{\overline{\beta }}+(\Gamma _{%
\underline{\alpha }\gamma }^{\overline{\beta }}+C_{\underline{\alpha }\gamma
}\text{ }^{\overline{\beta }}).\xi ^\gamma ]\}.e^\alpha = \\ 
=(p_{\alpha /\beta }.\xi ^\beta +\xi ^{\overline{\beta }}\text{ }_{/%
\underline{\alpha }}.p_\beta +T_{\gamma \underline{\alpha }}^{\,\,\,\,\,\,%
\overline{\beta }}.p_\beta .\xi ^\gamma ).e^\alpha \text{ ,}
\end{array}
\end{equation}

\noindent where 
\begin{equation}
\begin{array}{c}
\xi ^{\overline{\beta }}\text{ }_{/\underline{\alpha }}=f^\beta \text{ }%
_\delta .\xi ^\delta \text{ }_{/\gamma }.f_\alpha \text{ }^\gamma \text{ , }%
T_{\gamma \underline{\alpha }}^{\,\,\,\,\,\,\overline{\beta }}=f_\alpha 
\text{ }^\delta .T_{\gamma \delta }^{\,\,\,\,\,\,\sigma }.f^\beta \text{ }%
_\sigma \text{ ,} \\ 
T_{\beta \gamma }^{\,\,\,\,\,\,\,\alpha }=\Gamma _{\gamma \beta }^\alpha
-\Gamma _{\beta \gamma }^\alpha -C_{\beta \gamma }\text{ }^\alpha \text{ ,
\thinspace \thinspace \thinspace \thinspace \thinspace \thinspace \thinspace
\thinspace \thinspace \thinspace \thinspace \thinspace (in a non-co-ordinate
basis).}
\end{array}
\label{Ch 4 3.21}
\end{equation}

The action of the Lie differential operator on covariant tensor fields is
determined by its action on basic tensor fields.

In a co-ordinate basis 
\begin{equation}  \label{Ch 4 3.22}
\begin{array}{c}
\pounds _\xi W=\pounds _\xi (W_A.dx^A)=(\xi W_A).dx^A+W_A.\pounds _\xi dx^A=
\\ 
=(\pounds _\xi W_A).dx^A \text{ ,\thinspace \thinspace \thinspace \thinspace
\thinspace \thinspace \thinspace \thinspace \thinspace \thinspace \thinspace
\thinspace \thinspace \thinspace }W\in \otimes _k(M)\text{ ,} \\ 
\pounds _\xi dx^B=-k^m \text{ }_n(\xi ).S_{Am}\text{ }^{Bn}.dx^A\text{
,\thinspace \thinspace \thinspace \thinspace \thinspace \thinspace
\thinspace \thinspace }\pounds _\xi dx^m=k^m\text{ }_n(\xi ).dx^n\text{ ,}
\\ 
\pounds _\xi dx^B=[-\xi ^{\overline{k}}\text{ }_{,\underline{l}}.S_{Ak}\text{
}^{Bl}-S_{Am}\text{ }^{Bn}.(P_{nl}^m+\Gamma _{\underline{n}l}^{\overline{m}%
}).\xi ^l].dx^A\text{ .}
\end{array}
\end{equation}

After introducing the abbreviations 
\begin{equation}  \label{Ch 4 3.23}
\widetilde{\Gamma }\,_{Ak}^B=-S_{Ai}\text{ }^{Bj}.\Gamma _{\underline{j}k}^{%
\overline{i}}\text{ ,\thinspace \thinspace \thinspace \thinspace \thinspace
\thinspace \thinspace \thinspace \thinspace \thinspace }P_{Ak}^B=-S_{Ai}%
\text{ }^{Bj}.P_{jk}^i\text{ ,}
\end{equation}

$\pounds _\xi W$ can be written in the form 
\begin{equation}  \label{Ch 4 3.24}
\begin{array}{c}
\pounds _\xi W=(\pounds _\xi W_A).dx^A= \\ 
=[\xi ^k.W_A\text{ }_{,k}-\xi ^{\overline{k}}\text{ }_{,\underline{l}}.S_{Ak}%
\text{ }^{Bl}.W_B+(P_{Al}^B+\widetilde{\Gamma }_{Al}^B).W_B.\xi ^l].dx^A%
\text{ ,}
\end{array}
\end{equation}

\noindent where 
\begin{equation}
\begin{array}{c}
\pounds _\xi W_A=\xi ^k.W_{A,k}-\xi ^{\overline{k}}\text{ }_{,\underline{l}%
}.S_{Ak}\text{ }^{Bl}.W_B+(P_{Al}^B+\widetilde{\Gamma }\,_{Al}^B).W_B.\xi ^l=
\\ 
=\xi ^k.W_{A;k}-S_{A\overline{k}}\text{ }^{B\underline{l}}.W_B.(\xi ^k\text{ 
}_{;l}-T_{lj}^k.\xi ^j)= \\ 
=\xi ^k.W_{A;k}-S_{Ak}\text{ }^{Bl}.W_B.(\xi ^{\overline{k}}\text{ }_{;%
\underline{l}}-T_{\underline{l}j}^{\overline{k}}.\xi ^j)\text{ ,}
\end{array}
\label{Ch 4 3.25}
\end{equation}
\begin{equation}
\begin{array}{c}
\pounds _{\partial _j}W_A=W_{A,j}+(P_{Aj}^B+\widetilde{\Gamma }\,_{Aj}^B).W_B%
\text{ ,} \\ 
\pounds _{\partial _j}dx^B=-S_{Ai}\text{ }^{Bl}.(P_{lj}^i+\Gamma _{%
\underline{l}j}^{\overline{i}}).dx^A=(P\,_{Aj}^B+\widetilde{\Gamma }%
\,_{Aj}^B).dx^A\text{ .}
\end{array}
\label{Ch 4 3.26}
\end{equation}

The second Lie derivative of the components $W_A$ of the covariant tensor
field $W$ can be written in the form 
\begin{equation}  \label{Ch 4 3.27}
\pounds _\xi \pounds _uW_A=\xi ^k.(\pounds _uW_A)_{,k}-\xi ^{\overline{k}}%
\text{ }_{,\underline{l}}.S_{Ak}\text{ }^{Bl}.\pounds _uW_B+(P_{Al}^B+%
\widetilde{\Gamma }\,_{Al}^B).\xi ^l.\pounds _uW_B\text{ .}
\end{equation}

In a non-co-ordinate basis $\pounds _\xi W$ has the form 
\begin{equation}  \label{Ch 4 3.28}
\pounds _\xi W=(\xi W_A).e^A+W_B.(\pounds _\xi e^B)=(\pounds _\xi W_A).e^A%
\text{ ,}
\end{equation}

\noindent where 
\begin{equation}
\begin{array}{c}
\pounds _\xi W_A=\xi ^\beta .e_\beta W_A-S_{A\alpha }\text{ }^{B\beta
}.W_B.e_{\underline{\beta }}\xi ^{\overline{\alpha }}+(P_{A\gamma }^B+%
\widetilde{\Gamma }\,_{A\gamma }^B+\widetilde{C}_{A\gamma }\text{ }%
^B).W_B.\xi ^\gamma \text{ ,} \\ 
\widetilde{\Gamma }\,_{A\gamma }^B=-S_{A\alpha }\text{ }^{B\beta }.\Gamma _{%
\underline{\beta }\gamma }^{\overline{\alpha }}\text{ , \thinspace
\thinspace \thinspace \thinspace \thinspace \thinspace \thinspace \thinspace
\thinspace \thinspace \thinspace \thinspace \thinspace \thinspace }%
\widetilde{C}_{A\gamma }\text{ }^B=-S_{A\alpha }\text{ }^{B\beta }.C_{%
\underline{\beta }\gamma }\text{ }^{\overline{\alpha }}\text{ ,} \\ 
\pounds _{e_\beta }e^B=(P_{A\beta }^B+\widetilde{\Gamma }\,_{A\beta }^B+%
\widetilde{C}_{A\beta }\text{ }^B).e^A\text{ ,} \\ 
\pounds _{e_\beta }W_A=e_\beta W_A+(P_{A\beta }^B+\widetilde{\Gamma }%
\,_{A\beta }^B+\widetilde{C}_{A\beta }\text{ }^B).W_B\text{ .}
\end{array}
\label{Ch 4 3.29}
\end{equation}

The second Lie derivative of $W_A$ in a non-co--ordinate basis has the form 
\begin{eqnarray}
\pounds _\xi \pounds _uW_A &=&\xi ^\beta .e_\beta (\pounds _uW_A)-S_{A\alpha
}\text{ }^{B\beta }.(\pounds _uW_B).e_{\underline{\beta }}\xi ^{\overline{\alpha }}+  \label{Ch 4 3.30} \\
&&+(P_{A\gamma }^B+\widetilde{\Gamma }\,_{A\gamma }^B+\widetilde{C}_{A\gamma
}\text{ }^B).\xi ^\gamma .\pounds _uW_B\text{ .}  \nonumber
\end{eqnarray}
%

The Lie derivatives of covariant basic tensor fields can be given in terms
of the covariant derivatives of the components of the contravariant vector
field $\xi $ and the torsion tensor 
\begin{equation}  \label{Ch 4 3.31}
\pounds _\xi e^B=[-S_{A\alpha }\text{ }^{B\beta }.\xi ^{\overline{\alpha }}%
\text{ }_{/\underline{\beta }}+P_{A\gamma }^B.\xi ^\gamma +\widetilde{T}%
\,_{A\gamma }^{\,\,\,\,\,\,B}.\xi ^\gamma ].e^A\text{ ,}
\end{equation}

\noindent where $\widetilde{T}\,_{A\gamma }^{\,\,\,\,\,\,\,B}=S_{A\alpha }$ $%
^{B\beta }.T_{\underline{\beta }\gamma }^{\,\,\,\,\,\,\overline{\alpha }%
}\,=S_{A\overline{\alpha }}$ $^{B\underline{\beta }}.T_{\beta \gamma
}^{\,\,\,\,\,\,\alpha }$.

$\pounds _\xi W_A$ will then have the form 
\begin{equation}  \label{Ch 4 3.32}
\begin{array}{c}
\pounds _\xi W_A=\xi ^\beta .W_{A/\beta }-S_{A \overline{\alpha }}\text{ }^{B%
\underline{\beta }}.W_B.(\xi ^\alpha \text{ }_{/\beta }-T_{\beta \gamma
}^{\,\,\,\,\,\,\alpha }.\xi ^\gamma )= \\ 
=\xi ^\beta .W_{A/\beta }-S_{A\alpha }\text{ }^{B\beta }.W_B.(\xi ^{%
\overline{\alpha }}\text{ }_{/\underline{\beta }}-T_{\underline{\beta }%
\gamma }^{\,\,\,\,\,\,\overline{\alpha }}.\xi ^\gamma )\text{ .}
\end{array}
\end{equation}

The generalization of the Lie derivatives for mixed tensor fields is
analogous to that for covariant derivatives of mixed tensor fields.

\subsection{Classification of linear transports with respect to the
connections between contravariant and covariant affine connections}

By means of the Lie derivatives of covariant basis vector fields, a
classification can be proposed for the connections between the components $%
\Gamma _{jk}^i$ ($\Gamma _{\beta \gamma }^\alpha $) of the contravariant
affine connection $\Gamma $ and the components $P_{jk}^i$ ($P_{\beta \gamma
}^\alpha $) of the covariant affine connection $P$. On this basis, linear
transports (induced by the covariant differential operator or by
connections) and draggings-along (induced by the Lie differential operator)
can be considered as connected with each other through commutation relations
of both operators with the contraction operator.

\begin{center}
$
\begin{array}{ll}
\underline{Transport\ condition} & \underline{Type\ of\ dragging-along\ and\
transports} \\ 
\begin{array}{c}
P_{\beta \gamma }^\alpha +\Gamma _{\underline{\beta }\gamma }^{\overline{%
\alpha }}+C_{\underline{\beta }\gamma }\text{ }^{\overline{\alpha }}=%
\overline{F}\,_{\beta \gamma }^\alpha \text{ ,} \\ 
P_{jk}^i+\Gamma _{\underline{j}k}^{\overline{i}}=\overline{F}\,_{jk}^i\text{
.}
\end{array}
& 
\begin{array}{c}
\pounds _{e_\gamma }e^\alpha = \overline{F}\,_{\beta \gamma }^\alpha
.e^\beta \text{ ,} \\ 
\pounds _{\partial _k}dx^i= \overline{F}\,_{jk}^i.dx^j\text{ .} \\ 
\text{Transport with arbitrary dragging-along}
\end{array}
\\ 
\begin{array}{c}
P_{\beta \gamma }^\alpha +\Gamma _{\underline{\beta }\gamma }^{\overline{%
\alpha }}=\overline{A}_\gamma .g_\beta ^\alpha \text{ ,} \\ 
P_{jk}^i+\Gamma _{\underline{j}k}^{\overline{i}}=\overline{A}_k.g_j^i\text{ .%
}
\end{array}
& 
\begin{array}{c}
\pounds _{e_\gamma }e^\alpha = \overline{A}_\gamma .e^\alpha +C_{\underline{%
\beta }\gamma }\text{ }^{\overline{\alpha }}.e^\beta \text{ ,} \\ 
\pounds _{\partial _k}dx^i= \overline{A}_k.dx^i\text{ .} \\ 
\text{Transport with co-linear dragging-along}
\end{array}
\\ 
\begin{array}{c}
P_{\beta \gamma }^\alpha +\Gamma _{\underline{\beta }\gamma }^{\overline{%
\alpha }}=0\text{ ,} \\ 
P_{jk}^i+\Gamma _{\underline{j}k}^{\overline{i}}=0\text{ .}
\end{array}
& 
\begin{array}{c}
\pounds _{e_\gamma }e^\alpha =C_{\underline{\beta }\gamma }\text{ }^{%
\overline{\alpha }}.e^\beta \text{ ,} \\ 
\pounds _{\partial _k}dx^i=0 \text{ .} \\ 
\text{Transport with invariant dragging-along}
\end{array}
\end{array}
$

\textbf{Table 1. Relations between transport conditions and types of
draging-along}
\end{center}

The classification of the relations between the affine connections is
analogous to the classification proposed by Schouten \cite{Schouten} and
considered by Schmutzer \cite{Schmutzer}.

\section{Curvature operator. Bianchi identities}

\subsection{Curvature operator}

One of the well known operator constructed by means of the covariant and the
Lie differential operators which has been used in the differential geometry
of differentiable manifolds is the curvature operator.

%
\begin{definition}
{\it Curvature operator}. The operator 
\begin{equation}
R(\xi ,u)=\nabla _\xi \nabla _u-\nabla _u\nabla _\xi -\nabla _{\pounds _\xi
u}=[\nabla _\xi ,\nabla _u]-\nabla _{[\xi ,u]}\text{ , \thinspace \thinspace
\thinspace }\xi ,\,u\in T(M)\text{ ,}  \label{I.5.-1}
\end{equation}

is called {\it curvature operator} (or operator of the curvature).
\end{definition}
%

1. Action of the curvature operator on a function of a class $C^r(M)$, $%
r\geq 2$, over a manifold $M$%
\[
\lbrack R(\xi ,u)]f=0\text{ ,\thinspace \thinspace \thinspace \thinspace
\thinspace \thinspace }f\in C^r(M)\text{, \thinspace \thinspace }r\geq 2%
\text{ .} 
\]

2. $[R(\xi ,u)]fv=f.[R(\xi ,u)]v$, \thinspace \thinspace \thinspace $f\in
C^r(M)$,\thinspace \thinspace \thinspace \thinspace $r\geq 2$,\thinspace
\thinspace \thinspace \thinspace $v\in T(M)$.

3. Action of the curvature operator on a contravariant vector field 
\begin{equation}  \label{I.5.-2}
\begin{array}{c}
\lbrack R(\xi ,u)]v=\nabla _\xi \nabla _uv-\nabla _u\nabla _\xi v-\nabla
_{\pounds _\xi u}v= \\ 
=[(v^\delta \,_{/\beta /\gamma }-v^\delta \,_{/\gamma /\beta }).u^\beta .\xi
^\gamma +v^\delta \,_{/\alpha }.T_{\beta \gamma }\,^\alpha .\xi ^\beta
.u^\gamma ].e_\delta = \\ 
=[(v^i\,_{;j;k}-v^i\,_{;k;j}).u^j.\xi ^k+v^i\,_{;j}.T_{kl}\,^j.\xi
^k.u^l].\partial _i\text{ .}
\end{array}
\end{equation}

In such a way, we can find for $\forall \,\xi \in T(M)$ and $\forall \,u\in
T(M)$ the relation in a co-ordinate basis 
\begin{equation}  \label{I.5.-6}
v^i\,_{;k;l}-v^i\,_{;l;k}=-\,\,v^j.R^i\,_{jkl}+v^i\,_{;j}.T_{kl}\,^j\text{ ,}
\end{equation}

\noindent where 
\begin{equation}
R^i\,_{jkl}=\Gamma _{jl,k}^i-\Gamma _{jk,l}^i+\Gamma _{jl}^m.\Gamma
_{mk}^i-\Gamma _{jk}^m.\Gamma _{ml}^i  \label{I.5.-7}
\end{equation}

\noindent are called \textit{components of the (contravariant) curvature
tensor (Riemannian tensor) \ in a co-ordinate basis}.

4. Action of the curvature operator on contravariant tensor fields.

For $V=V^A.e_A=V^B.\partial _B$, $V\in \otimes ^l(M)$ and the bases $e_A$
and $\partial _B$ the following relations can be proved using the properties
of $S_{Ak}\,^{Bl}$ and $\Gamma _{Ai}^B$: 
\begin{equation}  \label{I.5.-15}
[R(\xi ,u)](f.V)=f.[R(\xi ,u)]V\text{ ,}
\end{equation}
\begin{equation}  \label{I.5.-16}
[R(\xi ,u)]V=V^A.[R(\xi ,u)]e_A=V^B.[R(\xi ,u)]\partial _B\text{ ,}
\end{equation}
\begin{equation}  \label{I.5.-17}
[R(\partial _j,\partial _i)]\partial _A=R^B\,_{Aji}.\partial
_B=-\,S_{Ak}\,^{Bl}.R^k\,_{lji}.\partial _B\text{ ,}
\end{equation}

\noindent where 
\begin{equation}
R^B\,_{Aji}=-\,S_{Ak}\,^{Bl}.R^k\,_{lji}\text{ ,\thinspace \thinspace
\thinspace \thinspace \thinspace \thinspace \thinspace \thinspace \thinspace
\thinspace \thinspace \thinspace \thinspace \thinspace \thinspace \thinspace
\thinspace }S_{Ak}\,^{Bl}\,_{,i}=0\text{ ,}  \label{I.5.-19}
\end{equation}
\begin{equation}
R^B\,_{Aji}=\Gamma _{Ai,j}^B-\Gamma _{Aj,i}^B+\Gamma _{Ai}^C.\Gamma
_{Cj}^B-\Gamma _{Aj}^C.\Gamma _{Ci}^B\text{ ,}  \label{I.5.-18}
\end{equation}
\begin{equation}
\lbrack R(\xi ,u)]V=-\,\,S_{Bk}\,^{Al}.V^B.R^k\,_{lij}.\xi ^i.u^j.\partial _A%
\text{ .}  \label{I.5.-20}
\end{equation}

On the other side, it follows from the explicit construction of $[R(\xi
,u)]V $%
\begin{equation}  \label{I.5.-21}
\begin{array}{c}
\lbrack R(\xi
,u)]V=(V^A\,_{;i;j}-V^A\,_{;j;i}+V^A\,_{;k}.T_{ji}\,^k).u^i.\xi ^j.\partial
_A \text{ ,} \\ 
V^A\,_{;i;j}-V^A\,_{;j;i}=-\,\,S_{Bk}\,^{Al}.V^B.R^k\,_{lji}-T_{ji}\,^k.V^A%
\,_{;k} \text{ ,} \\ 
\lbrack R(\partial _j,\partial _i)]-\nabla _{T(\partial _j,\partial
_i)}]V=(V^A\,_{;i;j}-V^A\,_{;j;i}).\partial _A\text{ .}
\end{array}
\end{equation}

5. The action of the curvature operator on covariant vector fields is
determined by its structure and by the action of the covariant differential
operator on covariant tensor field.

In a co-ordinate basis 
\begin{eqnarray}
\lbrack R(\xi ,u)]p &=&(\nabla _\xi \nabla _u-\nabla _u\nabla _\xi -\nabla
_{\pounds _\xi u})p=p_l.P^l\,_{ikj}.\xi ^k.u^j.dx^i=  \nonumber
\label{Ch 4 3.34a} \\
&=&(p_{i;j;k}-p_{i;k;j}+T_{kj}\,^l.p_{i;l}).u^j.\xi ^k.dx^i\text{ ,}
\label{Ch 4 3.34a}
\end{eqnarray}
%
\begin{equation}  \label{Ch 4 3.34b}
[R(\partial _k,\partial _l)]dx^i=P^i\,_{jkl}.dx^j\text{ ,}
\end{equation}

\noindent where 
\begin{equation}
P^i\,_{jkl}=P_{jl,k}^i-P_{jk,l}^i+P_{jk}^m.P_{ml}^i-P_{jl}^m.P_{mk}^i=-P^i%
\,_{jlk}\text{ }  \label{Ch 4 3.34c}
\end{equation}

\noindent are called components of the \textit{covariant curvature tensor}
in a co-ordinate basis.

\textit{Special case}: $S=C:f^i\,_j=g_j^i:P_{jk}^i+\Gamma _{jk}^i=0$. 
\begin{equation}  \label{Ch 4 3.34d}
P^i\,_{jkl}=-R^i\,_{jkl}\text{ .}
\end{equation}

In a non-co-ordinate basis: 
\begin{eqnarray}
\lbrack R(\xi ,u)]p &=&p_\delta .P^\delta \,_{\alpha \beta \gamma }.\xi
^\beta .u^\gamma .e^\alpha =  \nonumber \\
&=&(p_{\alpha /\gamma /\beta }-p_{\alpha /\beta /\gamma }+T_{\beta \gamma
}\,^\delta .p_{\alpha /\delta }).\xi ^\beta .u^\gamma .e^\alpha \text{ ,}
\label{Ch 4 3.34e}
\end{eqnarray}
%
\begin{equation}  \label{Ch 4 3.34f}
P^\alpha \,_{\delta \beta \gamma }=e_\beta P_{\delta \gamma }^\alpha
-e_\gamma P_{\delta \beta }^\alpha +P_{\delta \beta }^\sigma .P_{\sigma
\gamma }^\alpha -P_{\delta \gamma }^\sigma .P_{\sigma \beta }^\alpha
-C_{\beta \gamma }\,^\sigma .P_{\delta \sigma }^\alpha \text{ .}
\end{equation}

$P^\alpha \,_{\delta \beta \gamma }=-P^\alpha \,_{\delta \gamma \beta }$ are
called components of the \textit{covariant curvature tensor} in a
non-co-ordinate basis.

For a covariant tensor field $W=W_A.dx^A=W_C.e^C\in \otimes _k(M)$ we have
the relation in a co-ordinate basis

\begin{equation}  \label{Ch 4 3.34g}
W_{A;i;j}-W_{A;j;i}=S_{Am}\,^{Bn}.W_B.P^m\,_{nij}+W_{A;l}.T_{ij}\,^l\text{ ,}
\end{equation}

\noindent and in a non-co-ordinate basis 
\begin{equation}
W_{A/\beta /\gamma }-W_{A/\gamma /\beta }=S_{A\alpha }\,^{B\delta
}.W_B.P^\alpha \,_{\delta \beta \gamma }+W_{A/\delta }.T_{\beta \gamma
}\,^\delta \text{ .}  \label{Ch 4 3.34h}
\end{equation}

\subsection{Bianchi identities}

If we write down the cycle of the action of the curvature operator on
contravariant vector fields, i. e. if we write 
\begin{equation}  \label{I.5.-26}
<[R(\xi ,u)]v>\,=\,[R(\xi ,u)]v+[R(v,\xi )]u+[R(u,v)]\xi \text{ ,}
\end{equation}

\noindent and put the explicit form of every term in the cycle, then by the
use of the covariant and the Lie differential operator, after some (not so
difficult) calculations, we can find identities of the type written in the
form 
\begin{equation}
\begin{array}{c}
\lbrack R(\xi ,u)]v+[R(v,\xi )]u+[R(u,v)]\xi \equiv \\ 
\equiv T(T(\xi ,u),v)+T(T(v,\xi ),u)+T(T(u,v),\xi )+ \\ 
+\,(\nabla _\xi T)(u,v)+\,(\nabla _vT)(\xi ,u)+\,(\nabla _uT)(v,\xi )\text{ ,%
}
\end{array}
\label{I.5.-27}
\end{equation}

\noindent or in the form 
\begin{equation}
<[R(\xi ,u)]v>\,\equiv \,<T(T(\xi ,u),v)>+<(\nabla _\xi T)(u,v)>\text{ .}
\label{I.5.-28}
\end{equation}

The identities are called \textit{Bianchi identities of first type} (or 
\textit{of the type 1}.), where 
\[
\begin{array}{c}
<T(T(\xi ,u),v)>\,\equiv T(T(\xi ,u),v)+T(T(v,\xi ),u)+T(T(u,v),\xi ) \text{
,} \\ 
<(\nabla _\xi T)(u,v)>\,\equiv \,(\nabla _\xi T)(u,v)+\,(\nabla _vT)(\xi
,u)+\,(\nabla _uT)(v,\xi ) \text{ ,} \\ 
\nabla _\xi [T(u,v)]=(\nabla _\xi T)(u,v)+T(\nabla _\xi u,v)+T(u,\nabla _\xi
v)\text{ .}
\end{array}
\]

By the use of the curvature operator and the covariant differential operator
a new operator $(\nabla _wR)(\xi ,u)$ can be constructed in the form 
\begin{equation}  \label{I.5.-34}
(\nabla _wR)(\xi ,u)=[\nabla _w,R(\xi ,u)]-R(\nabla _w\xi ,u)-R(\xi ,\nabla
_wu)\text{ ,}
\end{equation}

\noindent where 
\[
\lbrack \nabla _w,R(\xi ,u)]=\nabla _w\circ R(\xi ,u)-R(\xi ,u)\circ \nabla
_w\text{ ,\thinspace \thinspace \thinspace \thinspace \thinspace \thinspace
\thinspace }w,\xi ,u\in T(M)\text{ .} 
\]

$(\nabla _wR)(\xi ,u)$ has the structure 
\begin{equation}  \label{I.5.-35}
\begin{array}{c}
(\nabla _wR)(\xi ,u)=\nabla _w\nabla _\xi \nabla _u-\nabla _w\nabla _u\nabla
_\xi +\nabla _u\nabla _\xi \nabla _w-\nabla _\xi \nabla _u\nabla _w+ \\ 
+\,\,\nabla _u\nabla _{\nabla _w\xi }-\,\nabla _{\nabla _w\xi }\nabla
_u+\,\nabla _{\nabla _wu}\nabla _\xi -\,\nabla _\xi \nabla _{\nabla _wu}+ \\ 
+\nabla _{\pounds _\xi u}\nabla _w-\nabla _w\nabla _{\pounds _\xi u}+\nabla
_{\pounds _\xi \nabla _wu}-\nabla _{\pounds _u\nabla _w\xi }\text{ .}
\end{array}
\end{equation}

This operator obeys the s. c. \textit{Bianchi identity of second type} (or 
\textit{of the type 2}.) 
\begin{equation}  \label{I.5.-36}
<(\nabla _wR)(\xi ,u)>\,\equiv \,<R(w,T(\xi ,u))>\text{ ,}
\end{equation}

\noindent where 
\[
\begin{array}{c}
<(\nabla _wR)(\xi ,u)>\,\equiv \,(\nabla _wR)(\xi ,u)+(\nabla _uR)(w,\xi
)+(\nabla _\xi R)(u,w)\text{ ,} \\ 
<R(w,T(\xi ,u))>\,\equiv \,R(w,T(\xi ,u))+R(u,T(w,\xi ))+R(\xi ,T(u,w))\text{
.}
\end{array}
\]

The Bianchi identity of type 2. can be written in a co-ordinate or in a
non-co-ordinate basis as an identity of the components of the contravariant
curvature tensor 
\begin{equation}  \label{I.5.-37}
R^i\,_{j<kl;m>}\equiv R^i\,_{j<kn}.T_{lm>}\,^n\equiv
-\,R^i\,_{jn<k}.T_{lm>}\,^n\text{ ,}
\end{equation}

\noindent where 
\begin{equation}
\begin{array}{c}
R^i\,_{j<kl;m>}\equiv R^i\,_{jkl;m}+R^i\,_{jmk;l}+R^i\,_{jlm;k}\text{ ,} \\ 
R^i\,_{j<kn}.T_{lm>}\,^n\equiv
R^i\,_{jkn}.T_{lm}\,^n+R^i\,_{jmn}.T_{kl}\,^n+R^i\,_{jlr}.T_{mk}\,^r\text{ .}
\end{array}
\label{I.5.-39}
\end{equation}

For the commutator 
\[
[\nabla _w,R(\xi ,u)]=\nabla _w\circ R(\xi ,u)-R(\xi ,u)\circ \nabla _w 
\]

\noindent the following commutation identity is valid: 
\begin{equation}
<[\nabla _w,R(\xi ,u)]>\,\equiv \,-\,<R(w,\pounds _\xi u)>\text{ ,}
\label{I.5.-40}
\end{equation}

\noindent where 
\begin{equation}
\begin{array}{c}
<[\nabla _w,R(\xi ,u)]>\,\equiv [\nabla _w,R(\xi ,u)]+[\nabla _u,R(w,\xi
)]+[\nabla _\xi ,R(u,w)]\text{ ,} \\ 
<R(w,\pounds _\xi u)>\,\equiv R(w,\pounds _\xi u)+R(u,\pounds _w\xi )+R(\xi
,\pounds _uw)\text{ .}
\end{array}
\label{I.5.-41}
\end{equation}

The curvature operator and the Bianchi identities have been applied in
differentiable manifolds with one affine connection. They can also find
applications in considerations concerning the characteristics of
differentiable manifolds with affine connections and metrics. The structure
of the curvature operator induces a construction of an other operator called
deviation operator.

\section{Deviation operator}

By means of the structure of the \textit{curvature operator, 
\begin{equation}  \label{Ch 3 4.1}
R(\xi ,u)=\nabla _\xi \nabla _u-\nabla _u\nabla _\xi -\nabla _{\pounds _\xi
u}=[\nabla _\xi ,\nabla _u]-\nabla _{[\xi ,u]}\text{ ,}
\end{equation}
}

\noindent the commutator $[\nabla _w,R(\xi ,u)]$ [$w,\xi ,u\in T(M)$] can be
presented in the form 
\begin{equation}
\lbrack \nabla _w,R(\xi ,u)]=[\nabla _w,\pounds \Gamma (\xi ,u)]+[\nabla
_w,[\nabla _\xi ,\nabla _u]]-[\nabla _w,[\pounds _\xi ,\nabla _u]]\text{ ,}
\label{Ch 3 4.2}
\end{equation}

\noindent where 
\begin{equation}
\pounds \Gamma (\xi ,u)=\pounds _\xi \nabla _u-\nabla _u\pounds _\xi -\nabla
_{\pounds _\xi u}=[\pounds _\xi ,\nabla _u]-\nabla _{[\xi ,u]}\text{ .}
\label{Ch 3 4.3}
\end{equation}

The operator $\pounds \Gamma (\xi ,u)$ appears as a new operator,
constructed by means of the Lie differential operator and the covariant
differential operator \cite{Manoff-1} - \cite{Manoff-4}.

\begin{definition}
{\bf \ }The operator $\pounds \Gamma (\xi ,u)$ is called {\it deviation
operator}. Its properties contain the relations:
\end{definition}
%

1. Action of the deviation operator on a function $f$ : $[\pounds \Gamma
(\xi ,u)]f=0$\thinspace \thinspace ,\thinspace \thinspace \thinspace $f\in
C^r(M)$, $r\geq 2$.

2. Action of the deviation operator on a contravariant vector field:

$[\pounds \Gamma (\xi ,u)](fv)=f[\pounds \Gamma (\xi ,u)]v$ , $\xi ,u,v\in
T(M)$,

$[\pounds \Gamma (\xi ,u)]v=v^\beta [\pounds \Gamma (\xi ,u)]e_\beta
=v^j[\pounds \Gamma (\xi ,u)]\partial _j=u^\gamma v^\beta [\pounds \Gamma
(\xi ,e_\gamma )]e_\beta =u^jv^i[\pounds \Gamma (\xi ,\partial _j)]\partial
_i$.

The connections between the action of the deviation operator and that of the
curvature operator on a contravariant vector field can be given in the form 
\begin{equation}  \label{Ch 3 4.4}
\begin{array}{c}
\lbrack \pounds \Gamma (\xi ,u)]v=[R(\xi ,u)]v+[\nabla _u\nabla _v-\nabla
_{\nabla _uv}]\xi - \\ 
-T(\xi ,\nabla _uv)+\nabla _u[T(\xi ,v)]\text{ .}
\end{array}
\end{equation}

In a co-ordinate basis $[\pounds \Gamma (\xi ,\partial _l)]\partial _k$ has
the form 
\begin{equation}  \label{Ch 3 4.5}
\lbrack \pounds \Gamma (\xi ,\partial _l)]\partial _k=[\xi ^i\text{ }%
_{;k;l}-R^i\text{ }_{klj}.\xi ^j+(T_{jk}\,^i.\xi ^j)_{;l}].\partial
_i=(\pounds _\xi \Gamma _{kl}^i).\partial _i\text{ ,}
\end{equation}

\noindent where 
\[
\nabla _{\partial _j}[T(\xi ,\partial _i)]-T(\xi ,\nabla _{\partial
_j}\partial _i)=(T_{li}\,^k.\xi ^l)_{;j}.\partial _k\text{ .} 
\]

$\pounds _\xi \Gamma _{kl}^i$ is called \textit{Lie derivative of
contravariant affine connection} along the contravariant vector field $\xi $%
. It can be written also in the form 
\begin{equation}  \label{Ch 3 4.6}
\pounds _\xi \Gamma _{kl}^i=\xi ^i\text{ }_{,k,l}+\xi ^j.\Gamma
_{kl,j}^i-\xi ^i\text{ }_{,j}.\Gamma _{kl}^j+\xi ^j\text{ }_{,k}.\Gamma
_{jl}^i+\xi ^j\text{ }_{,l}.\Gamma _{kj}^i\text{ .}
\end{equation}

By means of $\pounds _\xi \Gamma _{kl}^i$ the expression for $[\pounds
\Gamma (\xi ,u)]v$ can be presented in the form 
\begin{equation}  \label{Ch 3 4.7}
\begin{array}{c}
\lbrack \pounds \Gamma (\xi ,u)]v=v^k.u^l.(\pounds _\xi \Gamma
_{kl}^i).\partial _i= \\ 
=[\xi ^i\text{ }_{;k;l}.v^k.u^l-R^i\text{ }_{klj}.v^k.u^l.\xi
^j+(T_{jk}\,^i.\xi ^j)_{;l}.v^k.u^l].\partial _i\text{ .}
\end{array}
\end{equation}

In this way, the second covariant derivative $\nabla _u\nabla _v\xi $ of the
contravariant vector field $\xi $ can be presented by means of the deviation
operator in the form 
\begin{equation}  \label{Ch 3 4.8}
\begin{array}{c}
\nabla _u\nabla _v\xi =([R(u,\xi )]v)+\nabla _\xi \nabla _uv-\pounds _\xi
(\nabla _uv)-\nabla _u[T(\xi ,v)]+ \\ 
+[\pounds \Gamma (\xi ,u)]v= \\ 
=([R(u,\xi )]v)+\nabla _\xi \nabla _uv-\nabla _u\pounds _\xi v-\nabla
_{\pounds _\xi u}v-\nabla _u[T(\xi ,v)]\text{ .}
\end{array}
\end{equation}

For $v=u$ the last identity is called \textit{\ generalized} \textit{%
deviation identity} \cite{Manoff-1}. It is used for analysis of deviation
equations in spaces with affine connection and metric ($L_n$-spaces, $U_n$%
-spaces and $V_n $-spaces), where deviation equations are considered with
respect to their structure and solutions \cite{Bazanski} - \cite{Ciufolini-2}%
, and as a theoretical ground for gravitational wave detectors in (pseudo)
Riemannian spaces without torsion ($V_n$-spaces) \cite{Weber} - \cite
{Mashhoon}. Deviation equations of Synge and Schild and its generalization
for $(\overline{L}_n,g)$-spaces are considered in \cite{Manoff-6}.

2. Action of the deviation operator on contravariant tensor field 
\begin{equation}  \label{Ch 3 4.9}
\begin{array}{c}
\lbrack \pounds \Gamma (\xi ,u)]V=u^\gamma .V^A.[\pounds \Gamma (\xi
,e_\gamma )]e_A=u^\gamma .V^B.(\pounds _\xi \Gamma _{B\gamma }^A).e_A= \\ 
=-(S_{B\alpha }\text{ }^{A\beta }.V^B.\pounds _\xi \Gamma _{\beta \gamma
}^\alpha .u^\gamma )e_A\text{ , \thinspace \thinspace \thinspace \thinspace
\thinspace \thinspace \thinspace }V\in \otimes ^k(M)\text{ ,}
\end{array}
\end{equation}

\noindent where 
\begin{equation}
\begin{array}{c}
\pounds _\xi \Gamma _{B\gamma }^A=-S_{B\alpha }\text{ }^{A\beta }.\pounds
_\xi \Gamma _{\beta \gamma }^\alpha \text{ ,} \\ 
([\pounds \Gamma (\xi ,u)]e_B)=(\pounds _\xi \Gamma _{B\gamma }^A).u^\gamma
.e_A= \\ 
=-S_{B\alpha }\text{ }^{A\beta }.[\xi ^\alpha \text{ }_{/\beta /\gamma
}-R^\alpha \text{ }_{\beta \gamma \delta }.\xi ^\delta +(T_{\delta \beta
}\,^\alpha .\xi ^\delta )_{/\gamma }].u^\gamma .e_A\text{ .}
\end{array}
\label{Ch 3 4.10}
\end{equation}

3. The deviation operator obeys identity analogous to the 1. type Bianchi
identity for the curvature operator 
\begin{equation}  \label{Ch 3 4.11}
\begin{array}{c}
\langle [\pounds \Gamma (\xi ,u)]v\rangle \equiv \langle (\nabla _\xi \nabla
_u-\nabla _{\nabla _\xi u})v\rangle +\langle T(T(\xi ,u),v)\rangle - \\ 
-\langle T(u,\nabla _\xi v)\rangle \text{ , }\xi ,u,v\in T(M)\text{ ,}
\end{array}
\end{equation}

\noindent where 
\begin{equation}
\begin{array}{c}
\langle [\pounds \Gamma (\xi ,u)]v\rangle =[\pounds \Gamma (\xi
,u)]v+[\pounds \Gamma (v,\xi )]u+[\pounds \Gamma (u,v)]\xi \text{ ,} \\ 
\langle (\nabla _\xi \nabla _u-\nabla _{\nabla _\xi u})v\rangle =(\nabla
_\xi \nabla _u-\nabla _{\nabla _\xi u})v+(\nabla _v\nabla _\xi -\nabla
_{\nabla _v\xi })u+ \\ 
+(\nabla _u\nabla _v-\nabla _{\nabla _uv})\xi \text{ ,} \\ 
\langle T(u,\nabla _\xi v)\rangle =T(u,\nabla _\xi v)+T(v,\nabla _u\xi
)+T(\xi ,\nabla _vu)\text{ .}
\end{array}
\label{Ch 3 4.12}
\end{equation}

In a non-co-ordinate basis this identity obtains the form 
\begin{equation}  \label{Ch 3 4.13}
\begin{array}{c}
(\pounds _\xi \Gamma _{\alpha \beta }^\gamma ).v^\alpha .u^\beta +(\pounds
_u\Gamma _{\alpha \beta }^\gamma ).\xi ^\alpha .v^\beta +(\pounds _v\Gamma
_{\alpha \beta }^\gamma ).u^\alpha .\xi ^\beta \equiv \\ 
\equiv \xi ^\gamma \text{ }_{/\alpha /\beta }.v^\alpha .u^\beta +u^\gamma 
\text{ }_{/\alpha /\beta }.\xi ^\alpha .v^\beta +v^\gamma \text{ }_{/\alpha
/\beta }.u^\alpha .\xi ^\beta + \\ 
+T_{\langle \alpha \beta }\,^\kappa T_{\kappa \delta \rangle }\,^\gamma
.v^\alpha .\xi ^\beta .u^\delta - \\ 
-T_{\alpha \beta }\,^\gamma .(u^\alpha .v^\beta \text{ }_{/\delta }.\xi
^\delta +v^\alpha .\xi ^\beta \text{ }_{/\delta }.u^\delta +\xi ^\alpha
.u^\beta \text{ }_{/\delta }.v^\delta )\text{ .}
\end{array}
\end{equation}

The commutator of the covariant differential operator and the deviation
operator obeys the following identity 
\begin{equation}  \label{Ch 3 4.14}
\langle [\nabla _w,\pounds \Gamma (\xi ,u)]\rangle \equiv \langle [\nabla
_w,[\pounds _\xi ,\nabla _u]]\rangle -\langle R(w,\pounds _\xi u)\rangle 
\text{ ,}
\end{equation}

\noindent where 
\[
\begin{array}{c}
\langle [\nabla _w,\pounds \Gamma (\xi ,u)]\rangle =[\nabla _w,\pounds
\Gamma (\xi ,u)]+[\nabla _u,\pounds \Gamma (v,\xi )]+[\nabla _\xi ,\pounds
\Gamma (u,v)]\text{,} \\ 
\langle [\nabla _w,[\pounds _\xi ,\nabla _u]]\rangle =[\nabla _w,[\pounds
_\xi ,\nabla _u]]+[\nabla _u,[\pounds _w,\nabla _\xi ]]+[\nabla _\xi
,[\pounds _u,\nabla _w]]\text{ ,} \\ 
\langle R(w,\pounds _\xi u)\rangle =R(w,\pounds _\xi u)+R(u,\pounds _w\xi
)+R(\xi ,\pounds _uw)\text{ ,} \\ 
\xi ,u,w\in T(M)\text{.}
\end{array}
\]

4. The action of the deviation operator on covariant vector fields is
determined by its structure and especially by the Lie differential operator.

In a non-co-ordinate basis 
\begin{equation}  \label{Ch 4 4.37}
\lbrack \pounds \Gamma (\xi ,e_\gamma )]e^\alpha =\pounds _\xi \nabla
_{e_\gamma }e^\alpha -\nabla _{e_\gamma }\pounds _\xi e^\alpha -\nabla
_{\pounds _\xi e_\gamma }e^\alpha =(\pounds _\xi P_{\beta \gamma }^\alpha
).e^\beta \text{ ,}
\end{equation}

\noindent where 
\begin{eqnarray}
\pounds _\xi P_{\beta \gamma }^\alpha &=&\xi ^\delta .e_\delta P_{\beta
\gamma }^\alpha +P_{\delta \gamma }^\alpha .e_{\underline{\beta }}\xi ^{\overline{\delta }}+P_{\delta \gamma }^\alpha .(P_{\beta \rho }^\delta
+\Gamma _{\underline{\beta }\rho }^{\overline{\delta }}+C_{\underline{\beta }\rho }\,^{\overline{\delta }}).\xi ^\rho -  \nonumber \\
&&-e_\gamma (e_{\underline{\beta }}\xi ^{\overline{\alpha }})-e_\gamma
[(P_{\beta \rho }^\alpha +\Gamma _{\underline{\beta }\rho }^{\overline{\alpha }}+C_{\underline{\beta }\rho }\,^{\overline{\alpha }}).\xi ^\rho ]- 
\nonumber \\
&&-P_{\beta \gamma }^\delta .[e_{\underline{\delta }}\xi ^{\overline{\alpha }}+(P_{\delta \rho }^\alpha +\Gamma _{\underline{\delta }\rho }^{\overline{\alpha }}+C_{\underline{\delta }\rho }\,^{\overline{\alpha }}).\xi ^\rho ]+ 
\nonumber \\
&&+P_{\beta \delta }^\alpha .(e_\gamma \xi ^\delta -C_{\rho \gamma
}\,^\delta .\xi ^\rho )\text{ ,}  \label{Ch 4 4.38}
\end{eqnarray}
%
\begin{equation}
e_{\underline{\beta }}\xi ^{\overline{\delta }}=f_\beta \,^\sigma .e_\sigma
\xi ^\kappa .f^\delta \,_\kappa \text{ ,\thinspace \thinspace \thinspace
\thinspace \thinspace \thinspace \thinspace \thinspace \thinspace \thinspace
\thinspace \thinspace }\Gamma _{\underline{\beta }\rho }^{\overline{\delta }%
}=f^\delta \,_\kappa .f_\beta \,^\sigma .\Gamma _{\sigma \rho }^\kappa \text{
. }  \label{Ch 4 4.39}
\end{equation}

The expression for $\pounds _\xi P_{\beta \gamma }^\alpha $ can also be
written in the form 
\begin{equation}  \label{Ch 4 4.43}
\pounds _\xi P_{\beta \gamma }^\alpha =-P^\alpha \,_{\beta \gamma \delta
}.\xi ^\delta -\xi ^{\overline{\alpha }}\,_{/\underline{\beta }/\gamma }+T_{%
\underline{\beta }\delta }\,^{\overline{\alpha }}.\xi ^\delta \,_{/\gamma
}+T_{\underline{\beta }\delta }\,^{\overline{\alpha }}\,_{/\gamma }.\xi
^\delta \text{ .}
\end{equation}

$\pounds _\xi P_{\beta \gamma }^\alpha $ is called \textit{Lie derivative} 
\textit{of the components }$P_{\beta \gamma }^\alpha $\textit{\ of the
covariant affine connection }$P$\textit{\ in a non-co-ordinate basis.}

\textit{Special case}: $S=e^\varphi .C:f^\alpha \,_\beta =e^\varphi .g_\beta
^\alpha $,\thinspace \thinspace \thinspace \thinspace \thinspace $%
f^i\,_j=e^\varphi .g_j^i$,\thinspace \thinspace \thinspace \thinspace
\thinspace $f_i\,^j=e^{-\varphi }.g_i^j$. 
\begin{equation}  \label{Ch 4 4.45}
\pounds _\xi P_{jk}^i=-P^i\,_{jkl}.\xi ^l-\xi ^i\,_{;j;k}+T_{jl}\,^i.\xi
^l\,_{;k}+T_{jl}\,^i\,_{;k}.\xi ^l\text{ .}
\end{equation}

The Lie derivative of the components of the covariant affine connection $P$
could be used in considerations related to deviation equations for covariant
vector fields.

\section{Extended covariant differential operator. Extended derivative}

If $\Gamma _{jk}^i$ are components of a contravariant affine connection $%
\Gamma $ and $P_{jk}^i$ are components of a covariant affine connection $P$
in a given (here co-ordinate) basis in a $(\overline{L}_n,g)$-space, then $%
\overline{\Gamma }\,_{jk}^i$ and $\overline{P}\,_{jk}^i$%
\[
\begin{array}{c}
\overline{\Gamma }\,_{jk}^i=\Gamma _{jk}^i-\overline{A}\,_{\,\,\,\,jk}^i%
\text{ ,\thinspace \thinspace \thinspace \thinspace \thinspace \thinspace
\thinspace \thinspace }\overline{P}\,_{jk}^i=P_{\,jk}^i-\overline{B}%
\,^i\,_{jk}\text{ , \thinspace } \\ 
\text{\thinspace \thinspace }\overline{A}=\overline{A}\,^i\,_{jk}.\partial
_i\otimes dx^j\otimes dx^k\text{ , \thinspace \thinspace \thinspace
\thinspace }\overline{B}=\overline{B}\,^i\,_{jk}.\partial _i\otimes
dx^j\otimes dx^k\text{ , \thinspace \thinspace \thinspace }\overline{A}\text{
,\thinspace \thinspace }\overline{B}\in \otimes ^1\,_2(M)\text{ ,}
\end{array}
\]

\noindent are components (in the same basis) of a new contravariant affine
connection $\overline{\Gamma }$ and a new covariant affine connection $%
\overline{P}$ respectively.

$\overline{\Gamma }$ and $\overline{P}$ correspond to a new [extended with
respect to $\nabla _u$, $u\in T(M)$] covariant differential operator $%
^e\nabla _u$%
\[
\begin{array}{c}
^e\nabla _{\partial _k}\partial _j= \overline{\Gamma }\,_{jk}^i.\partial _i%
\text{ ,\thinspace \thinspace \thinspace \thinspace \thinspace \thinspace
\thinspace \thinspace \thinspace \thinspace \thinspace \thinspace \thinspace
\thinspace \thinspace \thinspace \thinspace \thinspace \thinspace \thinspace
\thinspace \thinspace \thinspace \thinspace \thinspace }^e\nabla _{e_\beta
}e_\alpha =\overline{\Gamma \,}_{\alpha \beta }^\gamma .e_\gamma \text{ ,}
\\ 
\text{\thinspace \thinspace \thinspace \thinspace \thinspace }^e\nabla
_{\partial _k}dx^i=\overline{P}\,_{jk}^i.dx^j\text{ ,\thinspace \thinspace
\thinspace \thinspace \thinspace \thinspace \thinspace \thinspace \thinspace
\thinspace \thinspace \thinspace \thinspace \thinspace \thinspace \thinspace
\thinspace }^e\nabla _{e_\gamma }e^\alpha =\overline{P\,}_{\beta \gamma
}^\alpha .e^\beta \text{ ,\thinspace }
\end{array}
\]

with the same properties as the covariant differential operator $\nabla _u$.

If we choose the tensors $\overline{A}$ and $\overline{B}$ with certain
predefined properties, then we can find $\overline{\Gamma }$ and $\overline{P%
}$ with predetermined characteristics. For instance, we can find $\overline{P%
}$ for which $^e\nabla _ug=0$, $\forall u\in T(M)$, $g\in \otimes _2(M)$,
although $\nabla _ug\neq 0$ for the covariant affine connection $P$. On the
other side, $\overline{A}\,^i\,_{jk}$ and $\overline{B}\,^i\,_{jk}$ are
related to each other on the basis of the commutation relations of $^e\nabla
_u$ and $\nabla _u$ with the contraction operator $S$. From 
\[
\nabla _u\circ S=S\circ \nabla _u\text{ ,\thinspace \thinspace \thinspace
\thinspace \thinspace \thinspace \thinspace \thinspace \thinspace \thinspace
\thinspace }^e\nabla _u\circ S=S\circ \,\,^e\nabla _u\text{ ,} 
\]

\noindent we have 
\[
(S\circ \,^e\nabla _{\partial _k})(\partial _j\otimes dx^i)=\Gamma
\,_{jk}^l.f^i\,_l-\overline{A}\,^l\,_{jk}.f^i\,_l+P\,_{lk}^i.f^l\,_j-%
\overline{B}\,^i\,_{lk}.f^l\,_j\text{ ,} 
\]
\[
\,(^e\nabla _{\partial _k}\circ S)(\partial _j\otimes dx^i)=\,^e\nabla
_{\partial _k}(S(\partial _j\otimes dx^i))=\,^e\nabla _{\partial
_k}(f^i\,_j)=\partial _k(f^i\,_j)=f^i\,_{j,k}\text{ .} 
\]

Therefore, 
\[
f^i\,_{j,k}=\Gamma \,_{jk}^l.f^i\,_l-\overline{A}\,^l\,_{jk}.f^i\,_l+P%
\,_{lk}^i.f^l\,_j-\overline{B}\,^i\,_{lk}.f^l\,_j\text{ .} 
\]

Since $(\nabla _{\partial _k}\circ S)(\partial _j\otimes dx^i)=(S\circ
\nabla _{\partial _k})(\partial _j\otimes dx^i)$ leads to the relation $%
f^i\,_{j,k}=\Gamma \,_{jk}^l.f^i\,_l+P\,_{lk}^i.f^l\,_j$, we obtain the
connection between $\overline{A}\,^i\,_{jk}$ and $\overline{B}\,^i\,_{jk}$
in the form $\overline{A}\,^l\,_{jk}.f^i\,_l+\overline{B}\,^i\,_{lk}.f^l%
\,_j=0$.

Therefore, $\overline{B}\,^i\,_{jk}=-\,\,\,\,\overline{A}\,^l\,_{mk}.f^i%
\,_l.f_j\,^m=-\,\,\,\overline{A}\,^{\overline{i}}\,_{\underline{j}k}$ and $%
\overline{A}\,^i\,_{jk}=-\,\,\,\overline{B}\,^m\,_{lk}.f^l\,_j.f_m\,^i=-\,\,%
\,\overline{B}\,^{\underline{i}}\,_{\overline{j}k}$.

We can write $^e\nabla _{\partial _k}$ in the form $^e\nabla _{\partial
_k}=\nabla _{\partial _k}-\overline{A}_{\partial _k}$ \thinspace \thinspace
\thinspace \thinspace with \thinspace \thinspace \thinspace $\overline{A}%
_{\partial _k}=\overline{A}\,^i\,_{jk}.\partial _i\otimes dx^j$.

$^e\nabla _u$ can also be written in the form $^e\nabla _u=\nabla _u-%
\overline{A}_u$\thinspace \thinspace \thinspace \thinspace \thinspace
\thinspace \thinspace with\thinspace \thinspace \thinspace \thinspace
\thinspace \thinspace \thinspace $\overline{A}_u=\overline{A}%
\,^i\,_{jk}.u^k.\partial _i\otimes dx^j$.

$\overline{A}_u$ appears as a mixed tensor field of second rank but acting
on tensor fields as a covariant differential operator because $^e\nabla _u$
is defined as covariant differential operator with the same properties as
the covariant differential operator $\nabla _u$.

\begin{definition}
{\it Extended to }$\nabla _u${\it \ covariant differential operator}. The
linear differential operator $^e\nabla _u:v\rightarrow \,^e\nabla _uv=\widetilde{v}$, $v,\widetilde{v}\in \otimes ^k\,_l(M)$, with the properties
of $\nabla _u$.
\end{definition}
%

From the properties of $^e\nabla _u$ and $\nabla _u$ the properties of the
operator $\overline{A}_u$ follow 
\[
\overline{A}_u:v\rightarrow \overline{A}_uv\text{ ,\thinspace \thinspace
\thinspace \thinspace \thinspace \thinspace \thinspace \thinspace \thinspace 
}u\in T(M)\text{ ,\thinspace \thinspace \thinspace \thinspace \thinspace
\thinspace \thinspace \thinspace \thinspace \thinspace \thinspace \thinspace
\thinspace }v,\overline{A}_uv\in \otimes ^k\,_l(M)\text{ .} 
\]

(a) $\overline{A}_u(v+w)=\overline{A}_uv+\overline{A}_uw$ , $v,w\in \otimes
^k\,_l(M)$.

(b) $\overline{A}_u(f.v)=f.\overline{A}_uv$ ,\thinspace \thinspace
\thinspace \thinspace \thinspace $f\in C^r(M)$.

(c) $\overline{A}_{u\,+\,v}w=\,\overline{A}_uw+\,\overline{A}_vw.$

(d) $\overline{A}_{f.u}v=f.\overline{A}_uv.$

(e) $\overline{A}_uf=0$.

(f) $\overline{A}_u(v\otimes w)=\,\overline{A}_uv\otimes w+v\otimes \,%
\overline{A}_uw$ , $v\in \otimes ^k\,_l(M)$, $w\in \otimes ^m\,_r(M)$.

(g) $\overline{A}_u\circ S=S\circ \overline{A}_u$ (commutation relation with
the contraction operator $S$).

All properties of $\overline{A}_u$ correspond to the properties of $^e\nabla
_u$ and $\nabla _u$ as well defined covariant differential operators. In
fact, $\overline{A}_u$ can be defined as $\overline{A}_u=\nabla
_u-\,^e\nabla _u$. If $\overline{A}_u$ is a given mixed tensor field, then $%
^e\nabla _u$ can be constructed in an unique way.

On the grounds of the above considerations we can formulate the following
proposition:

\begin{proposition}
To every covariant differential operator $\nabla _u$ and a given tensor
field $\overline{A}_u\in \otimes ^1\,_1(M)$ acting as a covariant
differential operator on tensor field in a $(\overline{L}_n,g)$-space
corresponds an extended covariant differential operator $^e\nabla _u=\nabla
_u-\overline{A}_u$.
\end{proposition}
%

In accordance to its property (c): $\overline{A}_{u\,+\,\,v}=\overline{A}_u+%
\overline{A}_v$, $\overline{A}_u$ has to be linear to $u$. On the other
side, $\overline{A}_u$ as a mixed tensor field of second rank can be
represented by the use of the existing in $(\overline{L}_n,g)$-space
contravariant and covariant metrics $\overline{g}$ and $g$ respectively in
the form $\overline{A}_u=\overline{g}(A_u)$, where $A_u$ is a covariant
tensor field of second rank constructed by the use of a tensor field $C$ and
a contravariant vector field $u$ in such a way that $A_u$ is linear to $u$.
There are at least three possibilities for construction of a covariant
tensor field of second rank $A_u$ in such a way that $A_u$ is linear to $u$,
i. e. $A_u=C(u)=C_{ij}(u).dx^i\otimes dx^j$ with

1. $A_u=C(u)=A(u)=A_{ij\overline{k}}.u^k.dx^i\otimes dx^j$, $%
A=A_{ijk}.dx^i\otimes dx^j\otimes dx^k\in \otimes _3(M)$, $u\in T(M)$.

2. $A_u=C(u)=\nabla _uB=B_{ij;k}.u^k.dx^i\otimes dx^j$, $B=B_{ij}.dx^i%
\otimes dx^j\in \otimes _2(M)$, $u\in T(M)$.

3. $A_u=C(u)=A(u)+\nabla _uB$, $A(u)=A_{ij\overline{k}}.u^k.dx^i\otimes dx^j$%
, $A=A_{ijk}.dx^i\otimes dx^j\otimes dx^k\in \otimes _3(M)$, $u\in T(M)$; $%
\nabla _uB=B_{ij;k}.u^k.dx^i\otimes dx^j$, $B=B_{ij}.dx^i\otimes dx^j\in
\otimes _2(M)$, $u\in T(M)$.

An extended covariant differential operator $^e\nabla _u=\nabla _u-\overline{%
A}_u$ can obey additional conditions determining the structure of the mixed
tensor field $\overline{A}_u$ (acting on tensor fields as a covariant
differential operator). One can impose given conditions on $^e\nabla _u$
leading to determined properties of $\overline{A}_u$ and vice versa: one can
impose conditions on the tensor field $\overline{A}_u$ leading to determined
properties of $^e\nabla _u$.

Every extended covariant differential operator as well as every covariant
differential operator have their special type of transports of covariant
vector fields.

\section{Metrics}

The notion contraction operator has been introduced, acting on two vectors
belonging to two different vector spaces with equal dimensions in a point of
a differentiable manifold $M$ and juxtaposing to them a function over $M$.
If a contraction operator is acting on two vectors belonging to one and the
same vector space, then this operator is connected with the notion metric.

\begin{definition}
{\it Metric.} Contraction operator $S$ acting on two vectors of one and the
same vector space and mapping them to an element of the field $F$ ($R$ or $C$).
\end{definition}
%

\begin{definition}
{\it Metric }over a differentiable manifold $M$. Contraction operator $S$,
acting on two vector fields which vectors in every given point $x\in M$
belong to one and the same vector space, i.e. $S:(u,v)\rightarrow S(u,v)\in
C^r(M)$, $u_x,v_x\in N_x(M)$.
\end{definition}
%

\subsection{Covariant metric}

\begin{definition}
{\bf \ }{\it Covariant metric.} Contraction operator $S$, acting on two
contravariant vector fields over a manifold $M$, which {\it action} is
identified with the {\it action} of a covariant symmetric tensor field of
rank two on the two vector fields, i.e. 
\begin{equation}
S(u,v)\equiv g(u,v):=S(g,q)=S(g,u\otimes v)=S(g\otimes (u\otimes v))\text{
,\thinspace \thinspace \thinspace \thinspace \thinspace }q=u\otimes v\text{.}
\label{Ch 5 0.1}
\end{equation}
\end{definition}
%

The tensor $g=g_{\alpha \beta }.e^\alpha .e^\beta =g_{ij}.dx^i.dx^j$ is
called \textit{covariant metric tensor field (covariant metric)} and $%
g(x)=g_x\in \otimes _{2/x}(M)$ is called \textit{covariant metric tensor\
(covariant metric)} at a point $x\in M$.

(a) Action of the covariant metric on two contravariant vector fields in a
co-ordinate basis 
\begin{equation}  \label{Ch 5 0.3}
\begin{array}{c}
g(u,v)=g_{kl}.f^k \text{ }_i.f^l\text{ }_j.u^i.v^j=g_{\overline{i}\overline{j%
}}u^i.v^j=g_{kl}.u^{\overline{k}}.v^{\overline{l}}=u_l.v^{\overline{l}}=u_{%
\overline{j}}.v^j\text{ ,} \\ 
g_{\overline{i}\overline{j}}=f^k\text{ }_i.f^l\text{ }_j.g_{kl}\text{
,\thinspace \thinspace \thinspace \thinspace \thinspace \thinspace
\thinspace \thinspace }u^{\overline{k}}=f^k\text{ }_i.u^i\text{ , } \\ 
u_{\overline{j}}=g_{\overline{i}\overline{j}}.u^i\text{ ,\thinspace
\thinspace \thinspace \thinspace \thinspace \thinspace \thinspace \thinspace
\thinspace \thinspace }u_l=g_{kl}.u^{\overline{k}}=g_{\overline{k}l}.u^k%
\text{ .}
\end{array}
\end{equation}

%
\begin{remark}
{\bf \ }$g(u,v)$ is also called {\it scalar product of the contravariant
vector fields }$u$ and $v$ over the manifold $M$. When $v=u$, then 
\begin{eqnarray}
g(u,u)=g_{\overline{\alpha }\overline{\beta }}.u^\alpha .u^\beta =g_{\alpha
\beta }.u^{\overline{\alpha }}.u^{\overline{\beta }}=u_{\overline{\alpha }}.u^\alpha =u_\alpha .u^{\overline{\alpha }}=  \label{Ch 5 0.5} \\
\ :=u^2=\pm \mid u\mid ^2:=\pm \,\,l_u^2\text{ ,}
\end{eqnarray}

and $g(u,u)=u^2=\pm \,\,l_u^2$ is called the {\it square of the length }of
the contravariant vector field $u$.
\end{remark}
%

(b) The action of the covariant metric on a contravariant vector field $u$
can be introduced by means of the contraction operator $S$ in a co--ordinate
basis as 
\begin{equation}  \label{Ch 5 0.6}
\begin{array}{c}
g(u):=S^j \text{ }_k(g,u)=S^j\text{ }_k(g_{ij}.dx^i.dx^j,u^k.\partial _k)=
\\ 
=g_{ij}.u^k.S^j \text{ }_k(dx^i.dx^j,\partial _k)=g_{ij}.u^k.f^j\text{ }%
_k.dx^i= \\ 
=g_{i\overline{k}}.u^k.dx^i=g_{ij}.u^{\overline{j}}.dx^i=u(g)\text{
,\thinspace \thinspace \thinspace \thinspace \thinspace \thinspace
\thinspace \thinspace \thinspace \thinspace \thinspace \thinspace \thinspace
\thinspace \thinspace \thinspace \thinspace }g_{i\overline{k}}=g_{ij}.f^j%
\text{ }_k\text{ .}
\end{array}
\end{equation}

\begin{remark}
The abbreviation $u(g)$ is equivalent to the abbreviation $(u)(g):=S(u,g)$.
It should not be considered as the result of the action of the contravariant
vector field $u$ on $g$. Such action of $u$ on $g$ is (until now) not
defined.
\end{remark}
%

The action of the covariant metric $g$ on a contravariant vector field $u$
considered in index form (in a given basis) is called \textit{lowering
indices }by means of $g$. The result of the action of $g$ on $u\in T(M)$ is
a covariant vector field $g(u)\in T^{*}(M)$. On this ground, $g$ can be
defined as linear mapping (operator) which maps every element of $T(M)$ in a
corresponding element of $T^{*}(M)$, i.e. $g:u\rightarrow g(u)\in T^{*}(M)$, 
$u\in T(M)$.

\subsubsection{Covariant symmetric affine connection}

In a non-co-ordinate basis the covariant affine connection $P$ will have the
form 
\begin{equation}  \label{Ch 5 0.19}
P_{\alpha \beta }^\gamma =\overline{P}\,_{\alpha \beta }^\gamma +\frac
12.U_{\alpha \beta }^{\,\,\,\,\,\,\gamma }\text{ ,}
\end{equation}

\noindent where 
\begin{equation}
\begin{array}{c}
\overline{P}\,_{\alpha \beta }^\gamma =\frac 12(P_{\alpha \beta }^\gamma
+P_{\beta \alpha }^\gamma +C_{\alpha \beta }\text{ }^\gamma )\text{ ,} \\ 
U_{\alpha \beta }^{\,\,\,\,\,\,\gamma }=P_{\alpha \beta }^\gamma -P_{\beta
\alpha }^\gamma -C_{\alpha \beta }\text{ }^\gamma =-U_{\beta \alpha
}^{\,\,\,\,\,\,\gamma }\text{ .}
\end{array}
\label{Ch 5 0.20}
\end{equation}

The components of the covariant derivative of covariant metric tensor field $%
g$ can be presented by means of the covariant symmetric affine connection.
If 
\begin{equation}  \label{Ch 5 0.21}
g_{\alpha \beta ;\gamma }=e_\gamma g_{\alpha \beta }+\overline{P}_{\alpha
\gamma }^\delta .g_{\delta \beta }+\overline{P}_{\beta \gamma }^\delta
.g_{\alpha \delta }
\end{equation}

\noindent is the covariant derivative of the components $g_{\alpha \beta }$
of the covariant metric tensor $g$ with respect to the covariant symmetric
affine connection $\overline{P}$ in a non-co-ordinate basis, then 
\begin{equation}
g_{\alpha \beta /\gamma }=g_{\alpha \beta ;\gamma }+\frac 12(U_{\alpha
\gamma }^{\,\,\,\,\,\,\delta }.g_{\delta \beta }+U_{\beta \gamma
}^{\,\,\,\,\,\,\delta }.g_{\alpha \delta })\text{ .}  \label{Ch 5 0.22}
\end{equation}

On the other side, the components of the covariant symmetric affine
connection $\overline{P}_{\alpha \beta }^\delta $ can be written in the form 
\begin{equation}  \label{Ch 5 0.23}
\begin{array}{c}
g_{\delta \gamma }. \overline{P}_{\alpha \beta }^\delta =-\{\alpha \beta
,\gamma \}+K_{\alpha \beta \gamma }+C_{\alpha \beta \gamma }+\frac
12(g_{\delta \alpha }.U_{\beta \gamma }^{\,\,\,\,\,\,\delta }+g_{\delta
\beta }.U_{\alpha \gamma }^{\,\,\,\,\,\delta })= \\ 
=-\{\alpha \beta ,\gamma \}+C_{\alpha \beta \gamma }+\frac 12(g_{\alpha
\gamma ;\beta }+g_{\beta \gamma ;\alpha }-g_{\alpha \beta ;\gamma })\text{ ,}
\end{array}
\end{equation}

\noindent where 
\begin{equation}
\begin{array}{c}
\{\alpha \beta ,\gamma \}=\frac 12(e_\alpha g_{\beta \gamma }+e_\beta
g_{\alpha \gamma }-e_\gamma g_{\alpha \beta })\text{ ,} \\ 
K_{\alpha \beta \gamma }=\frac 12(g_{\alpha \gamma /\beta }+g_{\beta \gamma
/\alpha }-g_{\alpha \beta /\gamma })\text{ ,} \\ 
C_{\alpha \beta \gamma }=\frac 12(g_{\delta \alpha }.C_{\beta \gamma }\text{ 
}^\delta +g_{\delta \beta }.C_{\alpha \gamma }\text{ }^\delta +g_{\delta
\gamma }.C_{\alpha \beta }\text{ }^\delta )\text{ .}
\end{array}
\label{Ch 5 0.24}
\end{equation}

By means of the last expressions $P_{\alpha \beta }^\gamma $ can be
represented in the form 
\begin{equation}  \label{Ch 5 0.25}
g_{\delta \gamma }.P_{\alpha \beta }^\delta =-\{\alpha \beta ,\gamma
\}+K_{\alpha \beta \gamma }+U_{\alpha \beta \gamma }+C_{\alpha \beta \gamma }%
\text{ ,}
\end{equation}

\noindent where 
\begin{equation}
U_{\alpha \beta \gamma }=\frac 12(g_{\delta \alpha }.U_{\beta \gamma
}^\delta +g_{\delta \beta }.U_{\alpha \gamma }^\delta +g_{\delta \gamma
}.U_{\alpha \beta }^\delta )\text{ .}  \label{Ch 5 0.26}
\end{equation}

In the special case, when the condition $g_{\alpha \beta ;\gamma }=0$ is
required, then the following proposition can be proved:

\begin{proposition}
The necessary and sufficient condition for $g_{\alpha \beta ;\gamma }=0$ is
the condition 
\begin{equation}
g_{\delta \gamma }.\overline{P}\,_{\alpha \beta }^\delta =-\{\alpha \beta
,\gamma \}+C_{\alpha \beta \gamma }.  \label{Ch 5 0.27}
\end{equation}
\end{proposition}
%

The proof follows imediately from (\ref{Ch 5 0.23}).

In a co-ordinate basis the covariant derivative of the components $g_{ij}$
of $g$ can in analogous way be presented by means of the components $%
P_{jk}^i $ of the covariant symmetric affine connection 
\begin{equation}  \label{Ch 5 0.28}
\begin{array}{c}
g_{ij;k}=g_{ij,k}+ \overline{P}\,_{ik}^l.g_{lj}+\overline{P}%
\,_{jk}^l.g_{il}+\frac 12(U_{ik}^l.g_{lj}+U_{jk}^l.g_{il})= \\ 
=g_{ij/k}+\frac 12(U_{ik}^l.g_{lj}+U_{jk}^l.g_{il})\text{ ,}
\end{array}
\end{equation}

\noindent where 
\begin{equation}
g_{ij/k}=g_{ij,k}+\overline{P}\,_{ik}^l.g_{lj}+\overline{P}\,_{jk}^l.g_{il}%
\text{ .}  \label{Ch 5 0.29}
\end{equation}

\subsubsection{Action of the Lie differential operator on the covariant
metric}

In a co-ordinate basis $\pounds _\xi g$ will take the form 
\begin{equation}  \label{Ch 5 0.48}
\begin{array}{c}
\pounds _\xi g=(\pounds _\xi g_{ij}).dx^i.dx^j= \\ 
=[g_{ij;k}.\xi ^k+g_{kj}.\xi ^{\overline{k}}\text{ }_{;\underline{i}%
}+g_{ik}.\xi ^{\overline{k}}\text{ }_{;\underline{j}}+(g_{kj}.T_{l\underline{%
i}}^{\,\,\,\,\overline{k}}+g_{ik}.T_{l\underline{j}}^{\,\,\,\,\overline{k}%
}).\xi ^l].dx^i.dx^j\text{ .}
\end{array}
\end{equation}

The following relations are also fulfilled: 
\begin{equation}  \label{Ch 5 0.49}
\begin{array}{c}
\pounds _\xi [g(u,v)]=\xi [g(u,v)]=(\pounds _\xi g)(u,v)+g(\pounds _\xi
u,v)+g(u,\pounds _\xi v) \text{ ,} \\ 
\pounds _\xi [g(u)]=(\pounds _\xi g)(u)+g(\pounds _\xi u)\text{ , \thinspace
\thinspace \thinspace \thinspace \thinspace \thinspace \thinspace \thinspace
\thinspace \thinspace \thinspace \thinspace \thinspace \thinspace \thinspace
\thinspace \thinspace \thinspace \thinspace \thinspace \thinspace \thinspace
\thinspace \thinspace \thinspace \thinspace }\xi ,u,v\in T(M)\text{ .}
\end{array}
\end{equation}

The action of the Lie differential operator is called \textit{dragging-along
a contravariant vector field}. On the basis of draggings-along the metric
tensor field $g$ notions as arbitrary (non-metric) draggings-along,
quasi-projective draggings-along, conformal motions and motions can be
defined and considered in analogous way as in ($L_n,g$)-spaces. Here we will
only define different types of draggings-along.

\textit{1. Arbitrary (non-metric) draggings-along } \textbf{
\[
\pounds _\xi g=q_\xi \text{ , }\forall \xi \in T(M)\text{ , }q_\xi \in
\otimes _{sym2}(M)\text{ ,} 
\]
}

\textit{2. Quasi-projective draggings-along} \textbf{
\[
\pounds _\xi g=\frac 12[p\otimes g(\xi )+g(\xi )\otimes p]\text{ ,\thinspace
\thinspace \thinspace \thinspace \thinspace \thinspace \thinspace \thinspace
\thinspace \thinspace \thinspace }\xi \in T(M)\text{ , }p\in T^{*}(M)\text{ .%
} 
\]
}

\textit{3. Conform-invariant draggings-along (conformal motions) }

\textbf{
\[
\pounds _\xi g=\lambda .g\text{ , \thinspace \thinspace \thinspace
\thinspace \thinspace \thinspace \thinspace \thinspace \thinspace \thinspace 
}\lambda \in C^r(M)\text{ ,\thinspace \thinspace \thinspace \thinspace
\thinspace \thinspace \thinspace \thinspace \thinspace \thinspace \thinspace
\thinspace \thinspace }\xi \in T(M)\text{ . } 
\]
}

\textit{4. Isometric draggings-along (motions) }

\textbf{
\[
\pounds _\xi g=0\text{ ,\thinspace \thinspace \thinspace \thinspace
\thinspace \thinspace \thinspace \thinspace \thinspace \thinspace }\xi \in
T(M)\text{ .} 
\]
}

For all types of draggings-along changes of the scalar product of two
contravariant vector fields and the changes of the length of these fields
can be found and used in the analogous way as in ($L_n,g$)-spaces.

\subsection{Covariant projective metric}

If a covariant metric field $g$ is given and there exists a contravariant
vector field $u$ which square of the length $g(u,u)=e\neq 0$, then a new
covariant tensor field can be constructed orthogonal to the vector field $u$%
. It possesses properties analogous to these of the covariant tensor field $%
g $ acting on contravariant vector fields in every orthogonal to $%
u(x)=u_x\in T_x(M)$ subspace $T_x^{\bot u}(M)$ in $T_x(M)$, where $%
(T_x^{\bot u}(M)=\{\xi _x\}:g_x(\xi _x,u_x)=0)$, $g_x\in \otimes
_{sym2/x}(M) $.

\begin{definition}
{\bf \ }{\it Covariant projective metric. }Covariant metric, orthogonal to a
given non-isotropic (non-null) vector field $u$ [$e=g(u,u)\neq 0$], i.e.
covariant metric $h_u$ satisfying the condition $h_u(u)=u(h_u)=0$ and
constructed by means of the covariant metric $g$ and $u$ in the form 
\begin{equation}
h_u=g-\frac 1{g(u,u)}.g(u)\otimes g(u)=g-\frac 1e.g(u)\otimes g(u)\text{ . }
\label{Ch 5 0.50}
\end{equation}
\end{definition}
%

The properties of the covariant projective metric follow from its
construction and from the properties of the covariant metric $g$:

(a) $h_u(u)=u(h_u)=0$, $[g(u)](u)=g(u,u)=e$.

(b) $h_u(u,u)=0$.

(c) $h_u(u,v)=h_u(v,u)=0$ , $\forall v\in T(M)$.

\subsection{Contravariant metric}

\begin{definition}
{\bf \ }{\it Contravariant metric. }Contraction operator $S$, acting on two
covariant vector fields over a manifold $M$ which {\it action} is identified
with the {\it action} of a contravariant symmetric tensor field of rank two
on the two vector fields, i.e. 
\[
S(p,q)\equiv \overline{g}(p,q):=S(\overline{g},w):=S(\overline{g},p\otimes
q)=S(\overline{g}\otimes (p\otimes q))\text{ , \thinspace \thinspace }w=p\otimes q\text{ ,} 
\]
\end{definition}
%

The tensor field $\overline{g}=g^{\alpha \beta }.e_\alpha .e_\beta
=g^{ij}.\partial _i.\partial _j$ is called \textit{contravariant metric
tensor field (contravariant metric)}. $\overline{g}(x)=\overline{g}_x\in
\otimes _x^2(M)$ is called \textit{contravariant metric tensor} in $x\in M$.

The properties of the contravariant metric are determined by the properties
of the contraction operator and its identification with the contravariant
symmetric tensor field of rank 2. On this basis the following properties can
be proved:

(a) Action of the contravariant metric on two covariant vector fields in a
co-ordinate basis 
\[
\begin{array}{c}
\overline{g}(p,q)=g^{kl}.f^i\text{ }_k.f^j\text{ }_l.p_i.q_j=g^{\overline{i}%
\overline{j}}.p_i.q_j=g^{kl}.p_{\overline{k}}.q_{\overline{l}}=p^{\overline{j%
}}.q_j=p_k.q^{\overline{k}}\text{ ,} \\ 
p_{\overline{k}}=f^i\text{ }_k.p_i\text{ ,\thinspace \thinspace \thinspace
\thinspace \thinspace \thinspace \thinspace \thinspace \thinspace \thinspace 
}q_{\overline{l}}=f^j\text{ }_l.q_j\text{ , \thinspace \thinspace \thinspace
\thinspace \thinspace \thinspace \thinspace \thinspace \thinspace \thinspace
\thinspace \thinspace \thinspace \thinspace \thinspace \thinspace \thinspace 
}p^{\overline{j}}=g^{\overline{j}\overline{i}}.p_i\text{ .}
\end{array}
\]

[When $q=p$, then $\overline{g}(p,p)=p^2=\pm \mid p\mid ^2$ is called 
\textit{square of the length }of the covariant vector field $p$.]

(b) Action of the contravariant metric $\overline{g\text{ }}$on covariant
vector field 
\[
\begin{array}{c}
\overline{g}(p)=p(\overline{g})=g^{ij}.p_k.f^k\text{ }_j.\partial
_i=g^{ij}.p_{\overline{j}}.\partial _i=g^{i\overline{k}}.p_k.\partial
_i=p^i.\partial _i\text{ ,} \\ 
\text{ }g^{i\overline{k}}=g^{il}.f^k\text{ }_l\text{ , \thinspace \thinspace
\thinspace \thinspace \thinspace \thinspace \thinspace \thinspace \thinspace
\thinspace \thinspace \thinspace \thinspace }p^i=g^{ij}.p_{\overline{j}}=g^{i%
\overline{j}}.p_j\text{ ,}
\end{array}
\]

The action of the contravariant metric $\overline{g}$ on covariant vector
field $p$ in a given basis is called \textit{raising of indices} by means of
the contravariant metric. The result of this action is a contravariant
vector field $\overline{g}(p)$. On this basis $\overline{g}$ can be defined
as a linear mapping (operator) which maps an element of $T^{*}(M)$ in an
element of $T(M)$: 
\[
\overline{g}:p\rightarrow \overline{g}(p)\in T(M)\text{ , }p\in T^{*}(M)%
\text{ .} 
\]

The connection between the contravariant and covariant metric can be
determined by the conditions 
\begin{equation}  \label{Ch 6 0.56}
\overline{g}[g(u)]=u\text{ ,\thinspace \thinspace \thinspace \thinspace
\thinspace \thinspace \thinspace \thinspace \thinspace \thinspace }u\in T(M)%
\text{ , \thinspace \thinspace \thinspace \thinspace \thinspace \thinspace
\thinspace \thinspace \thinspace \thinspace \thinspace }g[\overline{g}(p)]=p%
\text{ ,\thinspace \thinspace \thinspace \thinspace \thinspace \thinspace
\thinspace \thinspace \thinspace }p\in T^{*}(M)\text{ .}
\end{equation}

In a co-ordinate basis these conditions take the forms: 
\begin{equation}  \label{Ch 6 0.57}
g^{ij}.g_{\overline{j}\overline{k}}=g_k^i\text{ , }g_{ij}.g^{\overline{j}%
\overline{k}}=g_i^k\text{ .}
\end{equation}

From the last expressions the relation follows 
\begin{equation}  \label{Ch 6 0.59}
g[\overline{g}]=g_{ij}g^{\overline{i}\overline{j}}=n\text{ ,\thinspace
\thinspace \thinspace \thinspace \thinspace \thinspace \thinspace \thinspace
\thinspace }\overline{g}[g]=g^{\overline{i}\overline{j}}.g_{ij}=n\text{
,\thinspace \thinspace \thinspace \thinspace \thinspace \thinspace
\thinspace \thinspace \thinspace \thinspace \thinspace \thinspace \thinspace
\thinspace \thinspace \thinspace }\dim M=n\text{ .}
\end{equation}

\subsubsection{Contravariant symmetric affine connection}

From the transformation properties of the components of the contravariant
affine connection, it follows that the quantity 
\[
\frac 12(\Gamma _{\alpha \beta }^\gamma +\Gamma _{\beta \alpha }^\gamma
-C_{\alpha \beta }\text{ }^\gamma )\text{ or }\frac 12(\Gamma _{ij}^k+\Gamma
_{ji}^k) 
\]

\noindent has the same transformation properties as the contravariant affine
connection itself. This fact can be used as usual for representing the
contravariant affine connection by means of its symmetric and anti-symmetric
part in the form 
\begin{equation}
\begin{array}{c}
\Gamma _{ij}^k=\overline{\Gamma }\,_{ij}^k-\frac 12T_{ij}^{\,\,\,\,\,k}\text{
, \thinspace \thinspace \thinspace \thinspace \thinspace \thinspace
\thinspace \thinspace \thinspace \thinspace \thinspace \thinspace \thinspace
\thinspace \thinspace }\overline{\Gamma }\,_{ij}^k=\frac 12(\Gamma
_{ij}^k+\Gamma _{ji}^k)\text{ ,\thinspace \thinspace \thinspace \thinspace
\thinspace \thinspace \thinspace \thinspace \thinspace \thinspace }%
T_{ij}^{\,\,\,\,k}=\Gamma _{ji}^k-\Gamma _{ij}^k\text{ ,} \\ 
\text{(in a co-ordinate basis) ,} \\ 
\Gamma _{\alpha \beta }^\gamma =\overline{\Gamma }\,_{\alpha \beta }^\gamma
-\frac 12T_{\alpha \beta }^{\,\,\,\,\,\,\,\gamma }\text{ , \thinspace
\thinspace \thinspace \thinspace \thinspace \thinspace \thinspace \thinspace
\thinspace \thinspace \thinspace \thinspace \thinspace }\overline{\Gamma }%
\,_{\alpha \beta }^\gamma =\frac 12(\Gamma _{\alpha \beta }^\gamma +\Gamma
_{\beta \alpha }^\gamma -C_{\alpha \beta }\text{ }^\gamma )\text{ ,} \\ 
T_{\alpha \beta }^{\,\,\,\,\,\,\gamma }=\Gamma _{\beta \alpha }^\gamma
-\Gamma _{\alpha \beta }^\gamma -C_{\alpha \beta }\text{ }^\gamma \text{ ,}
\\ 
\text{(in a non-co-ordinate basis) .}
\end{array}
\label{Ch 6 0.69}
\end{equation}

$\overline{\Gamma }_{ij}^k$ ($\overline{\Gamma }_{\alpha \beta }^\gamma $)
are called components of the \textit{contravariant symmetric affine
connection} in a co-ordinate (respectively in a non-co-ordinate) basis.

The components of the covariant derivative of the contravariant metric
tensor field $\overline{g}$ can be represented by means of the contravariant
symmetric affine connection. If we introduce the abbreviations 
\begin{equation}  \label{Ch 6 0.70}
\begin{array}{c}
g^{\alpha \beta } \text{ }_{;\gamma }=e_\gamma g^{\alpha \beta }+\overline{%
\Gamma }\,_{\delta \gamma }^\alpha .g^{\delta \beta }+\overline{\Gamma }%
\,_{\delta \gamma }^\beta .g^{\alpha \delta }\text{ \thinspace \thinspace
\thinspace \thinspace \thinspace (in a non-co-ordinate basis) ,} \\ 
g^{ij}\text{ }_{/k}=g^{ij}\text{ }_{,k}+\overline{\Gamma }\,_{lk}^i.g^{lj}+%
\overline{\Gamma }\,_{lk}^j.g^{il}\text{ \thinspace \thinspace \thinspace
\thinspace \thinspace \thinspace \thinspace \thinspace \thinspace \thinspace
\thinspace \thinspace \thinspace \thinspace \thinspace \thinspace \thinspace
\thinspace \thinspace (in a co-ordinate basis) ,}
\end{array}
\end{equation}

\noindent where $g^{\alpha \beta }$ $_{;\gamma }$ is the covariant
derivative of the components of the contravariant metric tensor $\overline{g}
$ with respect to the contravariant symmetric affine connection $\overline{%
\Gamma }$ in a non-co-ordinate basis, then 
\begin{equation}
g^{\alpha \beta }\text{ }_{/\gamma }=g^{\alpha \beta }\text{ }_{;\gamma
}-\frac 12(T_{\delta \gamma }^{\,\,\,\,\,\,\alpha }.g^{\delta \beta
}+T_{\delta \gamma }^{\,\,\,\,\,\,\beta }.g^{\alpha \delta })\text{ .}
\label{Ch 6 0.71}
\end{equation}

By means of the explicit expression for $g^{\alpha \beta }$ $_{/\kappa
}.g^{\kappa \gamma }$ and the usual method for expressing the components of
the symmetric affine connection the components of the contravariant
symmetric affine connection can be represented in the form 
\begin{equation}  \label{Ch 6 0.72}
g^{\alpha \delta }.g^{\beta \kappa }.\overline{\Gamma }\,_{\delta \kappa
}^\gamma =\overline{K}\,^{\alpha \beta \gamma }-\{^{\alpha \beta ,\gamma }\}-%
\overline{C}\,^{\alpha \beta \gamma }-\frac 12g^{\gamma \delta }(g^{\beta
\kappa }T_{\kappa \delta }^{\,\,\,\,\,\,\alpha }+g^{\alpha \kappa }T_{\kappa
\delta }^{\,\,\,\,\,\beta })\text{ ,}
\end{equation}

\noindent where 
\begin{equation}
\begin{array}{c}
\{^{\alpha \beta ,\gamma }\}=\frac 12(g^{\alpha \kappa }.e_\kappa g^{\beta
\gamma }+g^{\beta \kappa }.e_\kappa g^{\alpha \gamma }-g^{\gamma \kappa
}.e_\kappa g^{\alpha \beta })\text{ ,} \\ 
\overline{K}\,^{\alpha \beta \gamma }=\frac 12(g^{\alpha \gamma }\text{ }%
_{/\kappa }.g^{\beta \kappa }+g^{\beta \gamma }\text{ }_{/\kappa }.g^{\alpha
\kappa }-g^{\alpha \beta }\text{ }_{/\kappa }.g^{\gamma \kappa })\text{ ,}
\\ 
\overline{C}\,^{\alpha \beta \gamma }=\frac 12(g^{\gamma \delta }.g^{\beta
\kappa }.C_{\kappa \delta }\text{ }^\alpha +g^{\gamma \delta }.g^{\alpha
\kappa }.C_{\kappa \delta }\text{ }^\beta +g^{\alpha \delta }.g^{\beta
\kappa }.C_{\delta \kappa }\text{ }^\gamma )\text{ .}
\end{array}
\label{Ch 6 0.73}
\end{equation}

$\{^{\alpha \beta ,\gamma }\}$ are called \textit{Christoffel symbols of the
first kind} for the contravariant symmetric affine connection in a
non-co-ordinate basis.

The components $\Gamma _{\alpha \beta }^\gamma $ of the contravariant affine
connection $\Gamma $ can be written by means of the last abbreviations in
the form 
\begin{equation}  \label{Ch 6 0.74}
g^{\alpha \delta }.g^{\beta \kappa }.\Gamma _{\delta \kappa }^\gamma
=-\{^{\alpha \beta ,\gamma }\}+\overline{K}^{\alpha \beta \gamma }-\overline{%
T}^{\alpha \beta \gamma }-\overline{C}^{\alpha \beta \gamma }\text{ ,}
\end{equation}

\noindent where 
\begin{equation}
\overline{T}^{\alpha \beta \gamma }=\frac 12(g^{\gamma \delta }.g^{\beta
\kappa }.T_{\kappa \delta }^{\,\,\,\,\,\,\alpha }+g^{\gamma \delta
}.g^{\alpha \kappa }.T_{\kappa \delta }^{\,\,\,\,\,\,\beta }+g^{\alpha
\delta }.g^{\beta \kappa }.T_{\delta \kappa }^{\,\,\,\,\,\gamma })\text{ .}
\label{Ch 6 0.75}
\end{equation}

By using the connections between the components of the covariant metric, the
components of the contravariant metric and their derivatives 
\begin{equation}  \label{Ch 6 0.76}
\begin{array}{c}
g_{\overline{\alpha }\overline{\beta }}.g^{\beta \gamma }=g_\alpha ^\gamma 
\text{ , }g_{\overline{\alpha }\overline{\delta }}.e_\kappa g^{\delta \gamma
}=-g^{\delta \gamma }.e_\kappa (g_{\overline{\alpha }\overline{\delta }})%
\text{ ,} \\ 
g_{\overline{\alpha }\overline{\delta }}.g^{\gamma \delta }\text{ }_{/\kappa
}=-g^{\gamma \delta }.g_{\overline{\alpha }\overline{\delta }/\kappa }\text{
,}
\end{array}
\end{equation}

\noindent the components of the contravariant affine connection $\Gamma $
can be represented in a non-co-ordinate basis in the form 
\begin{equation}
\Gamma _{\alpha \beta }^\gamma =\{_{\alpha \beta }^\gamma \}-\overline{K}%
_{\alpha \beta }{}^\gamma -\overline{S}_{\alpha \beta }\text{ }^\gamma -%
\overline{C}_{\alpha \beta }\text{ }^\gamma \text{ ,}  \label{Ch 6 0.77}
\end{equation}

\noindent where 
\begin{equation}
\begin{array}{c}
\{_{\alpha \beta }^\gamma \}=\frac 12g^{\gamma \delta }[e_\beta (g_{%
\overline{\alpha }\overline{\delta }})+e_\alpha (g_{\overline{\beta }%
\overline{\delta }})-e_\delta (g_{\overline{\alpha }\overline{\beta }})]= \\ 
=-g_{\overline{\alpha }\overline{\rho }}.g_{\overline{\beta }\overline{%
\sigma }}.\{^{\rho \sigma ,\gamma }\}\text{ , \thinspace \thinspace
\thinspace \thinspace \thinspace \thinspace \thinspace \thinspace \thinspace
\thinspace \thinspace \thinspace \thinspace \thinspace \thinspace \thinspace
\thinspace \thinspace \thinspace \thinspace \thinspace \thinspace \thinspace
\thinspace \thinspace \thinspace \thinspace \thinspace }\overline{K}_{\alpha
\beta }\text{ }^\gamma =-g_{\overline{\alpha }\overline{\rho }}.g_{\overline{%
\beta }\overline{\sigma }}.\overline{K}\,^{\rho \sigma \gamma }\text{ ,} \\ 
\overline{S}_{\alpha \beta }\text{ }^\gamma =-g_{\overline{\alpha }\overline{%
\rho }}.g_{\overline{\beta }\overline{\sigma }}.\overline{T}\,^{\rho \sigma
\gamma }\text{ , \thinspace \thinspace \thinspace \thinspace \thinspace
\thinspace \thinspace \thinspace \thinspace \thinspace \thinspace \thinspace
\thinspace \thinspace \thinspace \thinspace \thinspace \thinspace \thinspace 
}\overline{C}_{\alpha \beta }\text{ }^\gamma =-g_{\overline{\alpha }%
\overline{\rho }}.g_{\overline{\beta }\overline{\sigma }}.\overline{C}%
\,^{\rho \sigma \gamma }\text{ .}
\end{array}
\label{Ch 6 0.78}
\end{equation}

$\{_{\alpha \beta }^\gamma \}$ are called \textit{generalized} \textit{%
Christoffel symbols of the second kind }for the contravariant symmetric
affine connection in a non-co-ordinate basis.

In analogous way, when a contravariant and covariant metric fields are
given, the covariant affine connection can be presented by means of the both
types of tensor metric fields in the form 
\begin{equation}  \label{Ch 6 0.79}
P_{\alpha \beta }^\gamma =-\underline{\{_{\alpha \beta }^\gamma \}}+%
\underline{K}_{\alpha \beta }\text{ }^\gamma +\underline{U}_{\alpha \beta }%
\text{ }^\gamma +\underline{C}_{\alpha \beta }\text{ }^\gamma \text{ ,}
\end{equation}

\noindent where 
\begin{equation}
\begin{array}{c}
\underline{\{_{\alpha \beta }^\gamma \}}=g^{\overline{\gamma }\overline{%
\sigma }}.\{\alpha \beta ,\gamma \}\text{ ,\thinspace \thinspace \thinspace
\thinspace \thinspace \thinspace \thinspace \thinspace \thinspace \thinspace
\thinspace \thinspace \thinspace \thinspace \thinspace \thinspace \thinspace
\thinspace }\underline{K}_{\alpha \beta }\text{ }^\gamma =g^{\overline{%
\gamma }\overline{\sigma }}.K_{\alpha \beta \sigma }\text{ ,} \\ 
\underline{U}_{\alpha \beta }\text{ }^\gamma =g^{\overline{\gamma }\overline{%
\sigma }}.U_{\alpha \beta \sigma }\text{ , \thinspace \thinspace \thinspace
\thinspace \thinspace \thinspace \thinspace \thinspace \thinspace \thinspace
\thinspace \thinspace \thinspace \thinspace \thinspace \thinspace \thinspace
\thinspace \thinspace \thinspace \thinspace \thinspace \thinspace }%
\underline{C}_{\alpha \beta }\text{ }^\gamma =g^{\overline{\gamma }\overline{%
\sigma }}.C_{\alpha \beta \sigma }\text{ .}
\end{array}
\label{Ch 6 0.80}
\end{equation}

$\underline{\{_{\alpha \beta }^\gamma \}}$ are called \textit{generalized} 
\textit{Christoffel symbols of the second kind }for the covariant symmetric
affine connection in a non-co-ordinate basis.

The same expressions can be obtained also in a co-ordinate basis.

For the special case, when the condition of vanishing of the covariant
derivatives of the contravariant metric with respect to the contravariant
symmetric affine connection is required, i.e. $g^{\alpha \beta }$ $_{;\gamma
}=0$, then the components of the contravariant symmetric affine connection
can be written in the form $\overline{\Gamma }_{\alpha \beta }^\gamma
=\{_{\alpha \beta }^\gamma \}-\overline{C}_{\alpha \beta }$ $^\gamma $.

The last expression is the necessary and sufficient condition for $g^{\alpha
\beta }$ $_{;\gamma }=0$. In a co-ordinate basis the necessary and
sufficient condition for $g^{ij}$ $_{/k}=0$ takes the form $\overline{\Gamma 
}_{ij}^k=\{_{ij}^k\}$.

On the basis of the connection between the covariant derivative of the
contravariant tensor metric field and the covariant derivative of the
covariant tensor metric field 
\[
\nabla _\xi g=-g(\nabla _\xi \overline{g})g\text{ ,\thinspace \thinspace
\thinspace \thinspace \thinspace }(\nabla _\xi \overline{g})[g(u)]=-%
\overline{g}[(\nabla _\xi g)(u)]\text{ , }\forall \xi ,\forall u\in T(M)%
\text{ ,} 
\]
\begin{eqnarray*}
\nabla _\xi \overline{g} &=&-\overline{g}(\nabla _\xi g)\overline{g}\text{
,\thinspace \thinspace \thinspace \thinspace \thinspace }(\nabla _\xi g)[\overline{g}(p)]=-g[(\nabla _\xi \overline{g})(p)]\text{ , } \\
\text{ }\forall \xi &\in &T(M)\text{,}\,\,\,\,\,\,\,\,\,\,\,\forall p\in
T^{*}(M)\text{ ,}
\end{eqnarray*}
%

\noindent one can prove that there is one-to-one correspondence between the
transports of $g$ and $\overline{g}$. \textit{Every transport of the
covariant tensor metric field }$g$\textit{\ induces a corresponding
transport of the contravariant tensor metric field }$\overline{g}$\textit{\
and vice versa}.

\subsection{Contravariant projective metric}

The notion of contravariant projective metric with respect to a
non-isotropic (non-null) contravariant vector field $u$ can be introduced in
two different ways:

(a) by definition 
\begin{equation}  \label{Ch 6 0.86}
h^u=\overline{g}-\frac 1{g(u,u)}.u\otimes u=\overline{g}-\frac 1e.u\otimes u%
\text{ , }e=g(u,u)\neq 0\text{ ,}
\end{equation}

(b) by inducing from the covariant projective metric using the relations
between the covariant and contravariant metric 
\begin{equation}  \label{Ch 6 0.87}
h^u=\overline{g}(h_u)\overline{g}=\overline{g}-\frac 1e.u\otimes u\text{ ,
\thinspace \thinspace \thinspace }\overline{g}(g)\overline{g}=\overline{g}%
\text{ , \thinspace \thinspace }\overline{g}(g(u)\otimes g(u))\overline{g}%
=u\otimes u\text{ .}
\end{equation}

$h^u$ is called \textit{contravariant projective metric} with respect to the
non-isotropic contravariant vector field $u$.

The properties of the contravariant projective metric are determined by its
structure.

\section{Bianchi identities for the covariant curvature tensor}

\subsection{Bianchi identity of first type for the covariant curvature tensor
}

The existence of contravariant and covariant metrics allow us to consider
the action of the curvature operator on a covariant vector field $%
g(v)=g_{\alpha \beta }.v^{\overline{\beta }}.e^\alpha =g_{ij}.v^{\overline{j}%
}.dx^i$, constructed by the use of the covariant metric $g$ and a
contravariant vector field $v$.

The identity 
\begin{eqnarray}
&<&\overline{g}\{([R(\xi ,u)]g)(v)\}>\,\equiv \,<\overline{g}([R(\xi
,u)][g(v)])>-\,<[R(\xi ,u)]v>\equiv  \nonumber \\
&\equiv &<\overline{g}([R(\xi ,u)][g(v)])>-\,<T(T(\xi ,u),v)>-<(\nabla _\xi
T)(u,v)>  \label{Ch 6 0.98}
\end{eqnarray}
%

\noindent is called \textit{Bianchi identity of first type (of the type 1.)
for the covariant curvature tensor}.

In a co-ordinate basis the Bianchi identity of first type will have the
forms 
\begin{equation}  \label{Ch 6 0.110}
P^l\,_{<ijk>}\equiv \,-g^{m\overline{n}}.R^{\overline{l}}\,_{m<ij}.g_{k>n}%
\text{ ,}
\end{equation}
\begin{equation}  \label{Ch 6 0.111}
R^l\,_{<ijk>}\equiv \,-g^{l\overline{m}}.g_{mn}.P^n\,_{<\overline{i}%
jk>}\equiv \,T_{<ij}\,^l\,_{;k>}+T_{<ij}\,^m.T_{mk>}\,^l\text{ .}
\end{equation}

It is obvious that the form of the Bianchi identity of first type for the
components of the covariant curvature tensor is not so simple as the form of
the Bianchi identity for the components of the contravariant curvature
tensor.

\subsection{Bianchi identity of second type for the covariant curvature
tensor}

The action of the operator $(\nabla _wR)(\xi ,u)$ can be extended to an
action on covariant vector and tensor fields in an analogous way as in the
case of contravariant vector and tensor fields. By the use of the relation 
\[
\nabla _w\{[R(\xi ,u)]p\}=[(\nabla _wR)(\xi ,u)]p+[R(\nabla _w\xi ,u)]p\,+ 
\]
\begin{equation}  \label{Ch 6 0.112}
+[R(\xi ,\nabla _wu)]p+[R(\xi ,u)](\nabla _wp)\text{ ,\thinspace \thinspace
\thinspace \thinspace \thinspace }w,\xi ,u\in T(M)\text{ , }p\in T^{*}(M)%
\text{ ,}
\end{equation}

\noindent we can find the identity 
\begin{equation}
<(\nabla _wR)(\xi ,u)>p\equiv \,<R(w,T(\xi ,u))>p\text{ ,}
\label{Ch 6 0.116}
\end{equation}

\noindent where 
\begin{eqnarray}
&<&(\nabla _wR)(\xi ,u)>p=[(\nabla _wR)(\xi ,u)]p+[(\nabla _uR)(w,\xi
)]p+[(\nabla _\xi R)(u,w)]p\text{ ,}  \nonumber \\
&<&R(w,T(\xi ,u))>p=[R(w,T(\xi ,u))]p+[R(u,T(w,\xi ))]p+  \nonumber \\
&&+[R(\xi ,T(u,w))]p\text{ .}  \label{Ch 6 0.117}
\end{eqnarray}
%

The identity (\ref{Ch 6 0.116}) is called \textit{Bianchi identity of second
type} \textit{(of type 2.) for the covariant curvature tensor}.

The Bianchi identity of second type will have the form in a co-ordinate
basis 
\begin{eqnarray}
P^i\,_{j<kl;m>} &=&P^i\,_{jkl;m}+P^i\,_{jmk;l}+P^i\,_{jlm;k}\equiv  \nonumber
\\
&\equiv
&P^i\,_{j<kn}.T_{lm>}\,^n=P^i\,_{jkn}.T_{lm}\,^n+P^i\,_{jmn}.T_{kl}\,^n+P^i\,_{jlr}.T_{mk}\,^r\text{ .}  \label{Ch 6 0.119}
\end{eqnarray}
%

\section{Invariant volume element}

\subsection{Definition and properties}

The notion of volume element of a manifold $M$ can be generalized to the
notion of invariant volume element \cite{Lovelock}.

\begin{definition}
The {\it volume element of a manifold} $M$ $(\dim M=n)$ 
\[
\begin{array}{c}
d^{(n)}x=d^{(n)}x=dx^1\wedge ...\wedge dx^n\,\text{\thinspace \thinspace
\thinspace \thinspace \thinspace (in a co-ordinate basis),} \\ 
dV_n=e^1\wedge ...\wedge e^n\text{ \thinspace \thinspace \thinspace
\thinspace \thinspace \thinspace \thinspace \thinspace \thinspace \thinspace
\thinspace (in a non-co-ordinate basis).}
\end{array}
\]
\end{definition}
%

The properties of the volume element could be represented as follows: 
\begin{equation}  \label{VI.7.-85}
\begin{array}{c}
d^{(n)}x=\frac 1{n!}.\varepsilon _A.\omega ^A=\frac 1{n!}.\varepsilon _A.d 
\widehat{x}^A\text{ , \thinspace \thinspace \thinspace \thinspace \thinspace
\thinspace \thinspace \thinspace \thinspace \thinspace \thinspace }%
d^{(n^{\prime })}x^{\prime }=J^{-1}.d^{(n)}x\text{ ,} \\ 
dV_n=\frac 1{n!}.\varepsilon _A.\omega ^A=\frac 1{n!}.\varepsilon _A.%
\widehat{e}^A\text{ , \thinspace \thinspace \thinspace \thinspace \thinspace
\thinspace \thinspace \thinspace \thinspace \thinspace \thinspace \thinspace 
}dV_n^{\prime }=J^{-1}.dV_n\text{ ,}
\end{array}
\end{equation}

\noindent where $J=\det (A_{\alpha ^{\prime }}\,^\alpha )=\det (\partial
x^i/\partial x^{i^{\prime }})$,\thinspace \thinspace $dV_n^{\prime
}=e^{1^{\prime }}\wedge ...e^{n^{\prime }}$, $\varepsilon _A=\varepsilon
_{i_1...i_n}$, $\omega ^A=dx^{i_1}\wedge ...\wedge dx^{i_n}$, $\varepsilon
_A $ is the Levi-Civita symbol \cite{Lovelock}, 
\[
\begin{array}{c}
\varepsilon _{A^{\prime }}.\omega ^{A^{\prime }}=J^{-1}.\varepsilon
_A.\omega ^A\text{ , }\varepsilon _A.\omega ^A=J.\varepsilon _{A^{\prime
}}.\omega ^{A^{\prime }}\text{ ,} \\ 
\varepsilon _{A^{\prime }}.d\widehat{x}\,^{A^{\prime }}=J^{-1}.\varepsilon
_A.d\widehat{x}\,^A\text{ , }\varepsilon _A.d\widehat{x}\,^A=J.\varepsilon
_{A^{\prime }}.d\widehat{x}\,^{A^{\prime }}\text{ ,} \\ 
d^{(n)}x=\frac 1{n!}.\varepsilon _A.\omega ^A=J.\frac 1{n!}.\varepsilon
_{A^{\prime }}.\omega ^{A^{\prime }}=\frac 1{n!}.J.\varepsilon _{A^{\prime
}}.d\widehat{x}\,^{A^{\prime }}\text{ .}
\end{array}
\]

The transformation properties of the volume element are corresponding to
these of a tensor density of the weight $\omega =-\frac 12$. Therefore, for
the construction of an invariant volume element (keeping its form and
independent of the choice of a full anti-symmetric tensor basis) it is
necessary the volume element to be multiplied with a tensor density with the
weight $\omega =\frac 12$ and rank $0$. Since the covariant metric tensor
field is connected with the basic characteristics of contravariant (and
covariant) vector fields and determines along with them notions (such as
length of a contravariant vector, cosine of the angle between two
contravariant vectors) which in the Euclidean geometry are related to the
notion volume element, the covariant metric tensor density $\widetilde{Q}_g$
with a weight $\omega =\frac 12$ and rank $0$ $(\widetilde{Q}_g=\mid d_g\mid
^{\frac 12})\,$ appears as a suitable multiplier to a volume element \cite
{Manoff-10}.

\begin{definition}
The invariant volume element $d\omega $ of a manifold $M$ $(\dim M=n)$. 
\[
\begin{array}{c}
d\omega =\sqrt{-d_g}.d^{(n)}x:=\frac 1{n!}.\varepsilon _A.\overline{\omega }^A\text{ ,\thinspace \thinspace \thinspace \thinspace \thinspace \thinspace
\thinspace \thinspace \thinspace \thinspace \thinspace \thinspace }\overline{\omega }^A=\sqrt{-d_g}.\text{\thinspace }\omega ^A\text{ ,\thinspace
\thinspace \thinspace \thinspace \thinspace \thinspace \thinspace \thinspace
\thinspace \thinspace \thinspace \thinspace }d_g<0\text{ ,} \\ 
\text{(invariant volume element in a co-ordinate basis),} \\ 
d\omega =\sqrt{-d_g}.dV_n\text{ , \thinspace \thinspace \thinspace
\thinspace \thinspace \thinspace \thinspace \thinspace \thinspace \thinspace
\thinspace \thinspace \thinspace \thinspace \thinspace \thinspace }d_g<0\text{ ,} \\ 
\text{(invariant volume element in a non-co-ordinate basis).}
\end{array}
\]
\end{definition}
%

From the transformation properties of $\sqrt{-d_g}:\sqrt{-d_g^{\prime }}=\pm
J.\sqrt{-d_g}$ the invariance of the invariant volume element follows: $%
d\omega ^{\prime }=\pm \,d\omega $, where 
\begin{eqnarray}
d\omega ^{\prime } &=&\sqrt{-d_g^{\prime }}.d^{(n^{\prime })}x^{\prime }\text{ (in a co-ordinate basis),}  \nonumber  \label{VI.7.-89} \\
d\omega ^{\prime } &=&\sqrt{-d_g^{\prime }}.dV_n^{\prime }\text{ (in a
non-co-ordinate basis).}  \label{VI.7.-89}
\end{eqnarray}
%

\begin{remark}
The sign $(-)$ in $\pm \,d\omega $ can be omitted because of the identical
configuration (order, orientation) of the basic vector fields in the old and
in the new tensor basis.
\end{remark}
%

From the definition of the invariant volume element the relations connected
with its structure follow: 
\begin{equation}  \label{VI.7.-92}
d\omega ^{\prime }=\frac 1{n!}.\sqrt{-d_g^{\prime }}.\varepsilon _{A^{\prime
}}.\omega ^{A^{\prime }}=\frac 1{n!}.\sqrt{-d_g}.\varepsilon _A.d\omega
^A=d\omega \text{ .}
\end{equation}

\subsection{Action of the covariant differential operator on an invariant
volume element}

The action of the covariant differential operator on an invariant volume
element is determined by its action on the elements of the construction of
the invariant volume element (the Levi-Civita symbols, the full
anti-symmetric tensor basis, the metric tensor density). From $d\omega
=\frac 1{n!}.\varepsilon _A.\overline{\omega }^A$ and $\nabla _\xi (d\omega
) $, it follows 
\begin{equation}  \label{VI.8.-1}
\nabla _\xi (d\omega )=\nabla _\xi [\frac 1{n!}.(\varepsilon _A.\overline{%
\omega }^A)]=\frac 1{n!}[(\xi \varepsilon _A).\overline{\omega }%
^A+\varepsilon _A.\nabla _\xi \overline{\omega }^A]\text{ .}
\end{equation}

$\nabla _\xi (d\omega )$ can be written in the form 
\begin{equation}  \label{VI.8.-21}
\nabla _\xi (d\omega )=\frac 12.\overline{g}[\nabla _\xi g].\frac
1{n!}.\varepsilon _A.\overline{\omega }^A=\frac 12.\overline{g}[\nabla _\xi g%
].d\omega \text{ .}
\end{equation}

$\nabla _\xi (d\omega )$ is called \textit{covariant derivative of the
invariant volume element} $d\omega $ along the contravariant vector field $%
\xi $.

\subsection{Action of the Lie differential operator on an invariant volume
element}

The action of the Lie differential operator on an invariant volume element
is determined in analogous way as the action of the covariant differential
operator 
\begin{equation}  \label{VI.8.-22}
\begin{array}{c}
\pounds _\xi (d\omega )=\frac 1{n!}.\pounds _\xi (\varepsilon _A. \overline{%
\omega }^A)=\frac 1{n!}[(\xi \varepsilon _A).\overline{\omega }%
^A+\varepsilon _A.\pounds _\xi \overline{\omega }^A]= \\ 
=\frac 1{n!}.\varepsilon _A.\pounds _\xi \overline{\omega }^A\text{ .}
\end{array}
\end{equation}

After some computation, it follows for $\pounds _\xi (d\omega )$%
\begin{equation}  \label{VI.8.-33}
\begin{array}{c}
\pounds _\xi (d\omega )=\frac 1{n!}.\varepsilon _A.\frac 12. \overline{g}[%
\pounds _\xi g].\overline{\omega }^A=\frac 12.\overline{g}[\pounds _\xi g%
].\frac 1{n!}.\varepsilon _A.\overline{\omega }^A\text{ ,} \\ 
\\ 
\pounds _\xi (d\omega )=\frac 12.\overline{g}[\pounds _\xi g].d\omega \text{
.}
\end{array}
\end{equation}

$\pounds _\xi (d\omega )$ is called \textit{Lie derivative of the invariant
volume element }$d\omega $ along the contravariant vector field $\xi $.

\textit{Special case}: Metric transports $(\nabla _\xi g=0):\nabla _\xi
(d\omega )=0$.

\textit{Special case}: Isometric draggings along (motions) $(\pounds _\xi
g=0):\pounds _\xi (d\omega )=0$.

\begin{center}
\_ . \_
\end{center}

In some cases, when the conservation of the volume is required as an
additional condition, one can introduce a new covariant differential
operator or a new Lie differential operator which do not change the
invariant volume element, i. e. they act on $d\omega $ in an analogous way
as $\nabla _\xi $ and $\pounds _\xi $ act on constant functions.

\subsection{Covariant differential operator preserving the invariant volume
element}

The variation of the invariant volume element $d\omega $ under the action of
the covariant differential operator $\nabla _\xi $%
\[
\nabla _\xi (d\omega )=\frac 12.\overline{g}[\nabla _\xi g].d\omega 
\]

\noindent allows the introduction of a new covariant differential operator $%
^\omega \nabla _\xi $ preserving by its action the invariant volume element.

\begin{definition}
$^\omega \nabla _\xi $ is a {\it covariant differential operator preserving
the invariant volume element }$d\omega $ along a contravariant vector field $\xi $\[
^\omega \nabla _\xi =\nabla _\xi -\frac 12.\overline{g}[\nabla _\xi g]\text{
.} 
\]
\end{definition}
%

The properties of $^\omega \nabla _\xi $ are determined by the properties of
the covariant differential operator and the existence of a covariant metric
tensor field $g$ connected with its contravariant metric tensor field $%
\overline{g}$:

(a) Action on an invariant volume element $d\omega $: 
\begin{equation}  \label{VI.9.-1}
^\omega \nabla _\xi (d\omega )=0\text{ ,}
\end{equation}

It follows from the definition of $^\omega \nabla _\xi $ and (\ref{VI.8.-21}%
).

(b) Action on a contravariant basic vector field: 
\begin{equation}  \label{VI.9.-2}
^\omega \nabla _{\partial _j}\partial _i=(\Gamma _{ij}^k-\frac 12.g^{%
\overline{l}\overline{m}}.g_{lm;j}.g_i^k).\partial _k\text{ .}
\end{equation}

(c) Action on a covariant basic vector field: 
\begin{equation}  \label{VI.9.-4}
^\omega \nabla _{\partial _j}dx^i=(P_{kj}^i-\frac 12.g^{\overline{l}%
\overline{m}}.g_{lm;j}.g_k^i).dx^k\text{ .}
\end{equation}

(d) Action on a function $f$ over $M$%
\begin{equation}  \label{VI.9.-6}
^\omega \nabla _\xi f=\xi f-\frac 12.\overline{g}[\nabla _\xi g].f\text{ ,
\thinspace \thinspace \thinspace \thinspace \thinspace \thinspace \thinspace
\thinspace \thinspace \thinspace \thinspace \thinspace \thinspace \thinspace 
}f\in C^r(M)\text{ , }r\geq 1\text{ .}
\end{equation}

If we introduce the abbreviations 
\begin{equation}  \label{VI.9.-7}
Q_\beta =\overline{g}[\nabla _{e_\beta }g]=g^{\overline{\gamma }\overline{%
\delta }}.g_{\gamma \delta /\beta }\text{ , \thinspace \thinspace \thinspace
\thinspace \thinspace \thinspace \thinspace \thinspace \thinspace \thinspace
\thinspace \thinspace \thinspace \thinspace \thinspace \thinspace }Q_j=%
\overline{g}[\nabla _{\partial _j}g]=g^{\overline{k}\overline{l}}.g_{kl;j}%
\text{ ,}
\end{equation}
\begin{equation}  \label{VI.9.-9}
Q=Q_\beta .e^\beta =Q_j.dx^j\text{ ,}
\end{equation}
\begin{equation}  \label{VI.9.-10}
^\omega \Gamma _{\alpha \beta }^\gamma =\Gamma _{\alpha \beta }^\gamma
-\frac 12.g_\alpha ^\gamma .Q_\beta \text{ , \thinspace \thinspace
\thinspace \thinspace \thinspace \thinspace \thinspace }^\omega P_{\gamma
\beta }^\alpha =P_{\gamma \beta }^\alpha -\frac 12.g_\gamma ^\alpha .Q_\beta 
\text{ ,\thinspace \thinspace \thinspace \thinspace \thinspace }
\end{equation}
\begin{equation}  \label{VI.9.-12}
Q_\xi =\overline{g}[\nabla _\xi g]=Q_\beta .\xi ^\beta =Q_j.\xi
^j=2._c\theta _\xi \text{ ,}
\end{equation}

\noindent then $^\omega \nabla _\xi $, (\ref{VI.9.-2}), and (\ref{VI.9.-4})
can be written in the form 
\begin{equation}
^\omega \nabla _\xi =\nabla _\xi -\frac 12.Q_\xi \text{ ,}  \label{VI.9.-13}
\end{equation}
\begin{equation}
^\omega \nabla _{e_\beta }e_\alpha =\,^\omega \Gamma _{\alpha \beta }^\gamma
.e_\gamma \text{ ,\thinspace \thinspace \thinspace \thinspace \thinspace
\thinspace \thinspace \thinspace \thinspace \thinspace }^\omega \nabla
_{\partial _j}\partial _i=\,^\omega \Gamma _{ij}^k.\partial _k\text{ ,}
\label{VI.9.-14}
\end{equation}
\begin{equation}
^\omega \nabla _{e_\beta }e^\alpha =\,^\omega P_{\gamma \beta }^\alpha
.e^\gamma \text{ , \thinspace \thinspace \thinspace \thinspace \thinspace
\thinspace \thinspace \thinspace \thinspace \thinspace \thinspace \thinspace
\thinspace \thinspace }^\omega \nabla _{\partial _j}dx^i=\,^\omega
P_{kj}^i.dx^k\text{ .}  \label{VI.9.-15}
\end{equation}

$^\omega \Gamma _{\alpha \beta }^\gamma $ are called components of the 
\textit{contravariant affine connection }$^\omega \Gamma $ preserving the
invariant volume element $d\omega $ in a non-co-ordinate basis, $^\omega
P_{\alpha \beta }^\gamma $ are called components of the \textit{covariant
affine connection }$^\omega P$ preserving the invariant volume element $%
d\omega $ in a non-co-ordinate basis.

Since $^\omega \Gamma _{\alpha \beta }^\gamma $ and $^\omega P_{\alpha \beta
}^\gamma $ differ from $\Gamma _{\alpha \beta }^\gamma $ and $P_{\alpha
\beta }^\gamma $ respectively with the components of a mixed tensor field $%
\frac 12.g_\alpha ^\gamma .Q_\beta $ of rank $3$, $^\omega \Gamma $ and $%
^\omega P$ will have the same transformation properties as the affine
connections $\Gamma $ and $P$ respectively.

The action of $^\omega \nabla _\xi $ on a contravariant vector field $u$ can
be written in the form 
\begin{equation}  \label{VI.9.-16}
^\omega \nabla _\xi u=\nabla _\xi u-\frac 12.Q_\xi .u\text{ .}
\end{equation}

If $u$ is considered as a tangential vector field to a curve $x^i(\tau )$,
i. e. 
\begin{equation}  \label{VI.9.-17}
u=\frac d{d\tau }=u^\alpha .e_\alpha =u^i.\partial _i\text{ ,\thinspace
\thinspace \thinspace \thinspace \thinspace \thinspace \thinspace \thinspace 
}u^i=\frac{dx^i}{d\tau }\text{ ,}
\end{equation}
\begin{equation}  \label{VI.9.-18}
u^\alpha =A_i\text{ }^\alpha .u^i=A_i\text{ }^\alpha .\frac{dx^i}{d\tau }%
\text{ , \thinspace \thinspace \thinspace \thinspace \thinspace }e_\alpha
=A_\alpha \text{ }^k.\partial _k\text{ ,\thinspace \thinspace \thinspace
\thinspace \thinspace \thinspace \thinspace }A_i\text{ }^\alpha .A_\alpha 
\text{ }^k=g_i^k\text{ ,}
\end{equation}

\noindent and the parameter $\tau $ is considered as a function of another
parameter $\lambda $ [with one to one (injective) mapping between $\tau $
and $\lambda $], i. e. 
\begin{equation}
\tau =\tau (\lambda )\text{ ,\thinspace \thinspace \thinspace \thinspace
\thinspace \thinspace \thinspace \thinspace }\lambda =\lambda (\tau )\text{ ,%
}  \label{VI.9.-19}
\end{equation}
\begin{equation}
u=\frac d{d\tau }=\frac{d\lambda }{d\tau }.\frac d{d\lambda }=\frac{d\lambda 
}{d\tau }.v\text{ ,\thinspace \thinspace \thinspace \thinspace }v=\frac
d{d\lambda }\text{ ,}  \label{VI.9.-20}
\end{equation}

\noindent then $^\omega \nabla _\xi u$ can be represented by means of the
vector field $v$ and $\nabla _\xi v$ in the form 
\begin{equation}
^\omega \nabla _\xi u=\frac{d\lambda }{d\tau }.\nabla _\xi v+[\xi (\frac{%
d\lambda }{d\tau })-\frac 12.Q_\xi .\frac{d\lambda }{d\tau }].v\text{ .}
\label{VI.9.-21}
\end{equation}

If an additional condition for a relation between $\lambda $ and $\tau $ is
given in the form 
\begin{equation}  \label{VI.9.-22}
\xi (\frac{d\lambda }{d\tau })-\frac 12.Q_\xi .\frac{d\lambda }{d\tau }=0%
\text{ ,}
\end{equation}

\noindent then for an arbitrary vector field $\xi $ a solution for $\lambda
= $ $\lambda (\tau )$ exists in the form 
\begin{equation}
\lambda =\lambda _0+\lambda _1.\int [\exp (\frac 12\int Q_i.dx^i)].d\tau 
\text{ ,\thinspace \thinspace \thinspace \thinspace \thinspace \thinspace
\thinspace \thinspace \thinspace }Q_i=Q_i(x^k)\text{ ,\thinspace \thinspace
\thinspace \thinspace }\lambda _0,\lambda _1=\text{const.}  \label{VI.9.-23}
\end{equation}

\noindent and the connection between $^\omega \nabla _\xi u$ and $\nabla
_\xi v$ is obtained in the form 
\begin{equation}
^\omega \nabla _\xi u=\frac{d\lambda }{d\tau }.\nabla _\xi v=[\lambda
_1.\exp (\frac 12\int Q_i.dx^i)].\nabla _\xi v\text{ , \thinspace \thinspace
\thinspace \thinspace }\lambda _1=\text{const.}  \label{VI.9.-24}
\end{equation}

It follows from the last expression that there is a possibility the action
of $^\omega \nabla _\xi $ on a contravariant vector field $u$ (as a
tangential vector field to a given curve) to be juxtaposed to the action of $%
\nabla _\xi $ on the corresponding to vector field $u$ vector field $v$
(obtained after changing the parameter of the curve). If the vector field $v$
fulfils the condition for an auto-parallel transport along $\xi $, induced
by the covariant differential operator $\nabla _\xi $ ($\nabla _\xi v=0$),
then the vector field $u$ will also fulfil the auto-parallel condition along 
$\xi $ induced by the covariant differential operator $^\omega \nabla _\xi $
($^\omega \nabla _\xi u=0$).

The action of $^\omega \nabla _\xi $ on a metric tensor field $g$ can be
presented in the form 
\begin{equation}  \label{VI.9.-25}
^\omega \nabla _\xi g=\nabla _\xi g-\frac 12.Q_\xi .g\text{ .}
\end{equation}

After contraction of both components of $^\omega \nabla _\xi g$ with $%
\overline{g}$, i. e. for $\overline{g}[^\omega \nabla _\xi g]=g^{\overline{%
\alpha }\overline{\beta }}.(^\omega \nabla _\xi g)_{\alpha \beta }$, the
equality 
\begin{equation}  \label{VI.9.-27}
\overline{g}[^\omega \nabla _\xi g]=(1-\frac n2).Q_\xi \text{ }
\end{equation}

\noindent follows.

The trace free part of \thinspace $^\omega \nabla _\xi g$, 
\begin{equation}  \label{VI.9.-28}
^{\underline{\omega }}\nabla _\xi g=\,^\omega \nabla _\xi g-\frac 1n.%
\overline{g}[^\omega \nabla _\xi g].g\text{ ,}
\end{equation}

\noindent by means of (\ref{VI.9.-27}) can be written in the form 
\begin{equation}
^{\underline{\omega }}\nabla _\xi g=\,^\omega \nabla _\xi g+\frac{n-2}{2n}%
.Q_\xi .g\text{ .}  \label{VI.9.-29}
\end{equation}

Using this form, $^\omega \nabla _\xi g$ can be presented by means of its
trace free part and its trace part in the form 
\begin{equation}  \label{VI.9.-30}
^\omega \nabla _\xi g=\,^{\underline{\omega }}\nabla _\xi g-\frac{n-2}{2n}%
.Q_\xi .g\text{ , }
\end{equation}

\noindent where $\overline{g}[^{\underline{\omega }}\nabla _\xi g]=0$.

\textit{Special case}: $\dim M=n=2:\,^\omega \nabla _\xi g=\,^{\underline{%
\omega }}\nabla _\xi g$ , \thinspace \thinspace \thinspace \thinspace
\thinspace \thinspace \thinspace \thinspace \thinspace \thinspace \thinspace 
$\overline{g}[^\omega \nabla _\xi g]=0$.

\textit{Special case}: $\dim M=n=4:$\thinspace $\,\,\,^\omega \nabla _\xi
g=\,^{\underline{\omega }}\nabla _\xi g-\frac 14.Q_\xi .g$.

The covariant differential operator preserving the invariant volume element 
\textit{does not obey the Leibniz rule} when acting on a tensor product $%
Q\otimes S$ of two tensor fields $Q$ and $S$%
\begin{equation}  \label{VI.9.-33}
\begin{array}{c}
^\omega \nabla _\xi (Q\otimes S)=\,^\omega \nabla _\xi Q\otimes S+Q\otimes
\,^\omega \nabla _\xi S+\frac 12.Q_\xi .Q\otimes S \text{ ,} \\ 
Q\in \otimes ^k\text{ }_l(M)\text{ , }S\in \otimes ^m\text{ }_r(M)\text{ .}
\end{array}
\end{equation}

\subsection{Trace free covariant differential operator. Weyl's transport.
Weyl's space}

The description of the gravitational interaction and its unification with
the other types of interactions over differentiable manifolds with affine
connections and metric [$(L_n,g)$-spaces] induces \cite{Hehl-4} the
introduction of an affine connection with a corresponding covariant
differential operator $^s\nabla _\xi $ constructed by means of $\nabla _\xi $
and $Q_\xi $ in the form 
\begin{equation}  \label{VI.9.-52}
^s\nabla _\xi =\nabla _\xi -\frac 1n.Q_\xi \text{ , \thinspace \thinspace
\thinspace \thinspace \thinspace \thinspace \thinspace }\dim M=n\text{ .}
\end{equation}

The action of $^s\nabla _\xi $ on a covariant metric tensor field $g$ is
determined as 
\begin{equation}  \label{VI.9.-53}
^s\nabla _\xi g=\nabla _\xi g-\frac 1n.Q_\xi .g\text{ ,}
\end{equation}

\noindent obeying the condition 
\begin{equation}
\overline{g}[^s\nabla _\xi g]=0\text{ .}  \label{VI.9.-54}
\end{equation}

On the basis of this relation the covariant differential operator $^s\nabla
_\xi $ is called \textit{trace free covariant differential operator}.

If the transport of $g$ by the trace free covariant differential operator $%
^s\nabla _\xi $ obeys the condition 
\begin{equation}  \label{VI.9.-55}
^s\nabla _\xi g=0\text{ ,}
\end{equation}

\noindent equivalent to the condition for $\nabla _\xi g$%
\begin{equation}
\nabla _\xi g=\frac 1n.Q_\xi .g\text{ ,}  \label{VI.9.-56}
\end{equation}

\noindent then the transport is called \textit{Weyl's transport.}

The covariant vector field [s. (\ref{VI.9.-9})] 
\begin{equation}  \label{VI.9.-57}
\overline{Q}=\frac 1n.Q
\end{equation}

\noindent is called \textit{Weyl's covector field}.

A differentiable manifold $M$ $(\dim M=n)$ with affine connection and
metric, over which for every contravariant vector field $\xi \in T(M)$ the
transport of $g$ is a Weyl transport, is called \textit{Weyl's space with
torsion } (\textit{Weyl-Cartan space}) $Y_n$ \cite{Hehl-4}.

The trace free covariant differential operator $^s\nabla _\xi $ is connected
with the covariant differential operator $^\omega \nabla _\xi $ preserving
the invariant volume element $d\omega $ through the relation 
\begin{equation}  \label{VI.9.-58}
^\omega \nabla _\xi =\,^s\nabla _\xi -\frac{n-2}{2n}.Q_\xi =\nabla _\xi
-\frac 12.Q_\xi \text{ .}
\end{equation}

The action of the two operators $^\omega \nabla _\xi $ and $^s\nabla _\xi $
would be identical, if $\dim M=n=2$ ($Q_\xi \neq 0$) or if $Q_\xi =0$.

The components $\Gamma _{\beta \gamma }^\alpha $ of the affine connection $%
\Gamma $ can be represented by means of the components of the affine
connections corresponding to the operators $^\omega \nabla _\xi $ and $%
^s\nabla _\xi $.

$\nabla _{e_\beta }e_\alpha $ can be written in the form 
\begin{equation}  \label{VI.9.-59}
\nabla _{e_\beta }e_\alpha =\frac 12(\nabla _{e_\alpha }e_\beta +\nabla
_{e_\beta }e_\alpha -[e_\alpha ,e_\beta ])-\frac 12.T(e_\alpha ,e_\beta )%
\text{ ,}
\end{equation}

\noindent corresponding to the representation of $\Gamma _{\alpha \beta
}^\gamma $ in the form 
\begin{equation}
\Gamma _{\alpha \beta }^\gamma =\frac 12(\Gamma _{\alpha \beta }^\gamma
+\Gamma _{\beta \alpha }^\gamma -C_{\alpha \beta }\text{ }^\gamma )-\frac
12.T_{\alpha \beta }\,^\gamma =\overline{\Gamma }_{\alpha \beta }^\gamma
-\frac 12.T_{\alpha \beta }\,^\gamma \text{ .}  \label{VI.9.-60}
\end{equation}

If we introduce the abbreviations 
\begin{equation}  \label{VI.9.-61}
^g\nabla _{e_\beta }e_\alpha =\frac 12(\nabla _{e_\alpha }e_\beta +\nabla
_{e_\beta }e_\alpha -[e_\alpha ,e_\beta ])=\overline{\Gamma }_{\alpha \beta
}^\gamma .e_\gamma \text{ ,}
\end{equation}
\begin{equation}  \label{VI.9.-62}
^s\nabla _{e_\beta }e_\alpha =Q_{\alpha \beta }\text{ }^\gamma .e_\gamma
=(\Gamma _{\alpha \beta }^\gamma -\frac 1n.g_\alpha ^\gamma .Q_\beta
).e_\gamma \text{ ,}
\end{equation}

\noindent then 
\begin{equation}
\nabla _{e_\beta }e_\alpha =\,^s\nabla _{e_\beta }e_\alpha +\frac 1n.Q_\beta
.e_\alpha \text{ ,}  \label{VI.9.-63}
\end{equation}
\begin{equation}
\nabla _{e_\beta }e_\alpha =\,^g\nabla _{e_\beta }e_\alpha +\frac
12.T(e_\beta ,e_\alpha )\text{ .}  \label{VI.9.-64}
\end{equation}

From (\ref{VI.9.-59}), (\ref{VI.9.-61}), and (\ref{VI.9.-63}), it follows
that 
\begin{equation}  \label{VI.9.-65}
\nabla _{e_\beta }e_\alpha =\frac 12[^g\nabla _{e_\beta }e_\alpha +\frac
12.T(e_\beta ,e_\alpha )+\frac 1n.Q_\beta .e_\alpha +\,^s\nabla _{e_\beta
}e_\alpha ]\text{ .}
\end{equation}

The last equality corresponds to the representation of $\Gamma _{\alpha
\beta }^\gamma $ in the form 
\begin{equation}  \label{VI.9.-66}
\Gamma _{\alpha \beta }^\gamma \equiv \frac 12(\overline{\Gamma }_{\alpha
\beta }^\gamma -\frac 12.T_{\alpha \beta }\,^\gamma +\frac 1n.g_\alpha
^\gamma .Q_\beta +Q_{\alpha \beta }\text{ }^\gamma )\text{ .}
\end{equation}

In analogous way, using the relations 
\begin{equation}  \label{VI.9.-67}
\begin{array}{c}
^\omega \nabla _{e_\beta }e_\alpha =\,^\omega \Gamma _{\alpha \beta }^\gamma
.e_\gamma =(\Gamma _{\alpha \beta }^\gamma -\frac 12.g_\alpha ^\gamma
.Q_\beta ).e_\gamma = \\ 
=\nabla _{e_\beta }e_\alpha -\frac 12.Q_\beta .e_\alpha \text{ ,}
\end{array}
\end{equation}
\[
\begin{array}{c}
\nabla _{e_\beta }e_\alpha =\,^\omega \nabla _{e_\beta }e_\alpha +\frac
12.Q_\beta .e_\alpha \text{ ,} \\ 
\nabla _{e_\beta }e_\alpha =\,^g\nabla _{e_\beta }e_\alpha +\frac
12.T(e_\beta ,e_\alpha )\text{ ,}
\end{array}
\]

\noindent one can obtain for $\nabla _{e_\beta }e_\alpha $%
\begin{equation}
\nabla _{e_\beta }e_\alpha =\frac 12[^g\nabla _{e_\beta }e_\alpha +\frac
12.T(e_\beta ,e_\alpha )+\,^\omega \nabla _{e_\beta }e_\alpha +\frac
12.Q_\beta .e_\alpha ]\text{ .}  \label{VI.9.-68}
\end{equation}

The last equality is equivalent to the representation of $\Gamma _{\alpha
\beta }^\gamma $ in the form 
\begin{equation}  \label{VI.9.-69}
\Gamma _{\alpha \beta }^\gamma =\frac 12(\overline{\Gamma }_{\alpha \beta
}^\gamma -\frac 12.T_{\alpha \beta }\,^\gamma +\frac 12.g_\alpha ^\gamma
.Q_\beta +\,^\omega \Gamma _{\alpha \beta }^\gamma )\text{ .}
\end{equation}

From (\ref{VI.9.-63}) and (\ref{VI.9.-64}), the connection between $^g\nabla
_{e_\beta }e_\alpha $ and $^s\nabla _{e_\beta }e_\alpha $ follows in the
form 
\begin{equation}  \label{VI.9.-70}
^g\nabla _{e_\beta }e_\alpha =\,^s\nabla _{e_\beta }e_\alpha -\frac
12.T(e_\beta ,e_\alpha )+\frac 1n.Q_\beta .e_\alpha \text{ ,}
\end{equation}

\noindent equivalent to the connection between $\overline{\Gamma }_{\alpha
\beta }^\gamma $ and $Q_{\alpha \beta }{}^\gamma $%
\begin{equation}
\overline{\Gamma }_{\alpha \beta }^\gamma =Q_{\alpha \beta }\text{ }^\gamma
+\frac 12.T_{\alpha \beta }\,^\gamma +\frac 1n.g_\alpha ^\gamma .Q_\beta 
\text{ .}  \label{VI.9.-71}
\end{equation}

On the other side, there is a connection between $^g\nabla _{e_\beta
}e_\alpha $ and $^\omega \nabla _{e_\beta }e_\alpha $%
\begin{equation}  \label{VI.9.-72}
^g\nabla _{e_\beta }e_\alpha =\,^\omega \nabla _{e_\beta }e_\alpha -\frac
12.T(e_\beta ,e_\alpha )+\frac 12.Q_\beta .e_\alpha \text{ ,}
\end{equation}

\noindent corresponding to the connection between $\overline{\Gamma }%
_{\alpha \beta }^\gamma $ and $^\omega \Gamma _{\alpha \beta }^\gamma $%
\begin{equation}
\overline{\Gamma }_{\alpha \beta }^\gamma =\,^\omega \Gamma _{\alpha \beta
}^\gamma +\frac 12.T_{\alpha \beta }\,^\gamma +\frac 12.g_\alpha ^\gamma
.Q_\beta \text{ .}  \label{VI.9.-73}
\end{equation}

\subsection{Lie differential operator preserving the invariant volume element
}

The action of the Lie differential operator $\pounds _\xi $ on the invariant
volume element $d\omega $%
\[
\pounds _\xi (d\omega )=\frac 12.\overline{g}[\pounds _\xi g].d\omega \text{
,} 
\]

\noindent allows the construction of a new Lie differential operator
preserving the invariant volume element $d\omega $.

\begin{definition}
$^\omega \pounds _\xi $ := Lie differential operator preserving the
invariant volume element $d\omega $ along a contravariant vector field $\xi $\[
^\omega \pounds _\xi =\pounds _\xi -\frac 12.\overline{g}[\pounds _\xi g]\text{ .} 
\]
\end{definition}
%

The properties of $^\omega \pounds _\xi $ are determined by the properties
of the Lie differential operator and the existence of a covariant metric
tensor field $g$ connected with a contravariant metric tensor field $%
\overline{g}$:

(a) Action on the invariant volume element $d\omega $: 
\begin{equation}  \label{VI.9.-86}
^\omega \pounds _\xi (d\omega )=0\text{ .}
\end{equation}

It follows from the definition of $^\omega \pounds _\xi $ and (\ref{VI.8.-33}%
).

(b) Action on a contravariant basis vector field: 
\begin{equation}  \label{VI.9.-87}
\begin{array}{c}
^\omega \pounds _{e_\alpha }e_\beta =\pounds _{e_\alpha }e_\beta -\frac 12. 
\overline{g}[\pounds _{e_\alpha }g].e_\beta = \\ 
=(C_{\alpha \beta }\text{ }^\gamma -\frac 12.g^{\overline{\rho }\overline{%
\sigma }}.\pounds _{e_\alpha }g_{\rho \sigma }.g_\beta ^\gamma ).e_\gamma 
\text{ ,}
\end{array}
\end{equation}
\begin{equation}  \label{VI.9.-88}
^\omega \pounds _{\partial _i}\partial _j=-\frac 12.g^{\overline{k}\overline{%
l}}.\pounds _{\partial _i}g_{kl}.\partial _j\text{ .}
\end{equation}

(c) Action on a covariant basis vector field: 
\begin{equation}  \label{VI.9.-89}
\begin{array}{c}
^\omega \pounds _{e_\alpha }e^\beta =\pounds _{e_\alpha }e^\beta -\frac 12. 
\overline{g}[\pounds _{e_\alpha }g].e^\beta = \\ 
=k_{\gamma \alpha }\text{ }^\beta .e^\gamma -\frac 12.\overline{g}[\pounds
_{e_\alpha }g].e^\beta \text{ ,}
\end{array}
\end{equation}
\begin{equation}  \label{VI.9.-90}
^\omega \pounds _{\partial _i}dx^j=k_{mi}\text{ }^j.dx^m-\frac 12.\overline{g%
}[\pounds _{\partial _i}g].dx^j\text{ .}
\end{equation}

(d) Action on a function $f$: 
\begin{equation}  \label{VI.9.-91}
^\omega \pounds _\xi f=\xi f-\frac 12.\overline{g}[\pounds _\xi g].f\text{
,\thinspace \thinspace \thinspace \thinspace \thinspace \thinspace
\thinspace }f\in C^r(M)\text{ ,\thinspace \thinspace }r\geq 1\text{ .}
\end{equation}

If we introduce the abbreviations 
\begin{equation}  \label{VI.9.-92}
P_\beta =\overline{g}[\pounds _{e_\beta }g]=g^{\overline{\gamma }\overline{%
\delta }}.\pounds _{e_\beta }g_{\gamma \delta }\text{ ,}
\end{equation}
\begin{equation}  \label{VI.9.-93}
P_j=\overline{g}[\pounds _{\partial _j}g]=g^{\overline{k}\overline{l}%
}.\pounds _{\partial _j}g_{kl}\text{ ,}
\end{equation}
\begin{equation}  \label{VI.9.-94}
P=P_\beta .e^\beta =P_j.dx^j\text{ ,}
\end{equation}
\begin{equation}  \label{VI.9.-95}
P_\xi =\overline{g}[\pounds _\xi g]=2._l\theta _\xi \text{ ,}
\end{equation}
\begin{equation}  \label{VI.9.-96}
\widehat{C}_{\alpha \beta }\text{ }^\gamma =C_{\alpha \beta }\text{ }^\gamma
-\frac 12.P_\alpha .g_\beta ^\gamma \text{ , }\widehat{C}_{\alpha \beta }%
\text{ }^\gamma \neq -\widehat{C}_{\beta \alpha }\text{ }^\gamma \text{ ,}
\end{equation}

\noindent then $^\omega \pounds _\xi $ and (\ref{VI.9.-87}) $\div $ (\ref
{VI.9.-91}) can be written in the form 
\begin{equation}
^\omega \pounds _\xi =\pounds _\xi -\frac 12.P_\xi \text{ ,}
\label{VI.9.-97}
\end{equation}
\begin{equation}
\begin{array}{c}
^\omega \pounds _{e_\alpha }e_\beta =\pounds _{e_\alpha }e_\beta -\frac
12.P_\alpha .e_\beta = \\ 
=(C_{\alpha \beta }\text{ }^\gamma -\frac 12.g_\beta ^\gamma .P_\alpha
).e_\gamma =\widehat{C}_{\alpha \beta }\text{ }^\gamma .e_\gamma \text{ ,}
\end{array}
\label{VI.9.-98}
\end{equation}
\begin{equation}
^\omega \pounds _{\partial _i}\partial _j=-\frac 12.P_i.\partial _j\text{ ,}
\label{VI.9.-99}
\end{equation}
\begin{equation}
\begin{array}{c}
^\omega \pounds _{e_\alpha }e^\beta =\pounds _{e_\alpha }e^\beta -\frac
12.P_\alpha .e^\beta = \\ 
=k_{\gamma \alpha }\text{ }^\beta .e^\gamma -\frac 12.P_\alpha .e^\beta 
\text{ ,}
\end{array}
\label{VI.9.-100}
\end{equation}
\begin{equation}
^\omega \pounds _{\partial _i}dx^j=k_{mi}\text{ }^j.dx^m-\frac 12.P_i.dx^j%
\text{ ,}  \label{VI.9.-101}
\end{equation}
\begin{equation}
^\omega \pounds _\xi f=\xi f-\frac 12.P_\xi .f\text{ ,\thinspace \thinspace
\thinspace \thinspace \thinspace \thinspace \thinspace }f\in C^r(M)\text{
,\thinspace \thinspace \thinspace \thinspace \thinspace \thinspace
\thinspace }r\geq 1\text{ .}  \label{VI.9.-102}
\end{equation}

The commutator of two Lie differential operators preserving $d\omega $ has
the following properties:

(a) Action on a function $f$: 
\begin{equation}  \label{VI.9.-103}
\begin{array}{c}
\lbrack ^\omega \pounds _\xi ,\,^\omega \pounds _u]f=(\pounds _\xi u)f+\frac
12(uP_\xi -\xi P_u)f= \\ 
=[\pounds _\xi u+\frac 12(uP_\xi -\xi P_u)]f\text{ ,\thinspace \thinspace
\thinspace \thinspace \thinspace \thinspace \thinspace \thinspace \thinspace
\thinspace }f\in C^r(M)\text{ ,\thinspace \thinspace \thinspace \thinspace
\thinspace }r\geq 2\text{ .}
\end{array}
\end{equation}

(b) Action on a contravariant vector field: 
\begin{equation}  \label{VI.9.-104}
\begin{array}{c}
\lbrack ^\omega \pounds _\xi ,\,^\omega \pounds _u]v=[\pounds _\xi ,\pounds
_u]v+\frac 12(uP_\xi -\xi P_u)v= \\ 
=\{[\pounds _\xi ,\pounds _u]+\frac 12(uP_\xi -\xi P_u)\}v\text{ ,\thinspace
\thinspace \thinspace \thinspace \thinspace \thinspace \thinspace \thinspace
\thinspace \thinspace \thinspace \thinspace }\xi ,u,v\in T(M)\text{ .}
\end{array}
\end{equation}

(c) Satisfies the Jacobi identity 
\begin{equation}  \label{VI.9.-105}
\begin{array}{c}
<[[^\omega \pounds _\xi ,\,^\omega \pounds _u],\,^\omega \pounds
_v]>=[[^\omega \pounds _\xi ,\,^\omega \pounds _u],\,^\omega \pounds _v]+ \\ 
+[[^\omega \pounds _v,\,^\omega \pounds _\xi ],\,^\omega \pounds
_u]+[[^\omega \pounds _u,\,^\omega \pounds _v],\,^\omega \pounds _\xi
]\equiv 0\text{ .}
\end{array}
\end{equation}

The different types of differential operators acting on the invariant volume
element can be used for description of different physical systems and
interactions over a differentiable manifold with affine connections and
metric interpreted as a model of the space-time.

\section{Conclusions}

The main conclusions following from the obtained results could be grouped
together in the following statements:

\begin{enumerate}
\item[1]  A contraction operator $S$, commuting with the covariant
differential operator and with the Lie differential operator, for which the
affine connection $P$ determined by the covariant differential operator for
covariant tensor fields is different (not only by sign) from the affine
connection $\Gamma $ determined by the covariant differential operator for
contravariant tensor fields can be introduced over every differentiable
manifold. The components (in a co-ordinate or in a non-co-ordinate basis) of
the both affine connections $P_{jk}^i$ and $\Gamma _{jk}^i$ differ to each
other by the components $g_{j;k}^i$ of the covariant derivatives of the
Kronecker tensor. At that, at least three cases could be distinguished:
\end{enumerate}

\begin{description}
\item[(a)]  $g_{j;k}^i:=0:P_{jk}^i+\Gamma _{jk}^i=0$ , [$P_{jk}^i$ differs
only by sign from $\Gamma _{jk}^i$ (canonical case: $S:=C$)],

\item[(b)]  $g_{j;k}^i:=\varphi _{,k}.g_j^i:P_{jk}^i+\Gamma _{jk}^i=\varphi
_{,k}.g_j^i$ , $\varphi \in C^r(M)$, [$P_{jk}^i$ differs from $\Gamma
_{jk}^i $ by the derivative of an invariant function $\varphi \in C^r(M)$, $%
r\geq 2$, along a basis vector field $(\partial _k$ or $e_k)$, and the
components of the Kronecker tensor in the given basis],

\item[(c)]  $g_{j;k}^i=q_{jk}^i:P_{jk}^i+\Gamma _{jk}^i=q_{jk}^i$, $q\in
\otimes ^1$ $_2(M)$, [$P_{jk}^i$ differs from $\Gamma _{jk}^i$ by the
covariant derivative $g_{j;k}^i$ of the Kronecker tensor along a basis
vector field $\partial _k$ (or $e_k$)].
\end{description}

In the cases (b) and (c) the Lie derivatives of covariant tensor fields
depend also on structures determined by the affine connections in contrast
to the case (a), where the covariant derivative and the Lie derivative of
covariant tensor fields are independent of each other structures (although
the fact that the Lie derivatives can be expressed by means of the covariant
derivatives).

On the grounds of the obtained results the kinematics of vector fields has
been work out \cite{Manoff-6}, \cite{Manoff-7} - \cite{Manoff-10}. The
Lagrangian theory for tensor fields has been considered \cite{Manoff-11} and
applied to the Einstein theory of gravitation as a special case of a
Lagrangian theory of tensor fields \cite{Manoff-12} over $V_n$-spaces ($n=4$%
).

\begin{center}
\textsc{Acknowledgments}
\end{center}

The author is grateful to:

- Prof. Dr. N. A. Chernikov, Dr. N. S. Shavokhina and Dr. A. B. Pestov from
the Joint Institute for Nuclear Research (JINR) - Dubna (Russia) for the
helpful discussions;

- Prof. Dr. St. Dimiev from the Institute for Mathematics and Informatics of
the Bulgarian Academy of Sciences and Prof. Dr. K. Sekigawa from the
Department of Mathematics of the Faculty of Science at the Niigata
University (Japan) for their support of the considered topics;

- Prof. D. I. Kazakov for his kind hospitality during my stay at the JINR in
Dubna.

This work is supported in part by the National Science Foundation of
Bulgaria under Grants No. F-103, F-498, F-642.

\small %

\end{document}